\documentclass[fleqn,usenatbib]{mnras}

\usepackage{newtxtext,newtxmath}

\usepackage[T1]{fontenc}
\usepackage{ae,aecompl}

\usepackage{graphicx}	
\usepackage{amsmath}	
\usepackage{amssymb}	
\usepackage{subfigure}
\usepackage{comment}
\usepackage{threeparttable}

\title[On the $\alpha$-Intensity Correlation in GRBs]{On the $\alpha$-Intensity Correlation in Gamma-Ray Bursts: \\ Subphotospheric Heating with Varying Entropy}

\author[F. Ryde et al.]{Felix Ryde$^{1}$,
Hoi-Fung Yu$^{1}$,
H\"usne Dereli-B\'egu\'e$^{1,2}$,
Christoffer Lundman$^{1}$,
\newauthor
Asaf Pe'er$^{3, 4}$, 
Liang Li$^{1,5}$
\\
$^{1}$Department of Physics, KTH Royal Institute of Technology, and the Oskar Klein Centre for Cosmoparticle Physics,\\ 10691 Stockholm, Sweden\\
$^{2}$ Max-Planck-Institut f\"ur extraterrestrische Physik, Giessenbachstr. 1, 85748 Garching, Germany\\
$^{3}$Department of Physics, University College Cork, Cork, Ireland\\
$^{4}$Department of Physics, Bar-Ilan University, Ramat-Gan 52900, Israel\\
$^{5}$International Center of Relativistic Astrophysics Network, Piazza della Repubblica 10, 65122 Pescara, Italy
}

\date{Accepted ddmmyy. Received ddmmyy; in original form ddmmyy}

\pubyear{2018}

\hypersetup{draft}

\begin{document}
\label{firstpage}
\pagerange{\pageref{firstpage}--\pageref{lastpage}}
\maketitle

\begin{abstract}

The emission mechanism during the prompt phase in gamma-ray bursts (GRBs) can be investigated through correlations between spectral properties. Here, we revisit the correlation relating the instantaneous flux, $F$, and the photon index below the spectral break, $\alpha$, in individual emission pulses, by studying the 38 most prominent pulses in the {\it Fermi}/GBM GRB catalogue. First, we search for signatures of the bias in the determination of $\alpha$ due to the limited spectral coverage (window effect) expected in the synchrotron case. The absence of such a characteristic signature argues against the simplest synchrotron models. 
We instead find that the observed correlation between $F$ and $\alpha$ {  can, in general, be described by} the relation $F(t) \propto {\rm e}^{k\,\alpha(t)}$, for which the median $k = 3$. 
We suggest that this correlation is a manifestation of subphotospheric heating in a flow with a varying entropy. Around the peak of the light curve,  a large entropy causes the photosphere to approach the saturation radius, leading to an intense emission with a narrow spectrum. As the entropy decreases the photosphere secedes from  the saturation radius, and weaker emission with a broader spectrum is expected. This simple scenario naturally leads to a correlated variation of the intensity and spectral shape, covering the observed range.   
\end{abstract}




\section{Introduction}\label{sec:intro}

Even though most of the energy released by a gamma-ray burst (GRB) 
is emitted during the prompt emission phase, the emission mechanism 
is still not understood. In order to determine the emission 
process typically the low-energy spectral index, $\alpha$, 
of the GRB spectrum is analysed. This analysis has not given 
a conclusive answer since the $\alpha$-distribution is broad 
and has not been uniquely explained by a single emission process. 
{ For instance, the peak of the distribution has been used as an argument for synchrotron emission since it is close to the expected value. However, a large fraction ($\sim 28\%$) was found to be inconsistent with the theoretical limit of $-2/3$ 
\citep["line of death"; ][]{Preece1998, Ghirlanda2002, Goldstein2016, Guiriec2015_New_Model, Yu2016}; and only specific physical scenarios remain plausible \citep[large emitting radii and Lorentz factors of the flow, ][]{Beniamini&Piran2013, Iyyani2016, Beniamini2018, Burgess2018}}.  Alternatively, the spectral width or 
{ sharpness} angle has been used as a tool to characterise the 
spectrum, which also take into account the high-energy spectral 
slope \citep{Axelsson2015,Yu2015a}. { Around 80\% of the bursts were found to have fitted Band spectra that are narrower than what was expected for synchrotron emission. However, direct fitting with a synchrotron model decreases this fraction \citep{Burgess2017}}. 
Therefore, a firm conclusion based on the spectral width 
{ or sharpness angle} alone cannot either be reached. 
Yet another approach has been to study the correlation between spectral 
parameters. A well studied and strong correlation is the 
Golenetskii correlation \citep{golenetski83} which relates 
the instantaneous flux, $F$, and peak of the spectrum $E_{\rm pk}$ 
\citep{Kargatis1994,Borgonovo2001,Lu2012}. Another prominent 
correlation is between $E_{\rm pk}$ and $\alpha$ which is valid { in a fraction of bursts \citep{Crider1997,Kaneko2006}. Both of these correlations show a variety of behaviours. In some individual pulses the spectral parameters track each other, 
while in others the correlation is different during the rise and decay phases 
of the pulses. The variety of behaviours have complicated any physical explanation.} 

{Several early studies showed that the flux, $F$ 
and $\alpha$ have similar evolutions in GRBs} \citep[e.g.][]{Crider1997, Lloyd2002, Basak&Rao2014}. In particular, \citet{Ghirlanda2002} found that the synchtrotron limit is mainly violated around the emission peak. 
{ By analysing a sample of 38 individual pulses, \citet{Yu2018} found that the $F$-$\alpha$ relation is the one of all the relations that is most often valid.}  
In this paper, we therefore revisit and study this relation {within pulses,} in search of a functional correlation and a physical explanation. 


\section{Sample Selection and Method} \label{sec:data}

{ We study GRBs observed by the {\it Fermi} Gamma-ray Space Telescope 
and its Gamma-ray Burst Monitor (GBM). {We  revisit the sample defined in \citet{Yu2018}, which} makes use of the {\it Fermi} 
GBM burst catalogue published at HEASARC\footnote{www.heasarc.gsfc.gov/}. 
The sample consists of the 38 individual pulse structures that are presented in 
Table~\ref{tab:table1}. This sample is useful for investigation of spectral correlations, since it has been shown that the clearest parameter correlation 
appears in individual pulse structures in the light curve \citep[e.g.,][]{Kargatis1994, Crider1997,Borgonovo2001}.
In order to properly capture the temporal evolution of the properties of the emission,
the light curves were rebinned using the Bayesian block method \citep{Scargle2013}. 
This provides time bins over which any evolution is small, and 
therefore they can be integrated without the loss of ability to study the 
instantaneous emission properties. Furthermore, since detailed spectral 
and temporal analysis is performed, it is important to ensure enough photons are contained in each Bayesian block time bin. 
Therefore, it was required that the time bins have a statistical significance $S \geq 20$ and only the pulses that have at least 5 such time bins were selected \citep[see][for detailed discussion on various definition of statistical significance measures]{Vianello2018}.
}


\citet{Yu2018} used the Bayesian spectral fitting package {\tt 3ML} \citep{3ML} to perform time-resolved spectral fits, and followed the standard analysis procedure for data, energy, and background selections \citep{Goldstein2012,Gruber2014,Yu2016}. In particular, they used two empirical models {with a Poisson likelihood to fit the data.} The first model was a cutoff power law, 
\begin{equation}
N_{E}= A \, \left(\frac{E}{E_{\rm piv}}\right)^{\alpha}\, {\rm e}^{-E/E_{\rm c}},
\label{eq:cpl}
\end{equation}
where $N_{E}$ is the photon flux, $A$ is the normalisation for the 
spectral fit, $E_{\rm piv}$ is the pivot energy fixed at 100~keV,\footnote{The pivot energy is just where the fit is levered. It must be placed within and close to the centre 
of the fitting range. The GBM has an energy range of $\sim 8$~keV to 
$\sim 40$~MeV, therefore for all GBM related works it is fixed at 
100~keV in every fitting package available.} $E$ is the 
photon energy, and $E_{\rm c}$ is the break energy. { The priors used are as follows:}
\begin{equation} \label{eqn:priors}
\begin{cases}
	A \sim \log\mathcal{U}(10^{-11},10^2)\\
	\alpha \sim \mathcal{U}(-3,2)\\
	E_{\rm c} \sim \mathcal{U}(10,10^4)
\end{cases}
\end{equation}
{where $\mathcal{U}$ denotes an uniform distribution.\footnote{The symbol ``$\sim$'' denotes ``distributes as''.}} This model was chosen since, in the GBM GRB time-resolved spectral catalogue, it is the best model for a majority of bursts \citep{Yu2016}. The second model was the Band function \citep{Band1993}. Indeed, they found  for many time bins that the Band function is not constrained, even though the statistical significance $S \geq 20$ \citep{Vianello2018} was required. 

{ We also note that other models have been used in the past, in particular, including multiple spectral components \citep[e.g,][]{Ryde2005,RydePeer2009,Guiriec2011, Axelsson2012, Burgess2014b, Iyyani2015,Yu2015b,Nappo2017} and spectral breaks \citep[e.g.][]{Barat2000, Oganesyan2017,Oganesyan2018,Ravasio2018} for which the physical interpretation is different.}

\begin{figure*}
\centering

\subfigure{\includegraphics[width=0.33\linewidth]{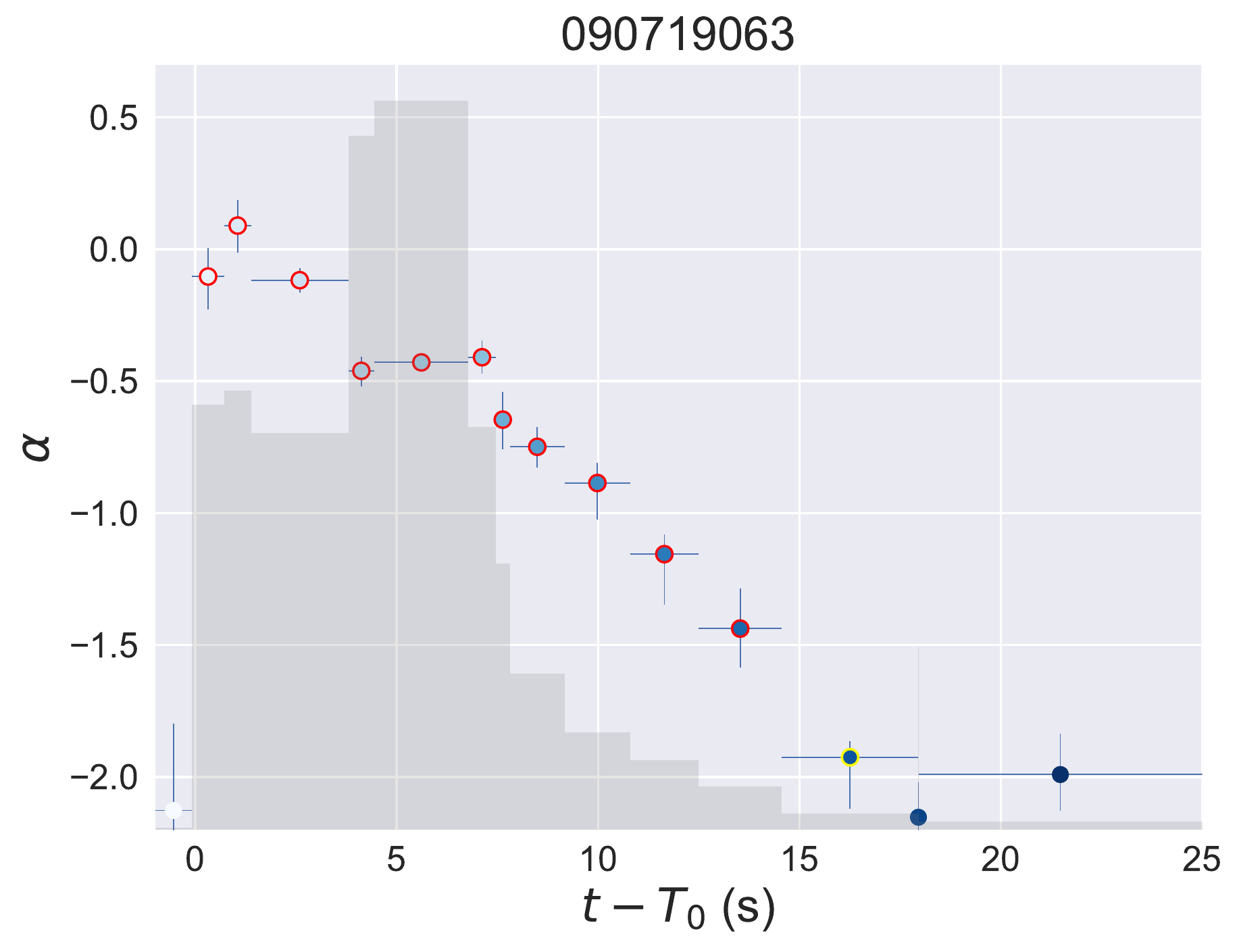}}
\subfigure{\includegraphics[width=0.33\linewidth]{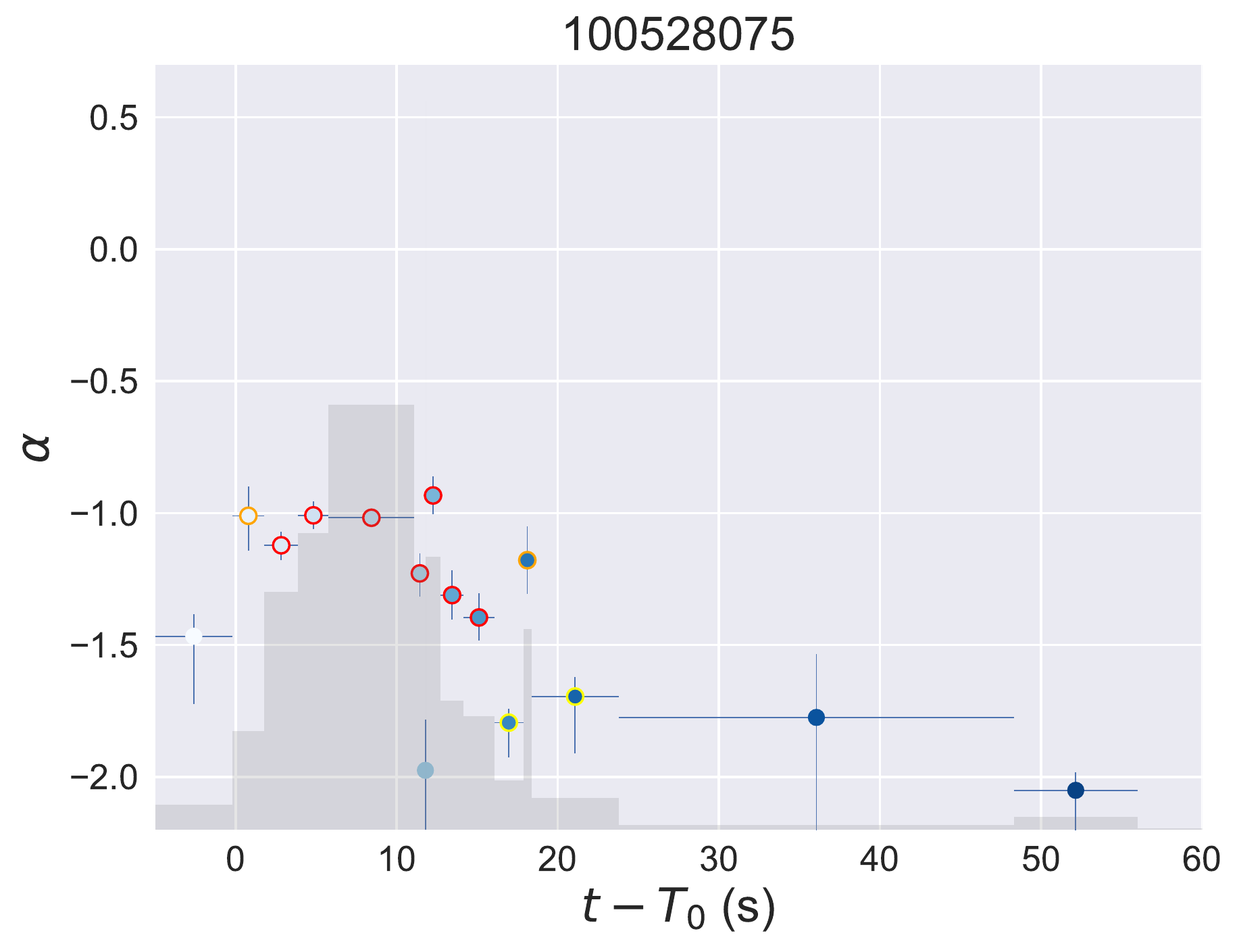}}
\subfigure{\includegraphics[width=0.33\linewidth]{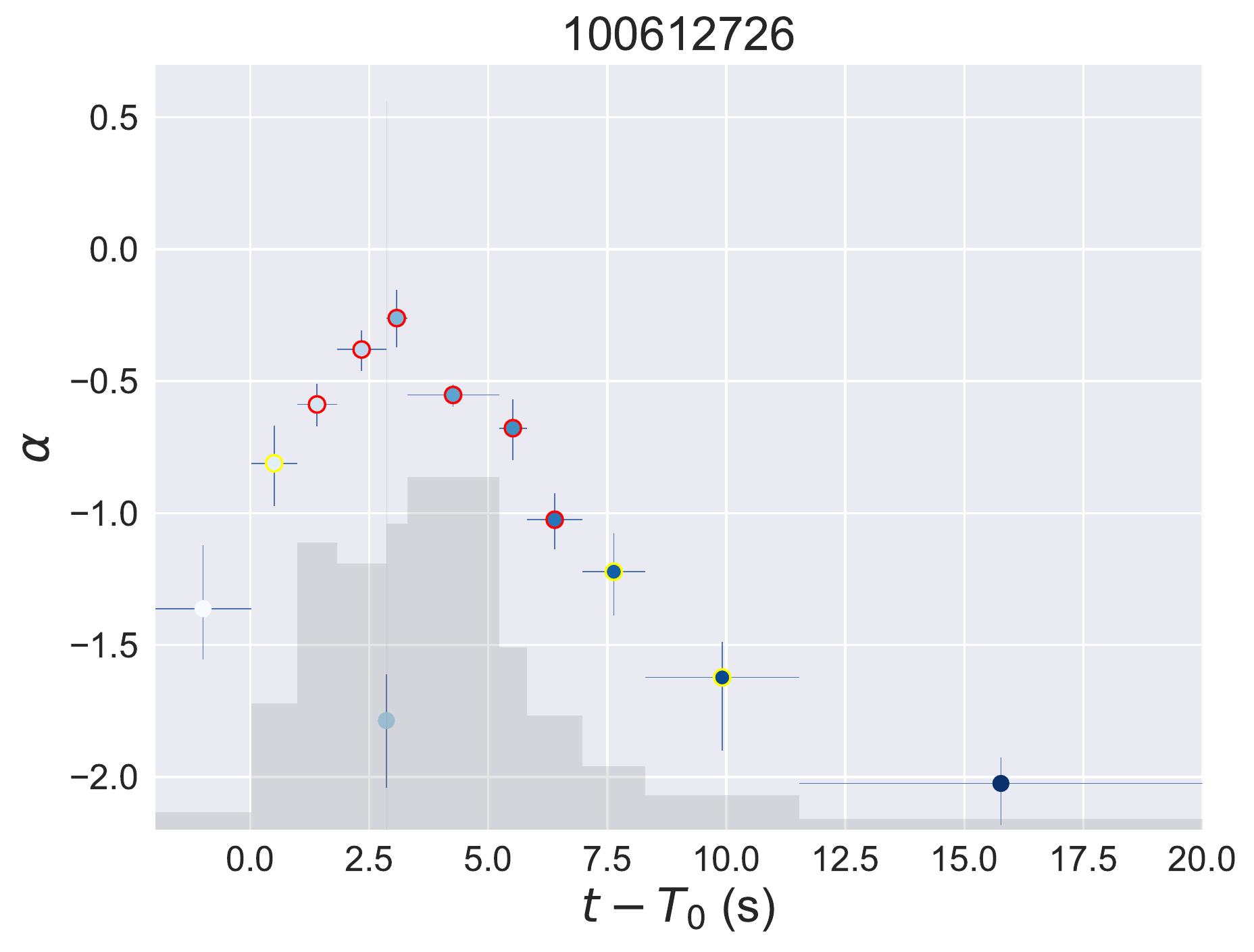}}

\subfigure{\includegraphics[width=0.33\linewidth]{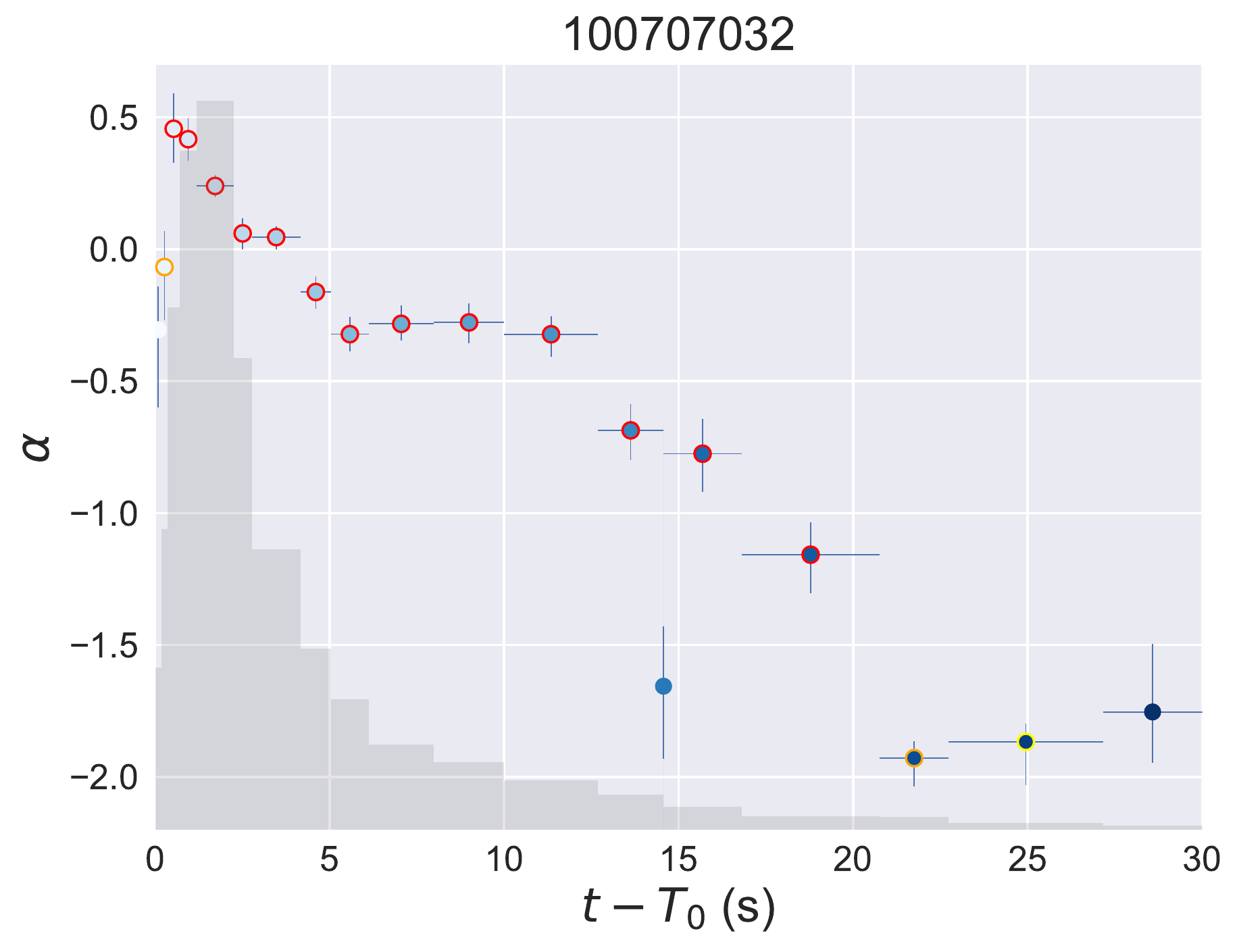}}
\subfigure{\includegraphics[width=0.33\linewidth]{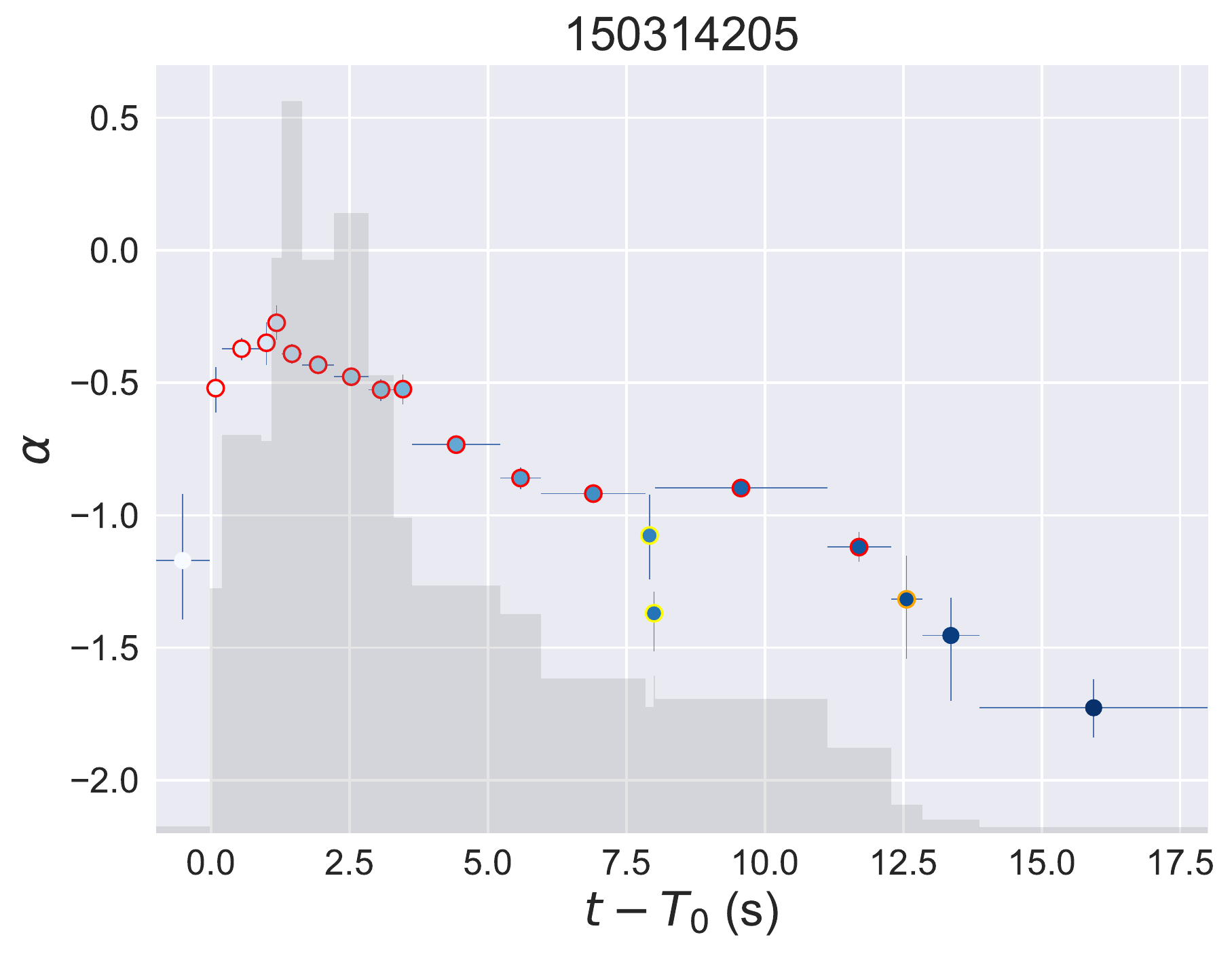}}
\subfigure{\includegraphics[width=0.33\linewidth]{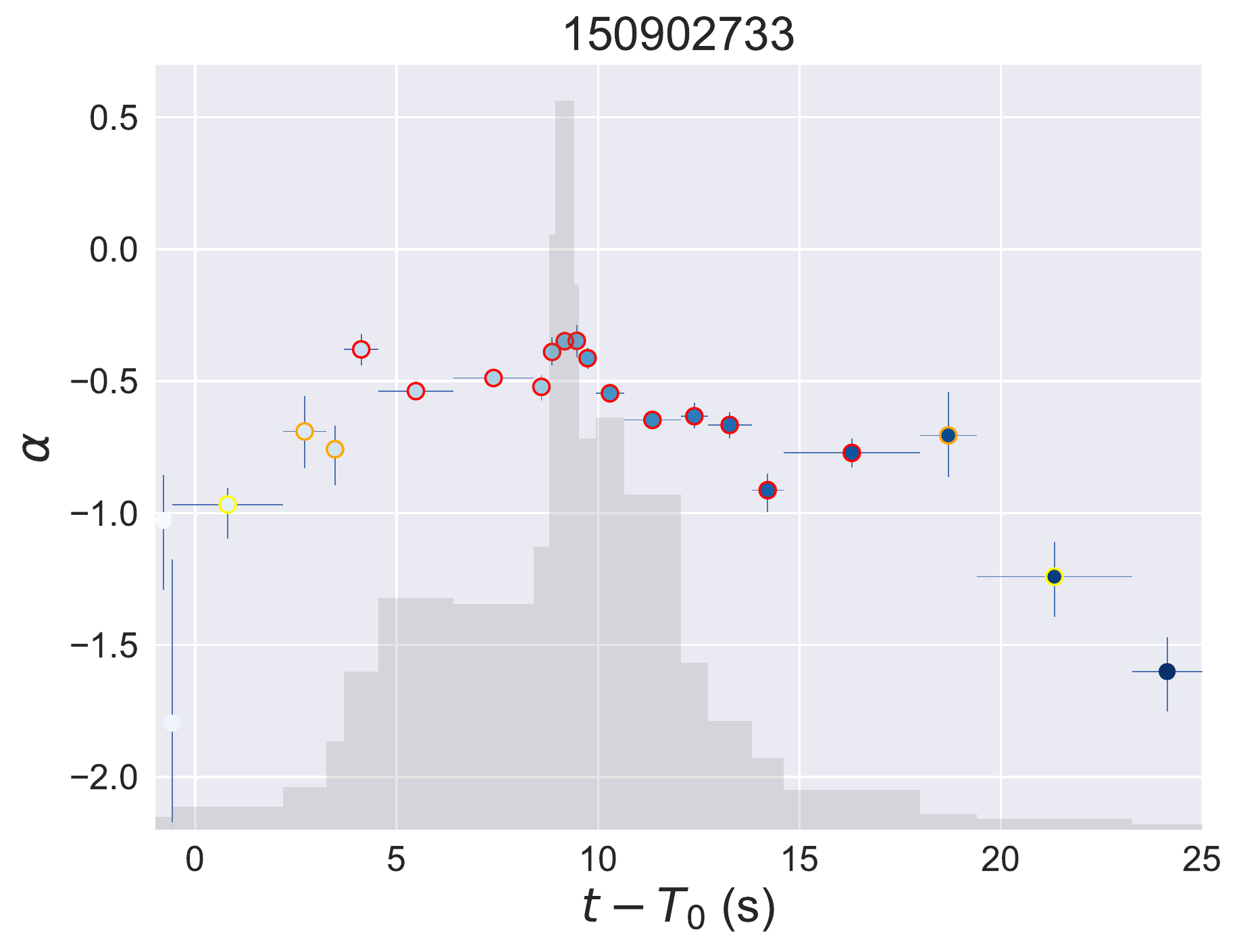}}

\subfigure{\includegraphics[width=0.33\linewidth]{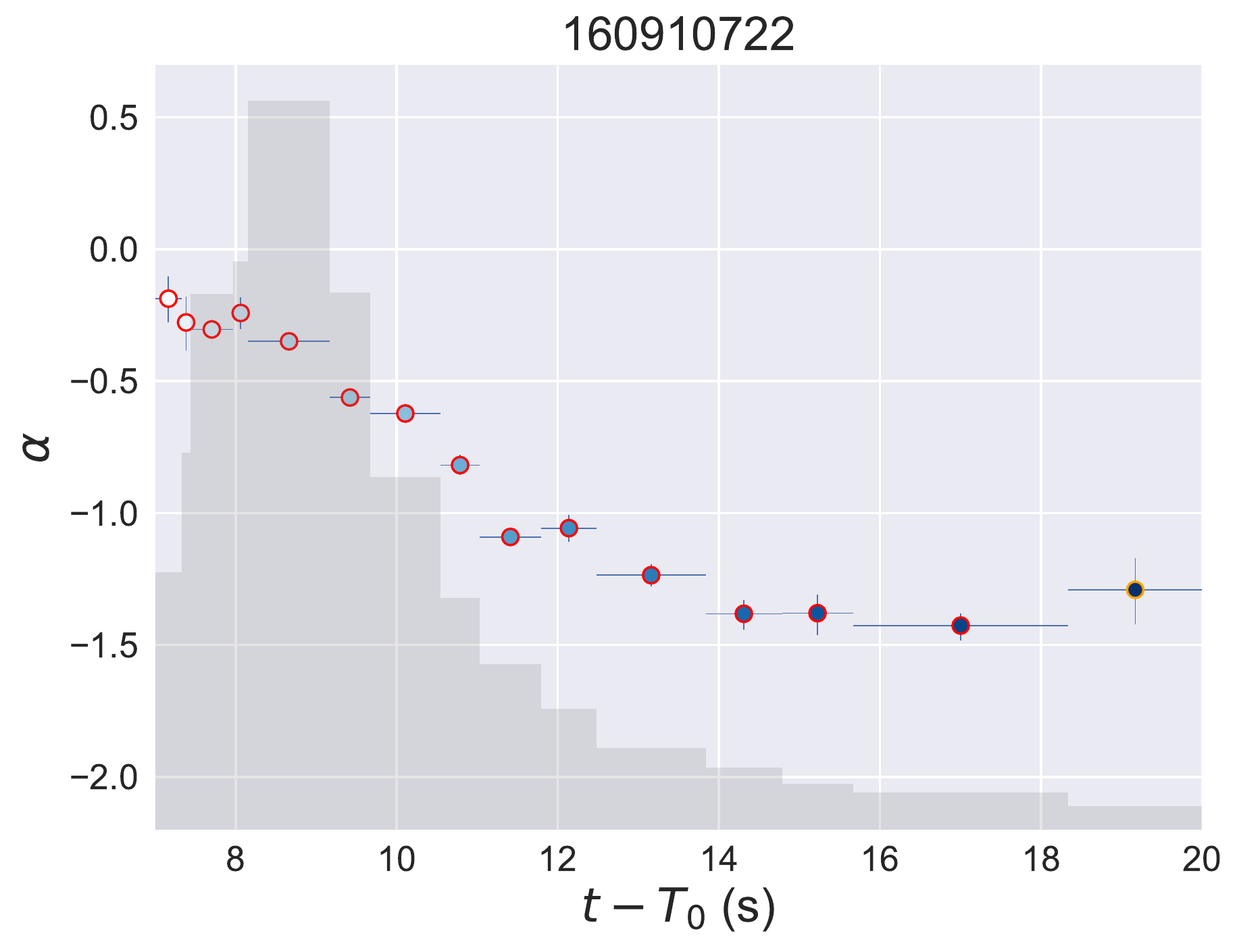}}
\subfigure{\includegraphics[width=0.33\linewidth]{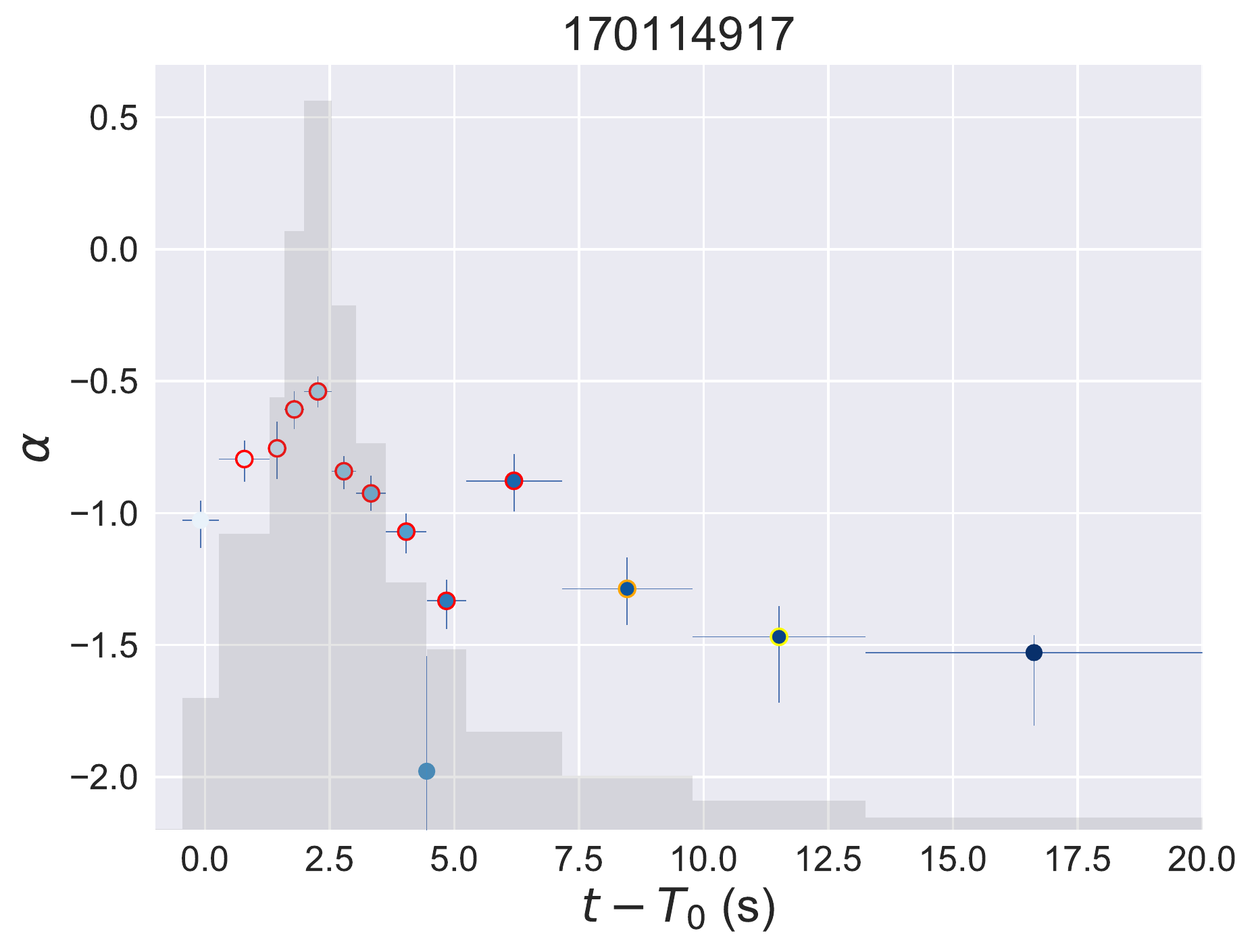}}
\subfigure{\includegraphics[width=0.33\linewidth]{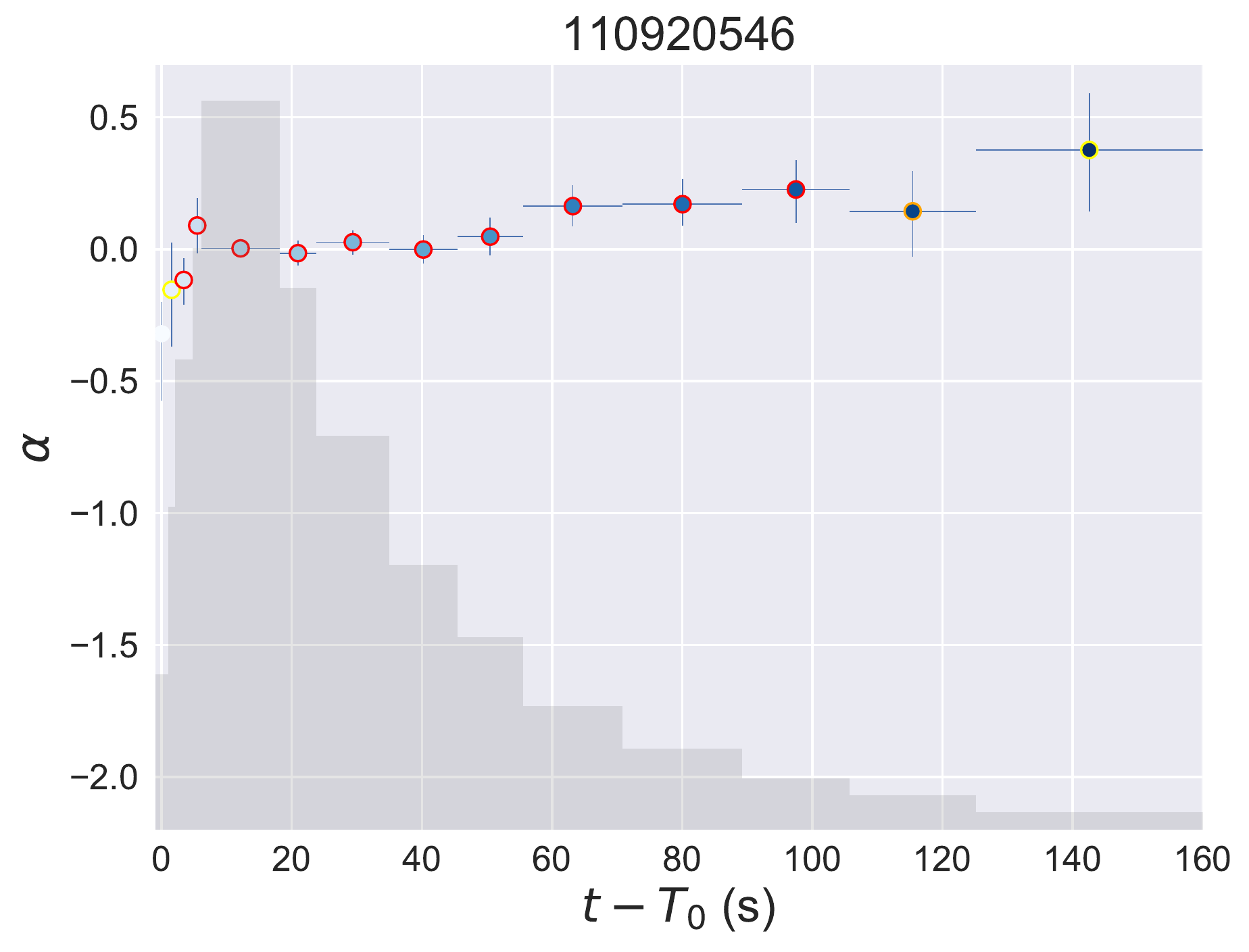}}

\caption{Examples of the temporal evolution of $\alpha$, with the photon fluxes overlaid. 
The color-coding of the data points reflects the temporal sequence (white to dark blue). The data points in red circle are for the time bins with significance $S \geq 20$ \citep{Vianello2018}. Data points in orange, yellow, and no circle are for bins with $15 \leq S < 20$, $10 \leq S < 15$, and $S < 10$, respectively. Time is plotted relative to the GRB trigger time $T_0$.
\label{fig:figure2}}
\end{figure*}

\begin{figure*}
\centering

\subfigure{\includegraphics[width=0.33\linewidth]{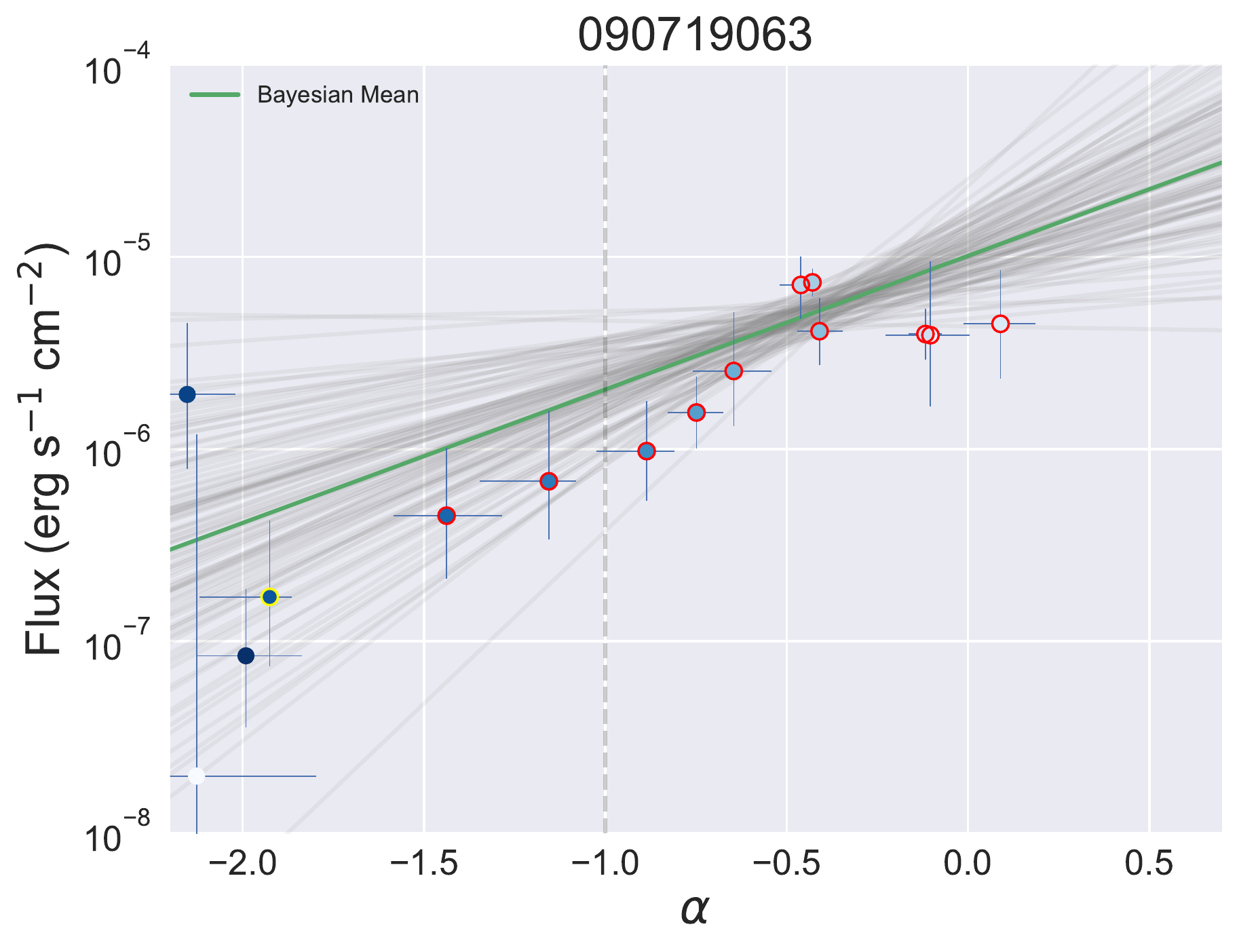}}
\subfigure{\includegraphics[width=0.33\linewidth]{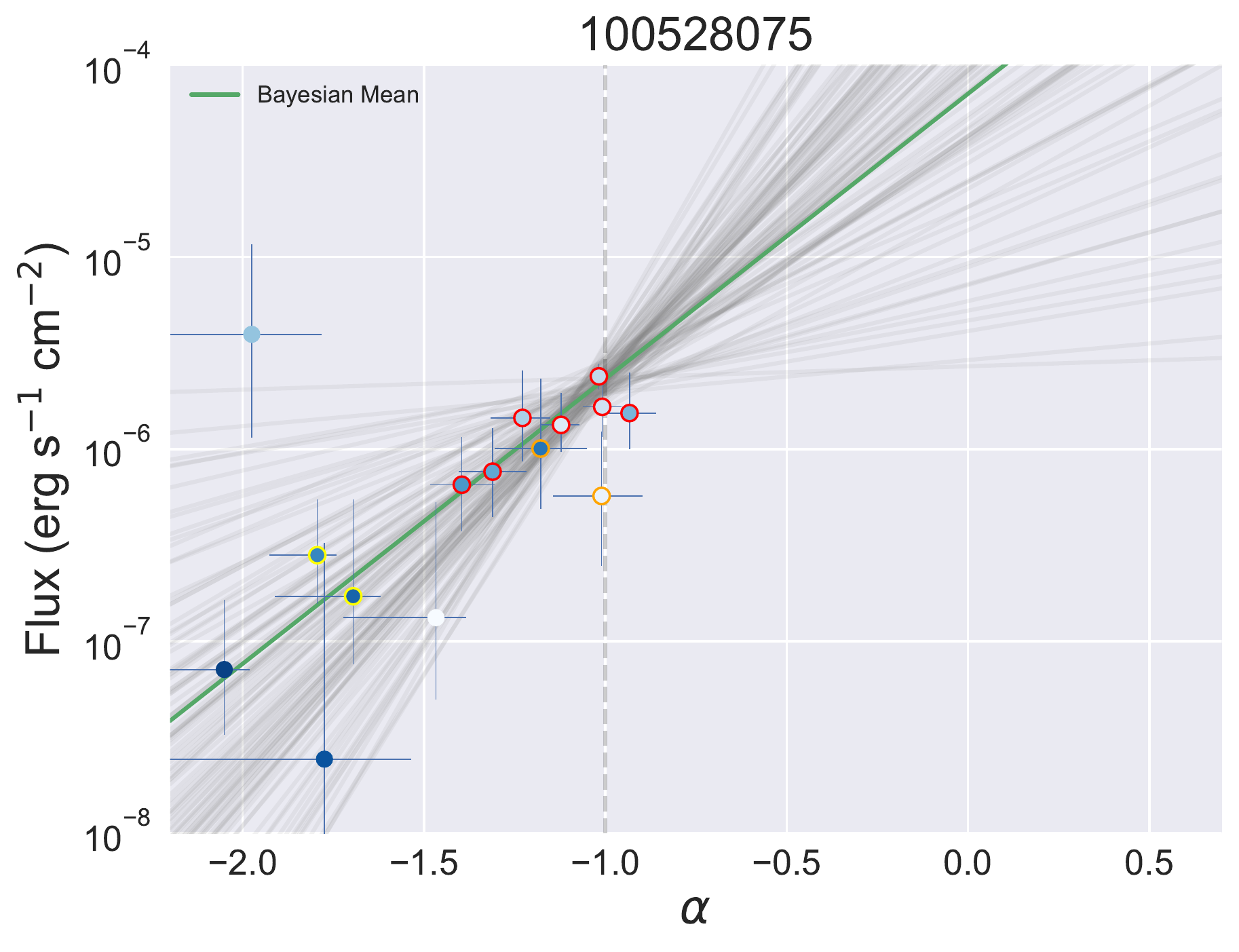}}
\subfigure{\includegraphics[width=0.33\linewidth]{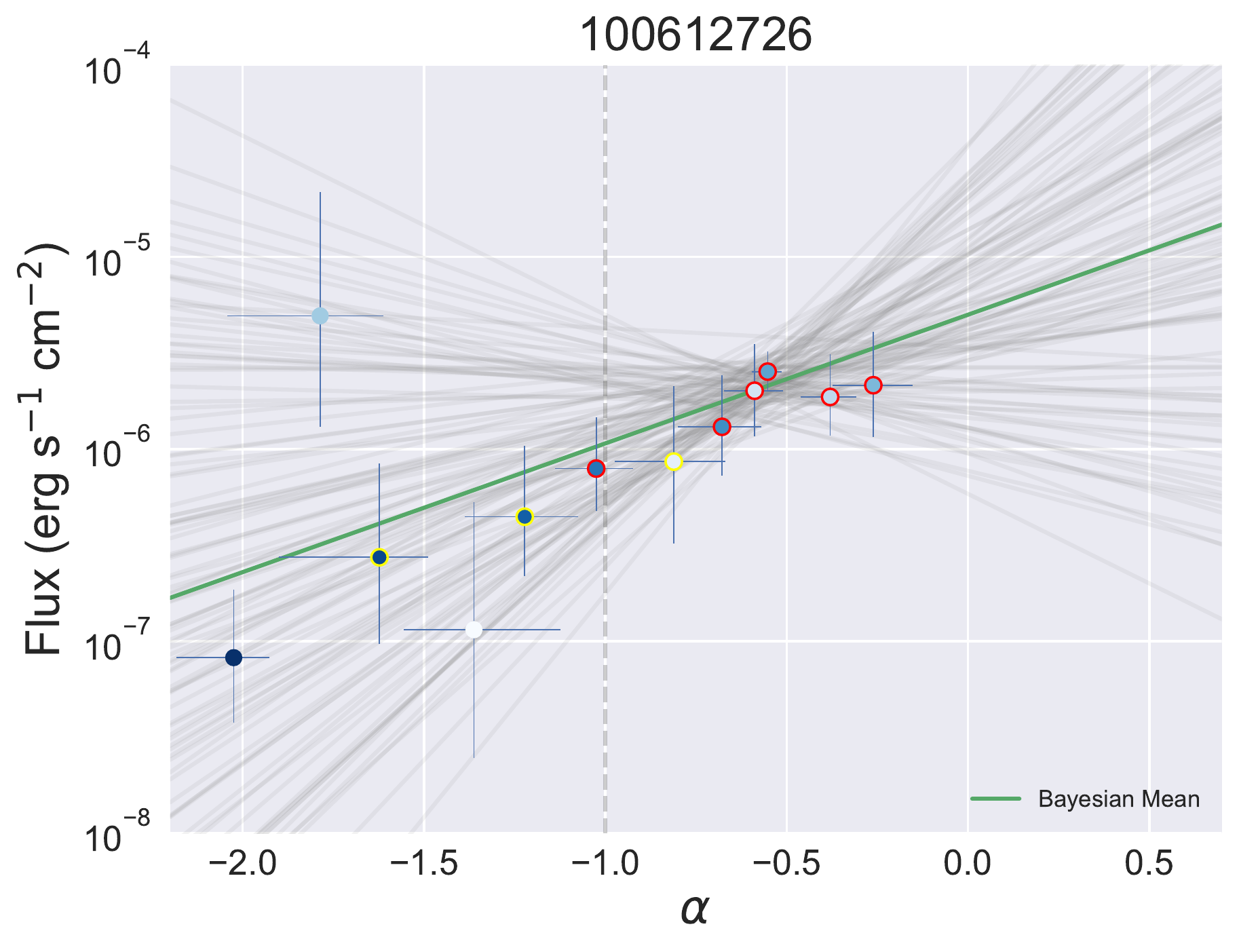}}

\subfigure{\includegraphics[width=0.33\linewidth]{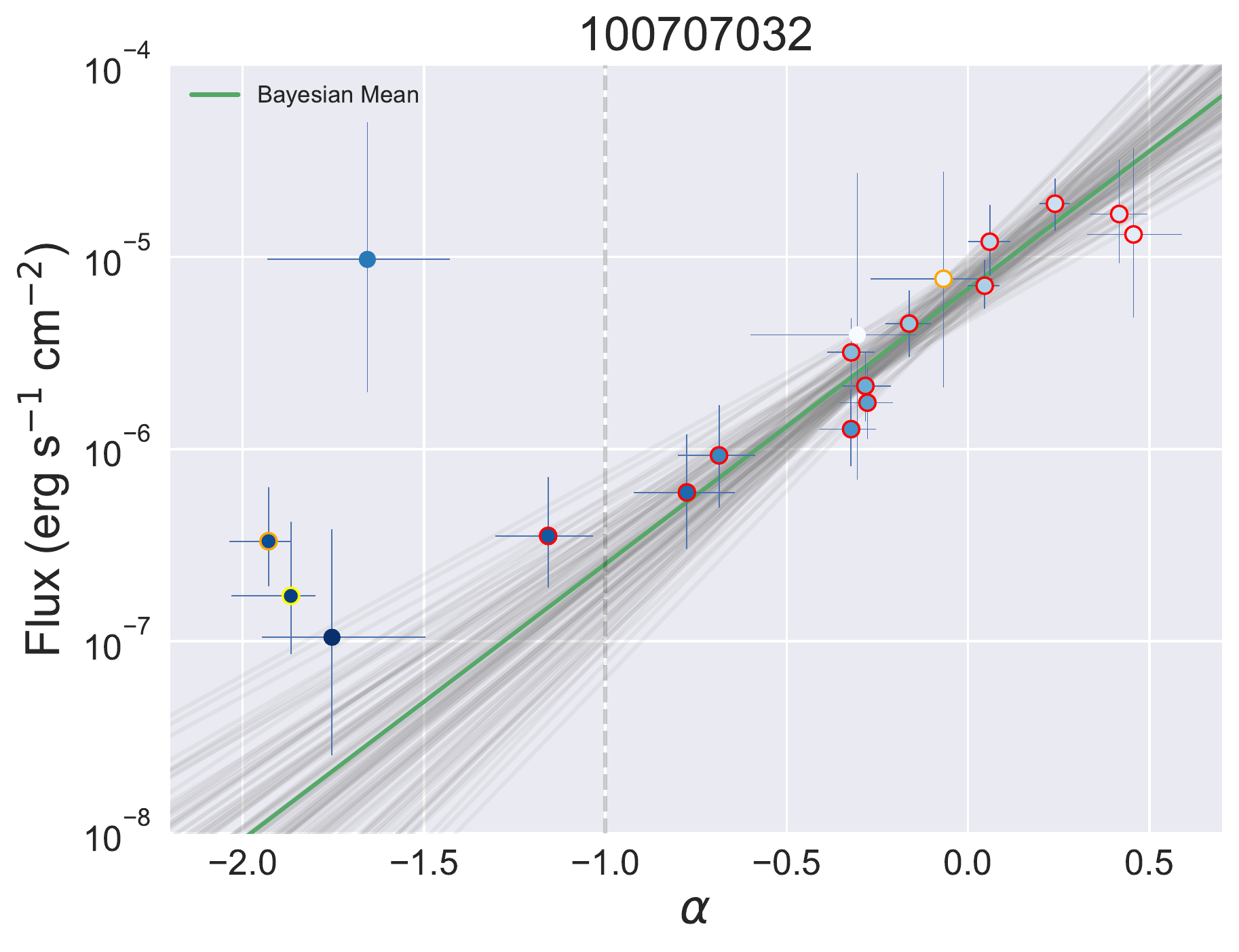}}
\subfigure{\includegraphics[width=0.33\linewidth]{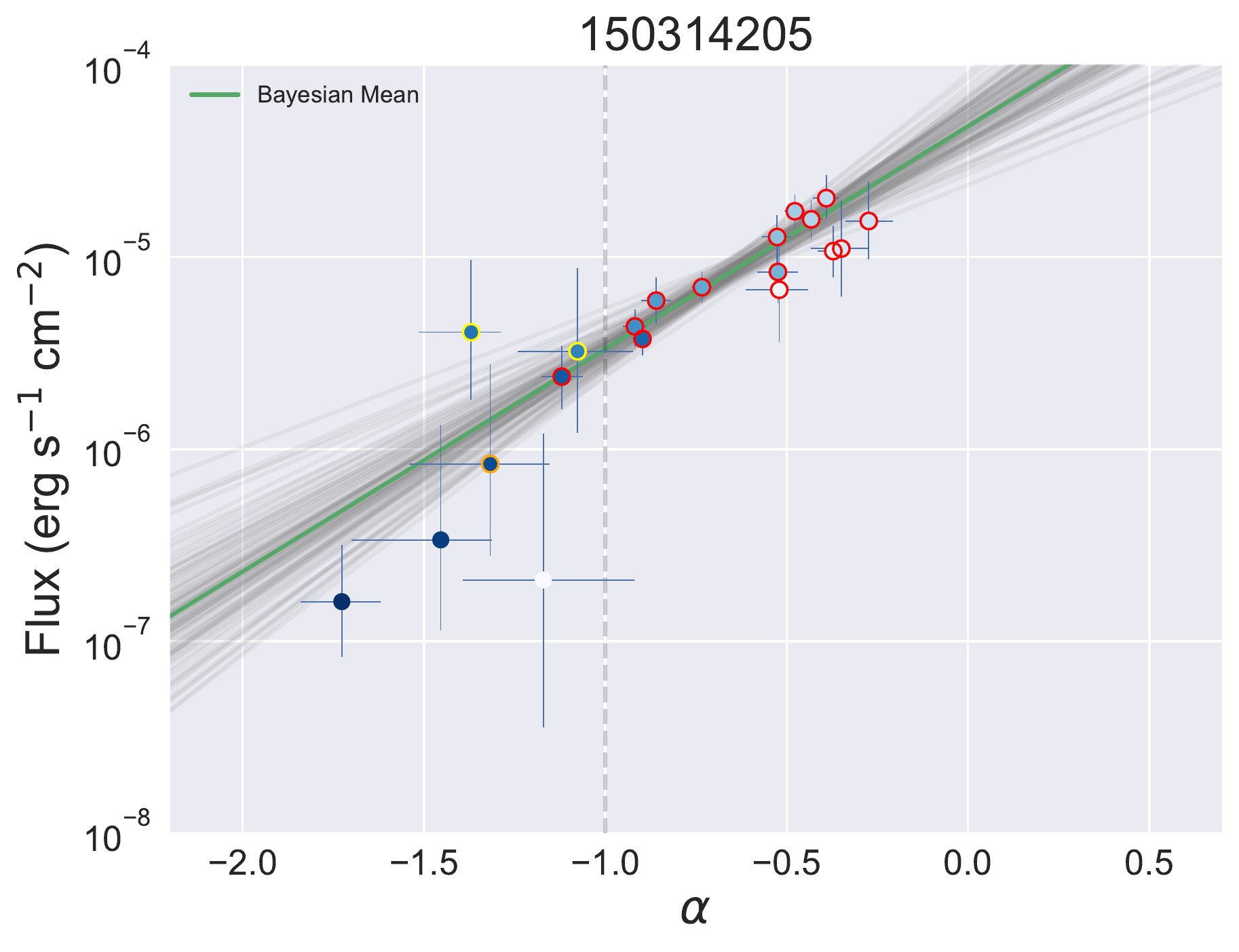}}
\subfigure{\includegraphics[width=0.33\linewidth]{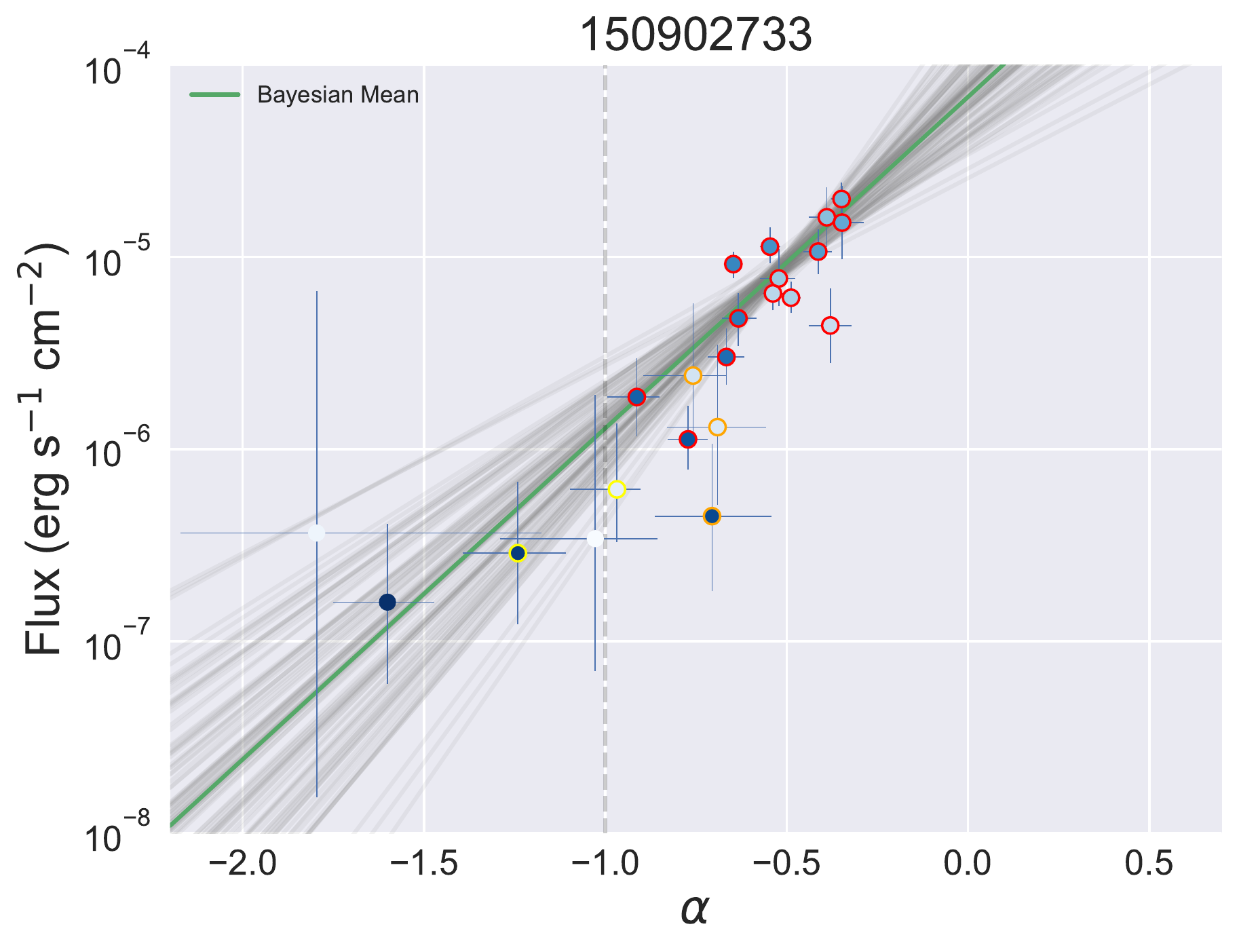}}

\subfigure{\includegraphics[width=0.33\linewidth]{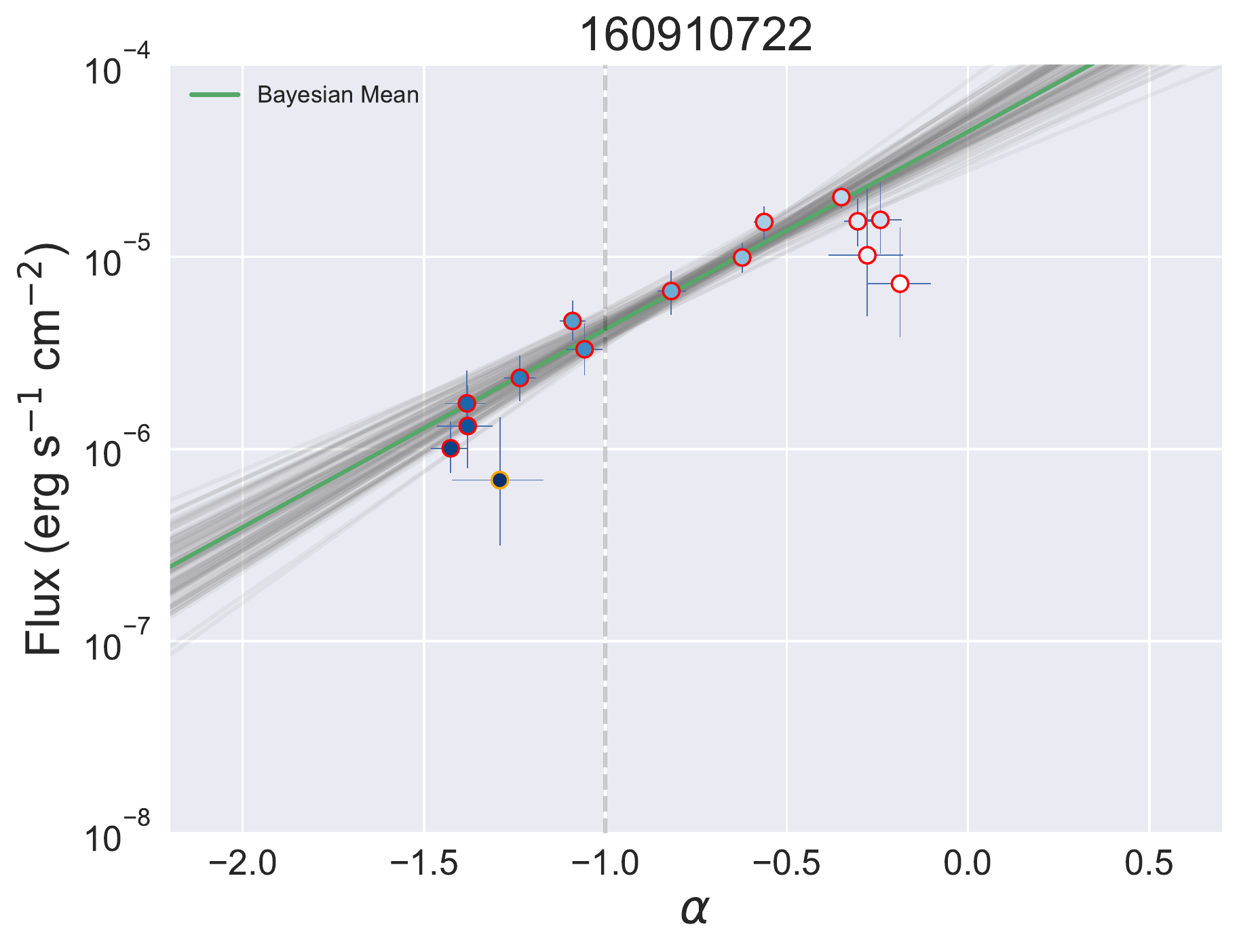}}
\subfigure{\includegraphics[width=0.33\linewidth]{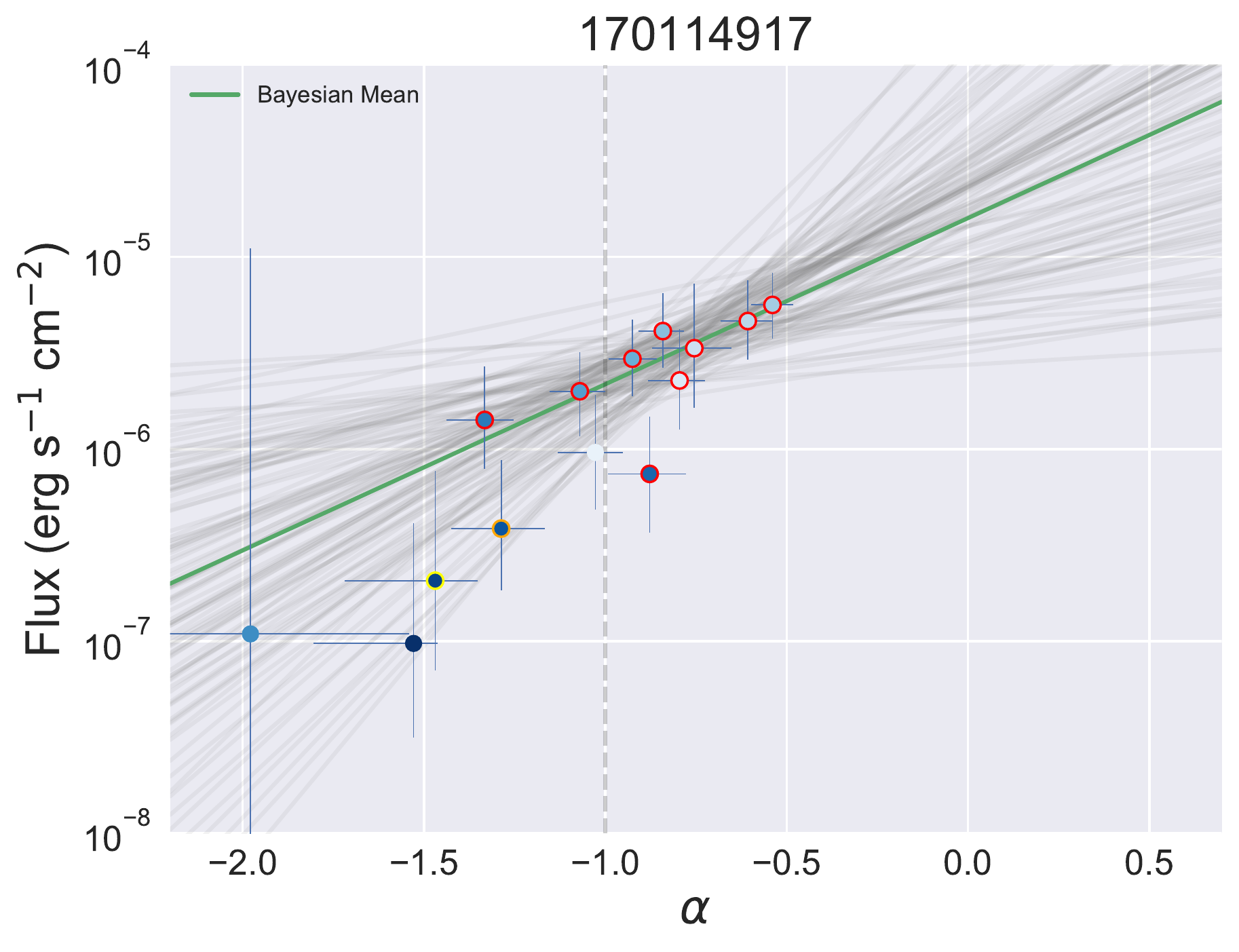}}
\subfigure{\includegraphics[width=0.33\linewidth]{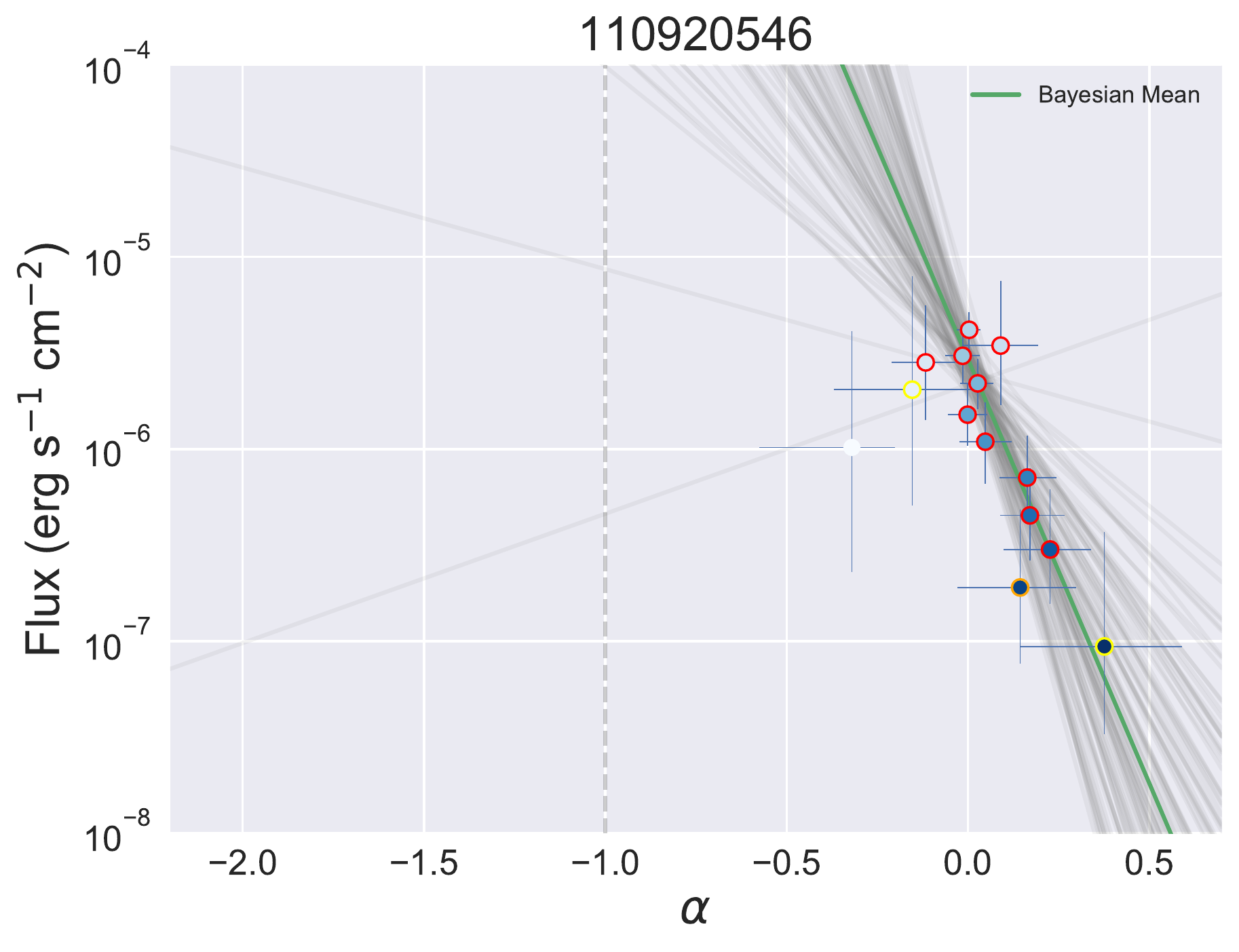}}

\caption{The $\alpha$-intensity correlation for the bursts in Fig.~\ref{fig:figure2}. The color-coding of the data points are the same as Fig.~\ref{fig:figure2}. Only the data points in red circle, with significance $S \geq 20$, are used in the Bayesian inference of the correlation.  The green line shows the mean of the posterior distribution and the grey lines are 100 randomly selected samples from the MCMC sampling. The vertical dashed line shows where $\alpha = -1 \sim \alpha_{\rm w}$. The $\alpha$-intensity correlation for the full sample is shown in Figs.~\ref{fig:figureB1}--\ref{fig:figureB3} in Appendix~\ref{sec:fullsmaple}. 
\label{fig:figure1}}
\end{figure*}

We employ Bayesian statistics instead of the conventional frequentist statistics in all of our current analysis, such as the fits of the $F$-$\alpha$ correlations (\S \ref{sec:analysis}). 
The most obvious advantage of the Bayesian method is that it infers directly the posterior probability density of any model parameter desired, enabling straightforward uncertainty (error bar) extraction, regardless the modality and skewness of the posterior distribution. To assess the ``goodness-of-fit'' of the $F$-$\alpha$ correlation fits (as well as of the spectral fits), we inspected the posterior and marginal distributions (the so-called Bayesian corner plot) for each pulse one by one. {Inconclusive fitting results (e.g., due to limited or noisy data) are revealed by unconstrained marginal distributions. Furthermore, posterior predictive checks were performed on the correlation fit results (see \S \ref{sec:analysis} for details). If the proposed model describes the data adequately (in our case, the linear fit in log-linear space of $F$-$\alpha$) then the posterior predicted values will be consistent with the average value of the data.}

\begin{table}
\caption{GRB pulses in the study.}
\label{tab:table1}
\def\arraystretch{1.5}
\begin{tabular}{cccc}
\hline
\hline
GRB trigger & $k$ $^a$ & $F_{\rm max}$ $^b$ & $\alpha_{\rm max}$ $^b$ \\
  &  & (${\rm erg~cm}^{-2}~{\rm s}^{-1}$) &  \\
\hline
081009140  &  $ 1.3 ^{+ 1.1}_{- 1.1  } $ & $ 1.0 ^{+ 0.4 }_{- 0.3 } \times 10^{ -5 }$ & $ -0.51 ^{+ 0.13 }_{- 0.14 }$ \\
081009140 (39 s)$^c$  &  $ 0.9 ^{+ 2.6 }_{- 2.7 } $ & $ 1.2 ^{+ 0.9 }_{- 0.5 } \times 10^{ -6 }$ & $ -1.05 ^{+ 0.34 }_{- 0.31 }$ \\
081125496  &  $ 1.7 ^{+ 4.6 }_{- 4.7 } $ & $ 5.0 ^{+ 4.0 }_{- 2.2 } \times 10^{ -6 }$ & $ -0.11 ^{+ 0.11 }_{- 0.11 }$ \\
081224887  &  $ 3.0 ^{+ 0.9 }_{- 1.0 } $ & $ 6.5 ^{+ 3.0 }_{- 2.2 } \times 10^{ -6 }$ & $ -0.13 ^{+ 0.05 }_{- 0.05 }$ \\
090530760  &  $ 1.3 ^{+ 0.4 }_{- 0.4 } $ & $ 1.2 ^{+ 1.0 }_{- 0.5 } \times 10^{ -6 }$ & $ -0.07 ^{+ 0.10 }_{- 0.09 }$ \\
090620400  &  $ 1.4 ^{+ 3.2 }_{- 3.0 } $ & $ 3.8 ^{+ 2.8 }_{- 1.6 } \times 10^{ -6 }$ & $ 0.06 ^{+ 0.08 }_{- 0.10 }$ \\
090626189 (34 s)$^c$  &  $ 3.1 ^{+ 1.5}_{- 1.7 } $ & $ 1.3 ^{+ 1.5 }_{- 0.7 } \times 10^{ -5 }$ & $ -0.63 ^{+ 0.08 }_{- 0.08 }$ \\
090719063  &  $ 1.6 ^{+ 0.6  }_{- 0.7  } $ & $ 7.4 ^{+ 1.4 }_{- 1.1 } \times 10^{ -6 }$ & $ 0.09 ^{+ 0.10 }_{- 0.10 }$ \\
090804940  &  $ 0.35 ^{+ 4.5  }_{- 4.5  } $ & $ 3.4 ^{+ 1.1 }_{- 0.8 } \times 10^{ -6 }$ & $ -0.22 ^{+ 0.15 }_{- 0.19 }$ \\
090820027  &  $ 3.2 ^{+ 0.4  }_{- 0.4  } $ & $ 2.6 ^{+ 0.4 }_{- 0.3 } \times 10^{ -5 }$ & $ -0.44 ^{+ 0.07 }_{- 0.07 }$ \\
100122616  &  $ 2.0 ^{+ 8.6  }_{- 7.0  } $ & $ 3.0 ^{+ 0.9 }_{- 0.6 } \times 10^{ -6 }$ & $ -1.42 ^{+ 0.07 }_{- 0.15 }$ \\
100528075  &  $ 3.4 ^{+ 1.5 }_{- 1.8  } $ & $ 2.4 ^{+ 0.4 }_{- 0.3 } \times 10^{ -6 }$ & $ -0.93 ^{+ 0.07 }_{- 0.07 }$ \\
100612726  &  $ 1.5 ^{+ 1.5  }_{- 1.8  } $ & $ 2.5 ^{+ 0.7 }_{- 0.5 } \times 10^{ -6 }$ & $ -0.26 ^{+ 0.11 }_{- 0.11 }$ \\
100707032  &  $ 3.3 ^{+ 0.5 }_{- 0.6  } $ & $ 1.9 ^{+ 0.6 }_{- 0.5 } \times 10^{ -5 }$ & $ 0.46 ^{+ 0.13 }_{- 0.13 }$ \\
101126198  &  $ 3.7 ^{+ 1.7 }_{- 2.1  } $ & $ 2.3 ^{+ 0.5 }_{- 0.4 } \times 10^{ -6 }$ & $ -1.04 ^{+ 0.04 }_{- 0.07 }$ \\
110721200  &  $ 4.1 ^{+ 0.6 }_{- 0.7  } $ & $ 7.7 ^{+ 0.8 }_{- 0.8 } \times 10^{ -6 }$ & $ -0.89 ^{+ 0.03 }_{- 0.03 }$ \\
110817191  &  $ 3.2 ^{+ 1.4 }_{- 1.9  } $ & $ 4.4 ^{+ 1.5 }_{- 1.2 } \times 10^{ -6 }$ & $ -0.42 ^{+ 0.08 }_{- 0.10 }$ \\
110920546  &  $ -10 ^{+ 4.0 }_{- 2.7  } $ & $ 4.2 ^{+ 1.1 }_{- 0.9 } \times 10^{ -6 }$ & $ 0.23 ^{+ 0.11 }_{- 0.13 }$ \\
111017657  &  $ 3.7 ^{+ 2.2 }_{- 2.4  } $ & $ 4.0 ^{+ 1.5 }_{- 0.9 } \times 10^{ -6 }$ & $ -0.75 ^{+ 0.05 }_{- 0.05 }$ \\
120919309  &  $ 5.1 ^{+ 2.1 }_{- 2.3  } $ & $ 4.8 ^{+ 1.4 }_{- 1.2 } \times 10^{ -6 }$ & $ -0.66 ^{+ 0.04 }_{- 0.04 }$ \\
130305486  &  $ 3.7 ^{+ 1.3  }_{- 1.6  } $ & $ 9.2 ^{+ 2.1 }_{- 1.8 } \times 10^{ -6 }$ & $ -0.34 ^{+ 0.05 }_{- 0.06 }$ \\
130612456  &  $ 2.9 ^{+ 1.4 }_{- 1.8 } $ & $ 3.4 ^{+ 1.0 }_{- 0.9 } \times 10^{ -6 }$ & $ -0.77 ^{+ 0.04 }_{- 0.05 }$ \\
130614997  &  $ 0.44 ^{+ 3.4 }_{- 3.9} $ & $ 1.9 ^{+ 1.1 }_{- 0.6 } \times 10^{ -6 }$ & $ -1.23 ^{+ 0.07 }_{- 0.08 }$ \\
130815660  &  $ 3.1 ^{+ 1.5 }_{- 1.8} $ & $ 3.1 ^{+ 0.9 }_{- 0.7 } \times 10^{ -6 }$ & $ -0.72 ^{+ 0.08 }_{- 0.09 }$ \\
140508128 (0-15 s)$^c$  &  $ 4.4 ^{+ 0.9 }_{- 1.0} $ & $ 2.1 ^{+ 0.8 }_{- 0.5 } \times 10^{ -5 }$ & $ -0.57 ^{+ 0.05 }_{- 0.05 }$ \\
141028455  &  $ 3.0 ^{+ 0.9 }_{- 1.0 } $ & $ 4.6 ^{+ 5.9 }_{- 2.4 } \times 10^{ -6 }$ & $ -0.57 ^{+ 0.09 }_{- 0.12 }$ \\
141205763  &  $ 1.8 ^{+ 1.7 }_{- 1.9 } $ & $ 2.4 ^{+ 1.3 }_{- 0.8 } \times 10^{ -6 }$ & $ -0.82 ^{+ 0.07 }_{- 0.07 }$ \\
150213001  &  $ 2.5 ^{+ 0.3 }_{- 0.4 } $ & $ 1.8 ^{+ 0.3 }_{- 0.2 } \times 10^{ -5 }$ & $ -0.94 ^{+ 0.03 }_{- 0.03 }$ \\
150306993  &  $ 1.8 ^{+ 2.0 }_{- 2.1 } $ & $ 4.0 ^{+ 4.1 }_{- 2.1 } \times 10^{ -6 }$ & $ -0.02 ^{+ 0.14 }_{- 0.17 }$ \\
150314205  &  $ 2.7 ^{+ 0.4 }_{- 0.38 } $ & $ 2.0 ^{+ 0.6 }_{- 0.5 } \times 10^{ -5 }$ & $ -0.27 ^{+ 0.07 }_{- 0.06 }$ \\
150510139  &  $ 3.3 ^{+ 0.3 }_{- 0.3 } $ & $ 1.6 ^{+ 0.4 }_{- 0.3 } \times 10^{ -5 }$ & $ -0.51 ^{+ 0.04 }_{- 0.05 }$ \\
150902733  &  $ 4.0 ^{+ 0.8 }_{- 0.8 } $ & $ 2.0 ^{+ 0.4 }_{- 0.3 } \times 10^{ -5 }$ & $ -0.35 ^{+ 0.06 }_{- 0.06 }$ \\
151021791  &  $ 3.7 ^{+ 2.8 }_{- 2.8 } $ & $ 5.0 ^{+ 1.90 }_{- 1.50 } \times 10^{ -6 }$ & $ -0.22 ^{+ 0.10 }_{- 0.10 }$ \\
160215773 (160 s)$^c$  &  $ -6.2 ^{+ 1.8 }_{- 2.2 } $ & $ 6.3 ^{+ 2.2 }_{- 1.4 } \times 10^{ -6 }$ & $ -0.77 ^{+ 0.05 }_{- 0.06 }$ \\
160530667  &  $ 8.6 ^{+ 1.0 }_{- 1.3 } $ & $ 1.9 ^{+ 0.2 }_{- 0.2 } \times 10^{ -5 }$ & $ -0.54 ^{+ 0.02 }_{- 0.02 }$ \\
160910722 (7-20 s)$^c$  &  $ 2.4 ^{+ 0.2}_{- 0.2} $ & $ 2.0 ^{+ 0.3 }_{- 0.2 } \times 10^{ -5 }$ & $ -0.19 ^{+ 0.08 }_{- 0.09 }$ \\
161004964  &  $ 2.5 ^{+ 2.4 }_{- 2.7 } $ & $ 2.4 ^{+ 1.0 }_{- 0.8 } \times 10^{ -6 }$ & $ -0.44 ^{+ 0.09 }_{- 0.12 }$ \\
170114917  &  $ 2.0 ^{+ 1.1}_{- 1.2 } $ & $ 5.6 ^{+ 2.9 }_{- 1.8 } \times 10^{ -6 }$ & $ -0.54 ^{+ 0.06 }_{- 0.06 }$ \\
\hline
\end{tabular}
\begin{tablenotes}
\item \item $^a$ Fits to Eq.~(\ref{eq:2}). {The mean of the posterior samples of $k$. The errors are the 68\% HPDIs, corresponding to conventional frequentist 1$\sigma$ level.}
\item $^b$The maximum instantaneous value over the pulse.
\item $^c$Time for the pulse.
\end{tablenotes}
\end{table}

\section{Temporal Variations of the low-energy index $\alpha$}
\label{sec:alpha}

In Fig.~\ref{fig:figure2} the evolution of the instantaneous  value of the low-energy power-law index, $\alpha$ is plotted for  a few representative examples in the sample, with the energy flux $F$ (in arbitrary scale of erg~s$^{-1}$~cm$^{-2}$) overlaid. These plots indicate a covariation of the value of $\alpha$ and the energy flux; in large $\alpha$ tracks the change in flux \citep[see similar plots in, e.g., ][]{Crider1997,Ghirlanda2002, Lloyd2002, Basak&Rao2014, Yu2018}.

{ Since the data we will be interpreting ($\alpha$, $F$, and $E_{\rm pk}$) are from empirical fits over a limited energy range, instrumental biases could give rise to spurious variations in the fitted parameters. In the next section we  therefore investigate such biases in order to validate these variations as being caused by the underlying physics or not.
}

\subsection{$E_{\rm pk}$-$\alpha$ correlations expected due to instrumental biases}
\label{sec:instrumental}

\citet{Preece1998} and \citet{Lloyd2002}  pointed out possible limitations in determining the true spectral shape due to the finite band width of the instrument used (known as the window effect). Since most physical emission spectra are smoothly curved, the characteristic power-law slopes are only reached asymptotically. Thus the characteristic slopes might not be detected within the finite energy band of the detector \citep[e.g.,][]{Sakamoto2009}. More importantly, though, is the ability of the empirical model to capture the curvature of the true spectrum.  In order to investigate this, \citet{Lloyd2000} studied various synchrotron models and fitted simulated spectra (for the {\it CGRO}/BATSE instrument, $E_{\rm min}\sim 25 \rm{keV}$) with the Band function (see also \citet{Ghirlanda2002}). For an electron distribution with a sharp minimum energy cut-off \citep[see, e.g., ][]{Tavani1996} they found that below 100 keV, $\alpha$ becomes significantly softer, $\alpha_{\rm obs} < \alpha_{\rm synch}$. Indeed, this is due to the broad curvature of the synchrotron function that is not captured correctly by the Band function (see their Fig. 5).   The induced correlation between $\alpha$ and $E_{\rm pk}$, which is  specific for the assumed type of synchrotron emission, thus becomes positive, below 100 keV. The window effect for slow-cooled synchrotron emission becomes even more pronounced if a broader electron distribution is assumed, since this would cause an even smoother curvature of the spectrum \citep{Lloyd2000}. Another effect of the inability of the Band function to fully characterise the curvature of slow-cooled synchrotron emission, is that the empirically fitted $\alpha$ typically is  $-0.8$ rather than the physically expected slope of $-2/3$ \citep{Burgess2015_alpha}. 

\begin{figure}
 \includegraphics[width=\columnwidth]{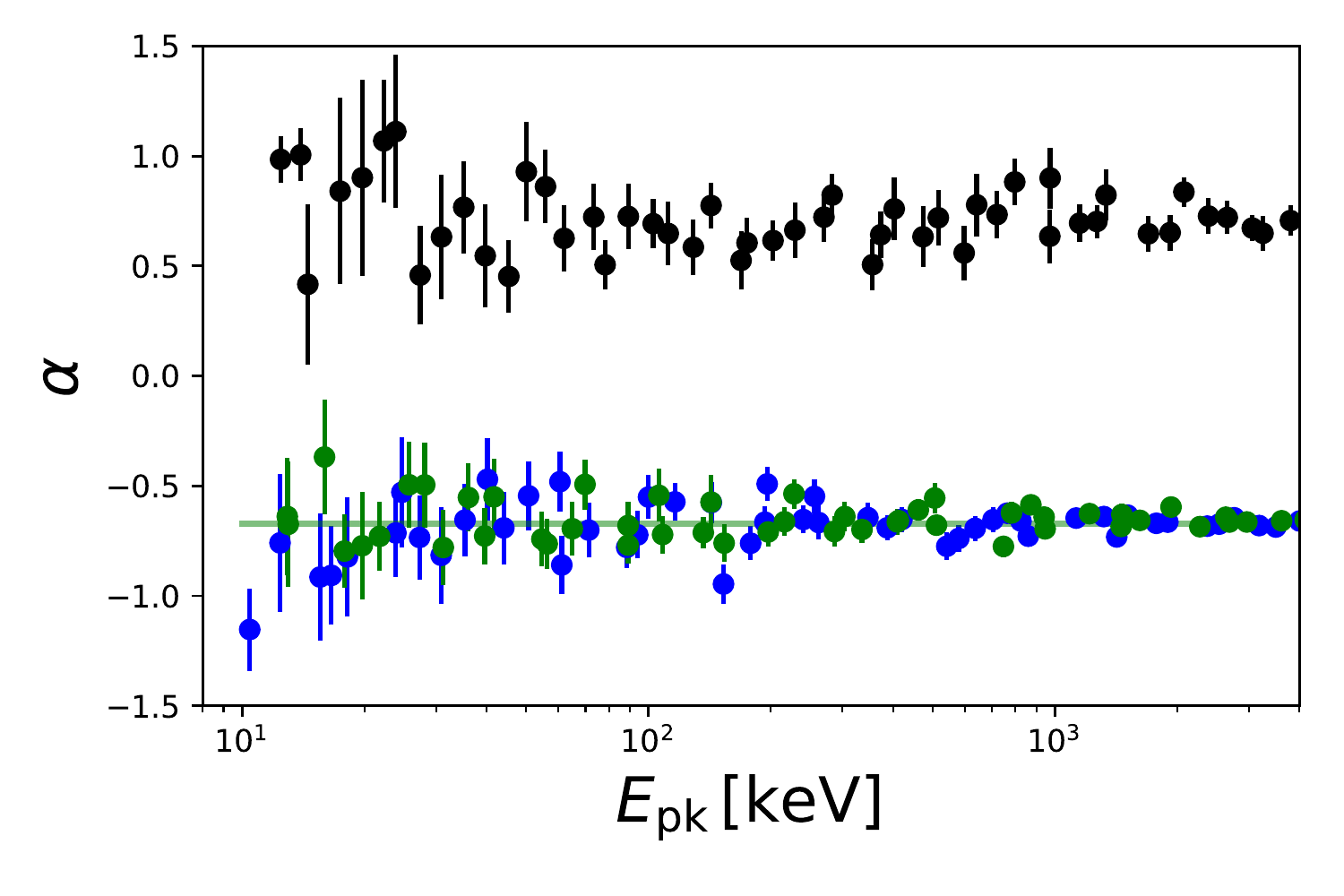}
 \caption{Results of fits to synthetic GBM spectra with $10 < E_{\rm pk}< 4 \times 10^3$ keV and all having a SNR $\sim 30$. The blue and green data points are from spectra produced from a cutoff power law function with $\alpha = -0.67$. The green points correspond to fits made with a Band function and the blue points correspond to fits made with a cutoff power-law function. 
The black points are for an acceleration phase photosphere, fitted with a cutoff power-law. No significant window effect appears in any of the cases, indicating that the functions used to fit the data can capture the curvature of the synthetic spectra well. Note that bursts observed by GBM are all above 40 keV, while the low energy limit of GBM is 8 keV.}
 \label{fig:bandband}
\end{figure}

{However, if the empirical (or physical) model that is used, matches the true, incoming emission spectrum, then the asymptotic power-law slope can still be determined, thereby eliminating such a window effect. This is the case even though the fit is made only over a limited range in energy. Examples of this are shown in Figure \ref{fig:bandband} where synthetic burst spectra are fitted, over the GBM energy range, with a cutoff power-law function. A range  of spectra are simulated for different peak energies. Two models have been used to produce the synthetic spectra of the true, incoming emission: a cutoff power-law function, with a low energy index $\alpha = -0.67$ and a photospheric spectrum expected from the acceleration phase \citep[Eq. (2) in][]{Ryde2017}. We use an arbitrary burst, observed by the GBM detector\footnote{The burst used is GRB150213, however, the results are not noticeably affected by the choice of burst.} as a template for the GBM detector response and directional information and background characteristics and require the simulated data to have s signal-to-noise ratio (SNR) of around 30, to closely resemble actual data observed by GBM\footnote{We use the routines available in {\tt 3ML} for generating synthetic spectra.}. In Figure \ref{fig:bandband} the coloured data points are for the bursts with the cutoff power-law as the true spectrum, fitted with a Band function (green) and a cutoff power-law (blue). In both cases the fitted values of $\alpha$ do not show any window effect, and recover the original value of $\alpha =-0.67$, albeit with larger dispersion below 100 keV\footnote{Note that bursts observed by GBM typically have $E_{\rm pk}>$ 40 keV \citep{Yu2016}}. Likewise, in Figure \ref{fig:bandband} the black data points correspond to synthetic GBM spectra produced from an acceleration phase photosphere. They are fitted with a cutoff power-law function. Again there is no significant window effect.
}

{ The reason for the absence of the window effect in these three cases is that the empirical functions used, correctly describe the curvature of the synthetic spectra. For the blue data points, this is obvious, since the incoming spectrum and the fitted spectrum are identical, but in the other cases it is, a priori, less clear. 
We also note that, in the case of fast-cooled synchrotron emission, the Band function typically captures the correct slope of $\alpha \sim -1.5$. This means that the Band function is better at describing the curvature of fast-cooled rather than slow-cooled synchrotron emission \citep{Burgess2015_alpha}. 
}

{ However, in general the empirical models that are typically used 
are not necessarily able {  to} describe the true curvature of any theoretical spectrum. Indeed, if these empirical models differ significantly from the true spectral shape, an instrumental bias could be introduced.
}

\subsection{Observed $E_{\rm pk}$-$\alpha$ Correlations}
\label{sec:instrumental2}

{\citet{Lloyd2002} found a large variation of $E_{\rm pk}$-$\alpha$ correlations, both between bursts and within bursts. Both positive and negative correlations are identified.  They therefore conclude that the physical correlation in many cases have to  overwhelm the instrumental effect, which predicts a purely positive correlation \citep{Lloyd2000}. They also point out that, in some cases, even though a positive correlation is observed, it can be explained by physical effects in the synchrotron model, such as a varying magnetic field strength.  
Likewise, \citet{Ghirlanda2002} found many types of relations.
However, they found that the observed $E_{\rm pk}$-$\alpha$-relation in many individual bursts, which have hard spectra,  do  not follow the expected relation for synchrotron emission.
}

{ Similarly, the fact that in the full sample \citep{Yu2018}, we do not either find any systematic behaviour of the $\alpha$-$E_{\rm pk}$ relations makes us draw the conclusion that 
the window effect should not have a dominant and general impact  on our determination of $\alpha$ in our sample. In particular, 
the fact that the $\alpha$-$E_{\rm pk}$ relations do not follow the one expected from the window effect in the synchrotron case, argues against the synchrotron model. On the contrary, the absence of any strong and systematic window effect could indicate that the true emission model is, in many cases, well approximated by a cutoff powerlaw, at least around the spectral peak. However, the window effect should still be considered, in particular, for low $E_{\rm pk}$ bursts, since we do not, a priori, know the true spectral shape, and thereby, magnitude of the effect, more than it probably has the largest effect  below 100 keV (see further discussion in \S\ref{sec:BBwindow}).
}

\section{The $\alpha$-Intensity Correlation}
\label{sec:analysis}


We now turn to investigating the functional correlation between $F(t)$ and $\alpha(t)$. A few examples of these are shown in Fig.~\ref{fig:figure1}, where the correlation is shown in {  log-linear} plots. The temporal sequence of the data points are {  colour}-coded from white (start) to dark blue (end). 
The correlations typically have a linear behaviour in the plots, which indicates an exponential relation.
We, therefore, fit the dependency of the observed instantaneous flux 
on $\alpha$ with the following function:
\begin{equation}
F(t) = F_0 \, {\rm e}^{k \, \alpha (t)},
\label{eq:2}
\end{equation}
where $F_0$ is the normalisation for the fit.

We perform Bayesian inference 
using the Bayesian statistical modeling package {\tt PyMC3} \citep{Salvatier2016} 
making use of Markov chain Monte Carlo (MCMC) algorithms { to explore the posterior distributions}. We used Gaussian 
likelihoods and uniform priors for $F_0$ and $k$. Unlike what is usually done 
in conventional line regression, in which only the errors of the ordinate (i.e., the ``$y$-error'') are considered, we also took the errors of the abscissa (i.e., the ``$x$-error'') into account (i.e., we consider in the fit both the errors on $F$ and $\alpha$). {  The Bayesian hierarchical structure is as follow:
\begin{equation} \label{eqn:priors}
\begin{cases}
	k \sim \mathcal{U}(-20,20)\\
	F_0 \sim \mathcal{U}(-100,100)\\
	\alpha^\textrm{true} \sim \mathcal{N}(\textrm{mean}(\alpha^\textrm{obs}),\textrm{std}(\alpha^\textrm{obs}))\\
    \log F^\textrm{true} \sim F_0 + k \alpha^\textrm{true}\\
    \mathcal{L}(\textrm{data}|\alpha^\textrm{true}) \propto \mathcal{N}(\alpha^\textrm{true}|\alpha^\textrm{obs},\sigma_\alpha^\textrm{obs})\\
    \mathcal{L}(\textrm{data}|\log F^\textrm{true}) \propto \mathcal{N}(\log F^\textrm{true}|\log F^\textrm{obs},\sigma_{\log F}^\textrm{obs}),
\end{cases}
\end{equation}
where $\mathcal{N}$ denotes a normal distribution, $\mathcal{L}$ denotes a likelihood function, mean($x$) and std($x$) represents the mean and standard deviation of the distribution of $x$, superscripts ``obs'' and ``true'' denotes observed value and latent variable, and $\sigma_x$ represents 1$\sigma$ error of $x$. If $\sigma_x$ is asymmetrical, we chose the larger value in order to be conservative.}

\subsection{Pulse-wise Correlation} 
\label{sec:pulsewise}

{  The fits to the data with Eq.~(\ref{eq:2}) over the individual pulses are also shown in Fig.~\ref{fig:figure1}, and for the full sample in Figs.~\ref{fig:figureB1}--\ref{fig:figureB3} in Appendix~\ref{sec:fullsmaple}. Only time bins with significance $S \geq 20$ (red circles) are included in the fits, since for these bins the spectral parameters are typically well constrained \citep{Yu2018}. In the figures, the green lines show  the mean of the posterior distributions, while the grey lines are randomly selected from the MCMC chains to show the degree of spread in the posterior distributions.  We find that for many pulses the posterior distributions are narrow which shows that the value $k$ is precisely constrained with only moderately small degree of uncertainty. We also note that in most bursts the data points with lower significance $S < 20$ (orange, yellow, and no circles) are consistent with the best fit, even though they were not included in the fits. This fact lends additional support to the correlation. 
}

{  In order to assess how well the ``straight line'' fits the correlation between $\log F$ and $\alpha$, 
we performed a posterior predictive check (PPC) using {\tt PyMC3}. This allows us to check whether the posterior distributions obtained could generate predictive data close to the observed ones. First, 5,000 random values of $k$ and $F_0$ are drawn from the MCMC sampling traces of the $F$-$\alpha$ correlation. The population of these samples contains not only the information of the best-fit values of $k$ and $F_0$, but also the errors on $F$ and $\alpha$ that were used in the inference of the range of $k$ and $F_0$. Then, 100 random values of $F$ and $\alpha$ were drawn from the normal distributions specified by each of the 5,000 values of $k$ and $F_0$. This results in predictive distributions of $F$ and $\alpha$, each consisting of 500,000 values. We found that for each of the 38 pulses the observed means of $F$ and $\alpha$ lie within these PPC distributions. This shows that the relation specified in Eq.~(\ref{eq:2}) is an adequate description to the observed behaviour between $F$ and $\alpha$.}

The fitted values of $k$ are shown in Table~\ref{tab:table1}, where the mean of the posterior distribution is shown together with the 68\% 
highest posterior density intervals (HPDIs, i.e., the Bayesian counterpart of $1\sigma$ error bars). The values of $k$ are typically positive, and range between 1 and 5. 

There are two clear outliers which have large, negative $k$-values.  We  note that one of these two bursts (GRB110920 in Fig.~\ref{fig:figure1}) has been suggested to have a significant, additional power-law component in the prompt spectrum \citep{Iyyani2015}. {Similarly, \citet{Axelsson2012} argue that GRB110721A (see Fig. \ref{fig:figureB2}) should be fitted with a multi-peaked spectrum. 
Since the analysis performed above uses a single spectral component, the instantaneous emission properties might not be determined befittingly for these particular cases. This fact raises the possibility that deviations from any underlying correlation might arise. 
In combination with the fact that the two pulses with negative $k$, significantly differ from $k$-values of the rest of the sample, entail that these bursts should be considered separately.}

\begin{figure}
 \includegraphics[width=\columnwidth]{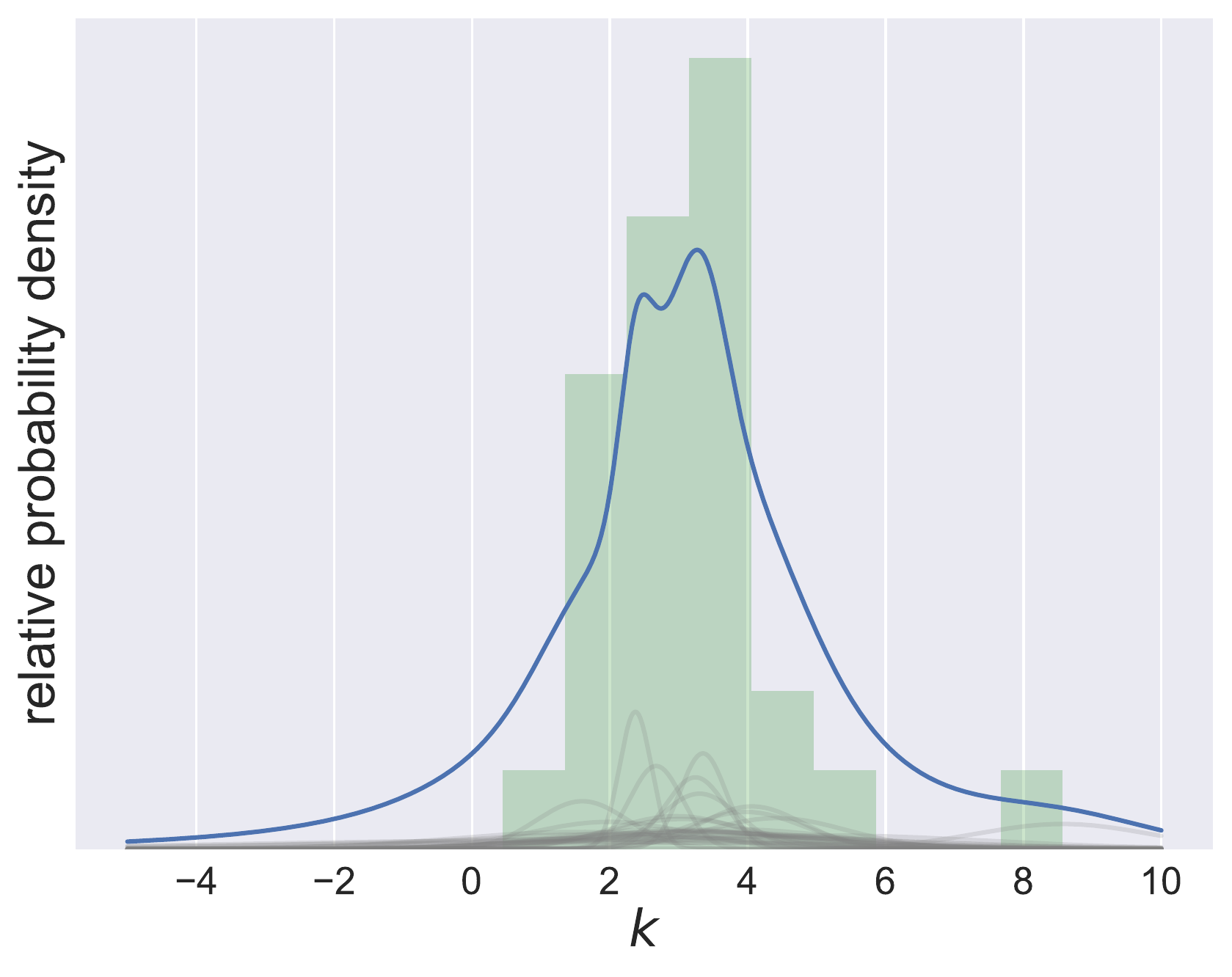}
 \caption{The green histogram represents the distribution of positive $k$-values from Table~\ref{tab:table1}. The blue curve is the overall KDE of $k$. The grey curves are individual marginal posterior distribution of $k$. {In order to be conservative, the maximum of the two asymmetrical errors of $k$ is used.}}
 \label{fig:kde}
\end{figure}

{The distribution of the positive $k$-values has median $k_{\rm mean} = 2.80$}. These are shown in Fig.~\ref{fig:kde} as a histogram (green) together with the probability density function (blue), found through a kernel density estimation (KDE) of individual $k$-values (grey). 
 In Table~\ref{tab:table1}, we also provide the peak value of the energy flux and the largest value of low energy index, $\alpha_{\rm max}$ for all the individual bursts. These values correspond to the maximal values on the two axes in Figures 
~\ref{fig:figureB1}--\ref{fig:figureB3}.

We also compared the correlations found from using Band function for the spectral fits, instead of Eq.~(\ref{eq:cpl}). However, since less spectra can be constrained with the Band function as aforementioned, not all correlations can be compared. For bursts which do have more than 5 data points from fits that are constrained, we find that the posterior distributions of the correlation overlap the ones in Figure \ref{fig:figure1}. Likewise the $k$-values of the mean of the posterior distributions are also similar. {Excluding negative values, the median value is $k^{\rm Band}_{\rm median} = 3.67$}, which is only slightly larger, {considering the dispersion between bursts}. We conclude that the results from the correlation study is not largely affected by the choice of spectral fitting function.

\subsection{Consistency Between Rise and Decay Phases}

{As noted in the previous section, the correlation between the intensity and $\alpha$ shown in Figures \ref{fig:figureB1}-\ref{fig:figureB3}, 
are typically valid throughout the pulse, seemingly covering both the rise and the decay phases, with only a few exceptions, such as GRB160910.  A clear example of a burst with many time bins covering both the rise and the decay phases is GRB160530. Here the correlation moves up and down the same trend. A fit to the rising and decaying phase of this bursts gives consistent slopes to within 1$\sigma$. Further five pulses have more than five time bins with $S>20$ during the rise phase and can thus be analysed in the same way\footnote{GRBs 081009, 090820, 150213, 150314, and 150927.}. In all these cases, the rise phase and decay phase $k$-values are consistent to within 1$\sigma$. However, typically the rise phase is short and only contain a few data points with $S>20$. This makes a quantitative analysis, for the whole sample, difficult. However, a visual inspection of the trend including the early time bins with $S<20$ (orange, yellow, and no circles) indicates that they do follow the general trend set by the later timebins, of which most are during the  decay phase.
}


\subsection{Parameter Covariance}\label{subsubsec:covariance}

For any type of correlation between two parameters 
{a possible concern is whether the covariance between the parameters for individual time bins might be related to the observed behaviour across time bins. This could give rise to a correlation between the parameters simply due to the parameters compensating for each other at the maximum of the posterior probability distribution. Such a correlation would thus not be physical and depending on its size it could affect the correlation across time bins, found over pulses. In order to investigate this we repeated the Bayesian analysis procedure with a different prescription of the cutoff power-law function in which the energy flux $F$ and $\alpha$ are both fitted parameters \citep{Calderone2015}. Then the covariance between $F$ and $\alpha$ can directly be investigated. If the shape of the contour of pairs of parameters are clearly non-circular this might indicate possible functional correlation. Typically, though, some degree of elongation from a circle is expected.
}

\begin{figure*}
\centering

\subfigure{\includegraphics[width=0.45\linewidth]{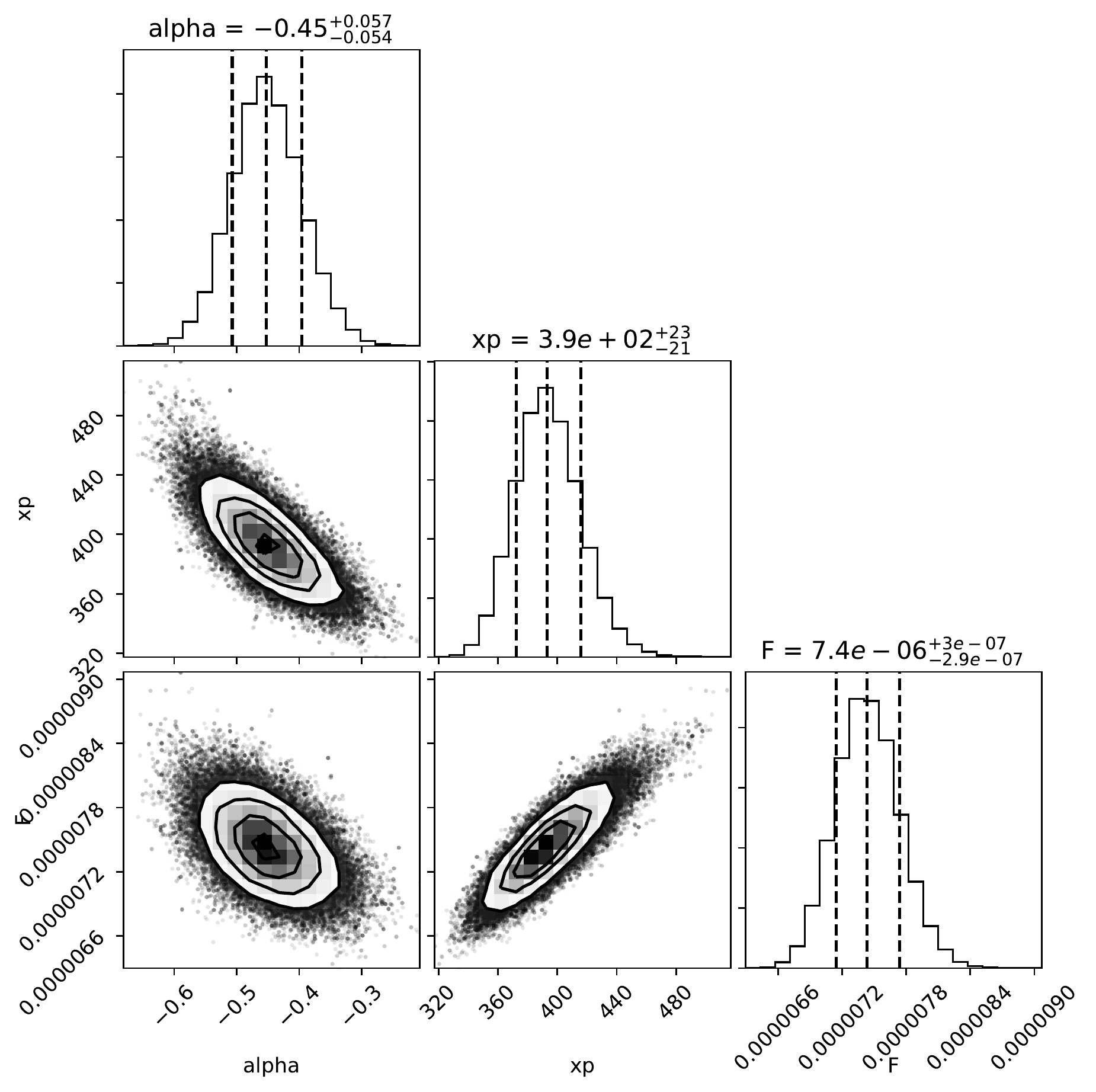}}
\subfigure{\includegraphics[width=0.45\linewidth]{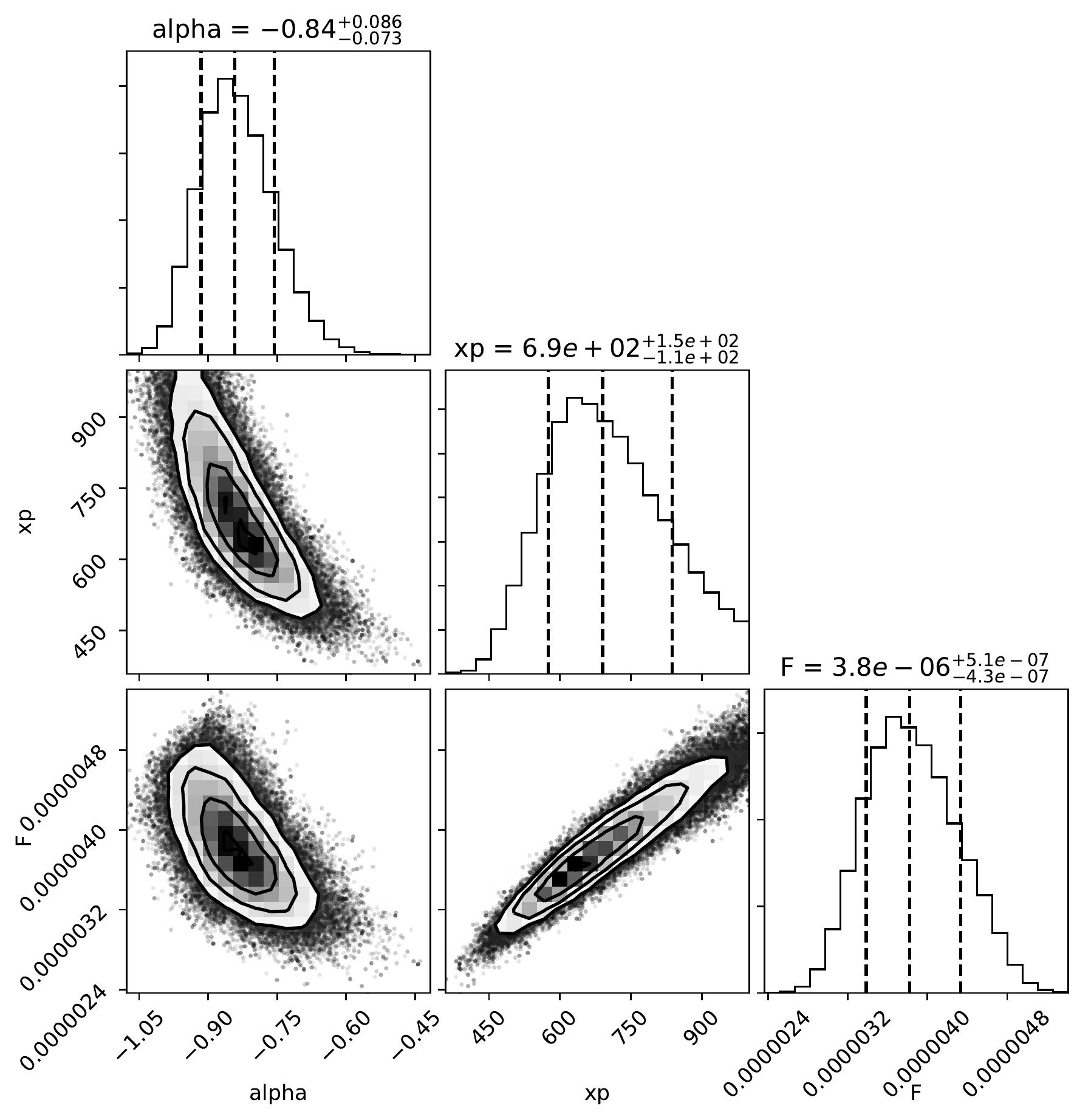}}

\caption{Bayesian Posterior corner plots for GRB090719 (left) and GRB160215 (right). The parameter functional correlations are show in the 2D contours, while the marginal distributions of the parameters are shown in histograms. The contour lines correspond to 68\%, 95\%, and 99.7\% highest posterior density intervals (HPDIs), and the vertical dashed lines in the marginal distributions show the mean and 68\% HPDIs. Note that xp means $E_{\rm p}$, alpha means $\alpha$, and $F$ is in units of erg~s$^{-1}$~cm$^{-2}$. 
\label{fig:corner}}
\end{figure*}

{ We illustrate these fits with the posterior corner plots of the peak time bin in two bursts in Fig.~\ref{fig:corner}: GRB090719 and GRB160215. We choose these two burst since they have opposite behviours in their $F$-$\alpha$ relations. We find that 
the two types of bursts do not show any significant difference in the parameter contours. Furthermore, among the parameter pairs, the elongation of the $F$-$\alpha$ contour is the smallest, for both bursts, which indicates the least covariance between the parameters. Furthermore, it is observed that the $F$-$\alpha$ contours have their  major axis along the $y=-x$ direction. This is in the opposite direction to the observed relation over the pulses, which is along the $y=x$ direction. This shows that the parameter covariance between $F$ and $\alpha$ due to the fitting procedure in individual time bins should not be greatly affected the observed $F$-$\alpha$ relation between time bins, over the pulses. 
}

\subsection{Indication of a Casual and Physical Relationship}

We have shown that the correlation between $F$ and $\alpha$ is valid throughout the pulse, both during the rise and the decay phases, with only a few exceptions. Moreover, the correlation is similar among the bursts.
This is in stark contrast to the two  correlations that have earlier been discussed for the prompt emission phase in GRBs, namely the $F$ versus $E_{\rm pk}$ \citep{golenetski83} and $E_{\rm pk}$ versus $\alpha$ \citep{Kaneko2006} correlations. Both of these {  correlations exhibit a diversity among bursts and} are typically only valid during the decay phase of GRB pulses. In \citet{Yu2018}, we show for the bursts in our sample that even though the $F$-$E_{\rm pk}$ and $E_{\rm pk}$-$\alpha$ correlations are diverse and sometimes complex, the $F$-$\alpha$ correlation remains relatively simple, and consistently a monotonous, increasing function. 
%
%
The monotoneous relationship between $F(t)$ and $\alpha(t)$ is suggestive of a direct casual and physical relationship between the intensity and spectral shape. 

\section{Discussion}
\label{sec:discussion}

In order to discuss a possible cause for this observed relationship, we first note that fit parameters such as $\alpha$ and $E_{\rm pk}$ vary smoothly across the pulses. It is therefore natural to assume that the same mechanism is responsible for producing the emission throughout the pulse. {  This assumption will be important for our following discussion.}

\subsection{Emission Mechanism}
\label{sec:emissionmechanism}

The low-energy power-law slope, $\alpha$, is an indicator for the emission mechanism.
Therefore, the fact that $\alpha$ varies over a large range of values within individual bursts {is of importance}. For instance, for GRB100707, $\alpha$ varies between $0.5$ and $-1.5$, while in GRB100528 $\alpha$ varies between $-1$ and $-2$. In most pulses (24/38) the maximal $\alpha$-value, $\alpha_{\rm max} > -0.67$, which is the value expected for slow-cooling synchrotron emission (see Table~\ref{tab:table1}). This fact is at odds with synchrotron emission being the emission mechanism for these pulses,  assuming the same emission mechanism throughout the pulse. 
Indeed, for bursts which have $\alpha_{\rm max} > 0$, such as GRB100707, the emission has to  originate from the photosphere, since no non-thermal emission mechanism can yield such a hard spectrum. 
Even though these bursts are initially hard, they  soften considerable at late times. Using similar arguments as in \citet{Ryde2011} for GRB090902B, the values of $\alpha \sim -1$ occurring towards the end of such pulses  suggest that the photospheric spectrum is broadened by significant heating of the jet \citep{Peer2006,Beloborodov2010,Ahlgren15}.





The 14 bursts which have $\alpha_{\rm max} \lesssim -0.67$  do, however, evolve within values that are allowed by synchrotron emission ($\sim -0.67$ for slow cooling and $\sim -1.5$ for fast cooling regimes, respectively). There are, though, two points that disfavour a synchrotron interpretation { involving a regime change,} for these cases. First, contrary to what is expected for the transition into the fast cooling regime, the observed intensity decreases as the spectra broaden. Second, 
such an interpretation would require that a change of emission regime always occur in GRB pulses in order to explain the evolution in $\alpha$. This would require a fine-tuning. 

{  On the other hand, as discussed in \S \ref{sec:instrumental}, the limited energy bandwidth of the instrument could give rise to a spurious evolution in $\alpha$, even though the true spectrum is due to slow-cooling synchrotron emission during the whole pulse. 
Under the assumption that there is no spectral evolution (such as a change in emission regime) and only the window effect is dominant, then the expected $\alpha$-$E_{\rm pk}$ relation should follow the curves in \citet[][see also \citet{Burgess2015_alpha}]{Lloyd2000}, being predominantly positive.
However,  only three bursts in our sample have an $\alpha$-$E_{\rm pk}$ relation that is consistent with such a scenario \citep[][see also \citet{Ghirlanda2002}]{Yu2018}.
Here we note that despite such an agreement for these three bursts, 
this scenario is still only physically viable in a limited range of outflow parameters, e.g., only at very large emission radii \citep{Beniamini2013,Beniamini2018}, which sets constraints on, among other things, the variability time-scale \citep{Burgess2016}. 
}

\begin{figure}
 \includegraphics[width=\columnwidth]{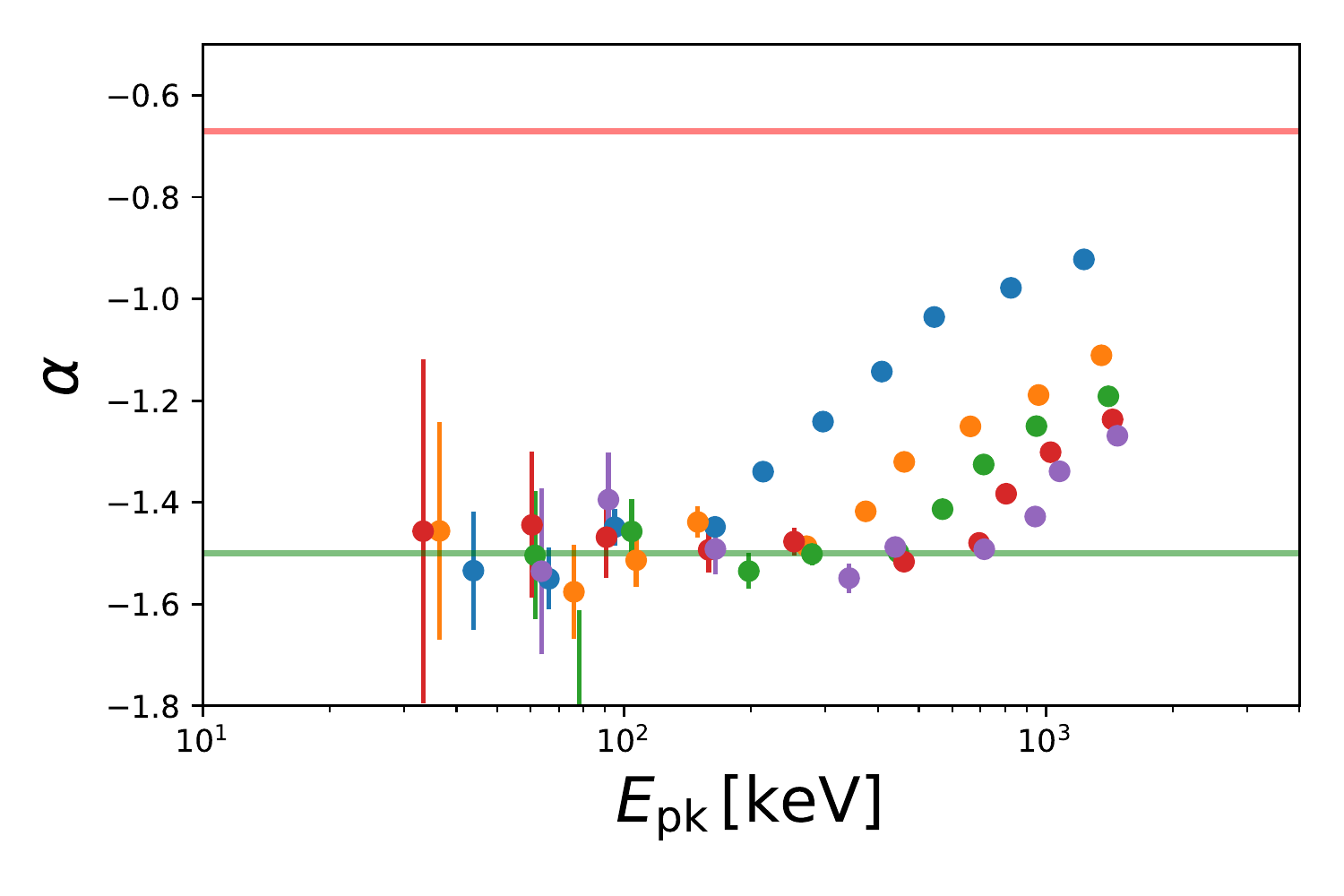}
 \caption{
 Results of spectral fits using a cutoff power-law function on synthetic GBM spectra. The synthetic spectra are produced from a model mimicking a marginally cooled synchrotron emission, letting the two breaks $E_{\rm pk}$ and $E_{\rm b}$ vary independently of each other. A SNR = 100 is used to high-light the trends. The different colors represent spectra with different ratios of $E_{\rm pk}/E_{\rm b}$; Blue: $E_{\rm pk}/E_{\rm b}=5$, Orange: $E_{\rm pk}/E_{\rm b}=18$, Green: $E_{\rm pk}/E_{\rm b}=31$, Red: $E_{\rm pk}/E_{\rm b}=44$, and Purple: $E_{\rm pk}/E_{\rm b}=57$. The red and green lines shows the expected slopes for synchrotron emission. 
}
 \label{fig:twobreaks}
\end{figure}

{ A further possibility for an apparent change in $\alpha$, is assuming marginally fast cooled synchrotron emission. 
In such a model, the two synchtrotron breaks appear close to each other in the spectrum:  $E_{\rm pk}$ corresponds to the minimum, injected electron energy and $E_{\rm b}$ is interpreted as the cooling break. Then $\alpha = -2/3$ and $-1.5$, below and above $E_{\rm b}$, respectively.
\citet{Yu2015b} used  a doubly broken power-law to approximate such a situation and found 
a ratio between the two breaks generally smaller than 10, with a peak in the distribution $E_{\rm pk}/E_{\rm b} \sim 4$. Similarly, \citet{Oganesyan2018} modeled such spectra with a Band function (with $E_{\rm pk}$) multiplied with a high-energy cutoff at $E_{\rm b}$ \citep[see, also, ][]{Oganesyan2017, Ravasio2018}.
 In the fits that they perform, the ratio $E_{\rm pk}/E_{\rm b}$ lies in the range $\sim 5$ to $\sim 70$.  In such scenarios, an evolution of $\alpha$ (from a fit with a cutoff power-law function) could be caused by particular evolutions of $E_{\rm pk}$ and $E_{\rm b}$. Depending on the position of $E_{\rm pk}$ and $E_{\rm b}$ relative to the observation window, the different spectral slopes ($\alpha = -2/3$ and $-1.5$) could dominate the GBM window. If such a spectrum is fitted by a cutoff power-law, the fitted $\alpha$ could then appear to evolve. In order to investigate such a scenario, we again make synthetic GBM observations, but now assuming that the true incoming spectrum is described by a Band function with a high-energy cut-off, with $\alpha_{\rm band} = -0.67$ and $\beta_{\rm band} = -1.5$ \citep{Oganesyan2017}.  In Figure \ref{fig:twobreaks}, we show the results of the fitted values of $E_{\rm pk}$ and $\alpha$ using a cut-off power-law function. For the synthetic spectra, we allow for a range of $E_{\rm pk}$-values and for several different values of the ratio of $E_{\rm pk}/E_{\rm b}$ in the range 5 -- 70. We find that in no case does the spectra reach the asymptotic slope of $\alpha=-2/3$, and the spectral variation is limited to above 200 keV. In only cases in which $E_{\rm pk}/E_{\rm b}\lesssim 20$ do the fits yield $\alpha > -1.2$. Theoretically, it is not obvious why the cooling break and the minimum electron energy should line up to within a factor of 10 \citep{Beniamini&Piran2013}. 
Finally, we note that there is no evidence of the synthetic spectral evolutions found in Figure \ref{fig:twobreaks} in the data of our sample \citep{Yu2018}.
}

{  Most importantly, though,} we note that the values of $k$, found above in \S \ref{sec:analysis}, are similar for many of the bursts and, in particular, independent of the range of $\alpha$-values in individual bursts. Even though the two examples given above (GRB100707 and GRB100528) cover different ranges in $\alpha$ the $k$ values are similar. 
{  Likewise, the three bursts which are consistent with slow-cooled synchrotron emission have similar $k$-values to the rest of bursts.} This suggests that the mechanism giving rise to the correlation should be the same in all bursts. We, therefore, argue that these facts together point towards that the same emission mechanism operate in all bursts, and that it is emission from the photosphere, including heating at high optical depths.

\subsection{A qualitative photospheric emission scenario}
\label{sec:scenario}

The jet properties are in general expected to be variable. The ratio of the photospheric radius, $r_{\rm ph}$, to the saturation radius, $r_{\rm s}$, has a strong dependence on the dimensionless entropy $\eta = L/{\dot{M}}c^2$, where $L$ and ${\dot{M}}$ are the kinetic luminosity and the baryon load, respectively. The ratio is $r_{\rm s}/r_{\rm ph} \propto L/\eta^4$ \citep[for $r_{\rm ph} > r_{\rm s}$, e.g.,][]{Rees&Meszaros2005, Ryde2017}. Thus, a small variation in $\eta$ can change the location of the saturation radius, relative to the photospheric radius. Since the $\eta$-dependence is so strong, we neglect variations in the other quantities for the qualitative discussion below.

Within hydrodynamically dominated models, dissipation is in general expected to be ineffective below $r_{\rm s}$. This is because the jet kinetic luminosity is much smaller than the radiation luminosity, so that dissipating a fraction of the kinetic energy will not affect the radiation much. We define $\eta_c$ as the critical value of $\eta$ that $r_{\rm ph} = r_{\rm s}$. Jets with $\eta \gg \eta_c$, which have $r_{\rm ph} \ll r_{\rm s}$, are expected to be luminous (as essentially all energy is carried by radiation) and appear almost thermal (as any dissipation is weak as compared to the already existing radiation energy). The fact that the jet is still accelerating as it becomes optically thin means that the observed spectrum becomes very narrow; see Eq.~(2) in \citet{Ryde2017}.

For $\eta \sim \eta_c$, we find $r_{\rm ph} \sim r_{\rm s}$. The jet kinetic energy is then comparable to the radiation energy, and very strong dissipation event might be able to disturb the radiation spectrum. At the same time, geometrical broadening of the spectrum makes the low energy slope somewhat softer \citep{Beloborodov2011, Lundman2013}.

If $\eta \ll \eta_c$ we find $r_{\rm ph} \gg r_{\rm s}$. The jet kinetic energy then dominates the radiation energy, making the emission weaker. At the same time, the kinetic energy reservoir is large as compared to the radiation energy, and dissipation can easily affect the radiation spectrum. For instance, in the continuous dissipation models of \citet{Vurm2011, Vurm2016}, the dissipation produces relativistic electrons that cool partially by emission of low-energy synchrotron photons. Unsaturated Comptonisation of the synchrotron photons forms a soft power law below the spectral peak at moderate optical depths. In summary, $\eta < \eta_c$ translates to weaker emission with a soft spectrum below the peak, while $\eta > \eta_c$ gives strong emission with an almost thermal spectrum. This simple scenario is therefore qualitatively in agreement with the findings of this paper.

As the ratio of $r_{\rm ph} / r_{\rm s} = \tau_{\rm sat}$ will affect both the spectral shape and the intensity, this also means that if this ratio evolves in time, then both $\alpha$ and $F$ evolve.
To estimate the range of these evolutions, we note that if $r_{\rm s}$ varies from $r_{\rm ph}$ to $r_{\rm w}$ then $\alpha$ should vary from $0$ to $-1$, where $r_{\rm w}$ is the radius within the Wien zone \citep{Beloborodov2010,Vurm2016}.
This is because when the photosphere transitions from the accelerating to the coasting phase then the measured $\alpha \sim 0$ \citep[see ][]{Ryde2017}, or equivalently,  $\alpha_{\rm sat} \equiv \alpha(\tau_{\rm sat} = 1) \simeq 0$. Furthermore, \citet{Vurm2016} showed that when $r_{\rm w}>r_{\rm s}$ the softening gives a typical value of $\alpha \sim -1$ (marked by dashed lines in Fig.~\ref{fig:figure1}). Therefore, $\alpha_{\rm w} \equiv  \alpha(\tau_{\rm w} \sim 10^2) \sim -1$.  This variation in $\alpha$ will be accompanied by a variation in flux. 
If we assume that the fraction of the kinetic energy that is dissipated in the coasting phase is only a small fraction, but still large enough to affect the spectral shape, the flow can be approximated as being adiabatic.
In such a case the corresponding flux variation, due to adiabatic cooling, will be  $F(\tau_{\rm sat}=1)/F(\tau_{\rm sat} = \tau_{\rm w}) = ( r_{\rm ph} / r_{\rm w})^{2/3}= \tau_{\rm w}^{2/3}$.  Finally, these estimations can be used to determine $k$  in equation (\ref{eq:2}),  which gives $\alpha_{\rm sat} - \alpha_{\rm w} = k^{-1} \, 2/3 \, \ln \tau_{\rm w}$. Therefore, $k = 2/3 \, \ln \tau_{\rm w} \approx 3.07$, with $\tau_{\rm w} = 10^2$.   The expected value of the slope of the correlation for this scenario is thus $k \sim 3$.

\subsection{Explaining the $\alpha$-intensity correlation}
\label{sec:aI}

A variation in $\eta$ thus provides a natural explanation of the qualitative behavior of the pulses in GRBs and, in particular, of the $\alpha$-intensity correlation. As mentioned above, large values of $\eta$ at the pulse peak would cause intense, narrow spectra.
During the decay phase of the pulse, the assumed decrease in $\eta$ will cause the photosphere 
to secede from the saturation radius, thereby making the emission weaker and, at the same time, the spectrum broader. The reverse happens during the rise phase of the pulse. 

{GRBs that are emitting close to their saturation radius has been discussed before.
\citet{Ryde2017} reanalysed the two {\it Fermi}/GBM bursts with the narrowest reported spectra (GRBs 100507 and 101219). They fitted a physical model for emission from a non-dissipative photosphere to the time resolved spectral data. The fits showed that the photosphere occurs close to the saturation radius of the flow.  
In both these cases 
the free jet expansion has to begin at $r_0$ of a few $\times \,  10^9$ cm. Such a radius is interpreted to be just within or comparable to the radius of the core of the progenitor star \citep{Thompson2007, DeColle2017}. While in these two cases the spectra are very hard throughout their duration, many of the burst spectra in the \citet{Yu2018}-sample evolve significantly over the pulse, which therefore suggest a large variation in $\eta$, and heating below the photosphere.}

The entropy $\eta$ (or equivalently the Lorentz factor in saturated flows) is expected to vary during a burst. For instance, \citet{Lopez-Camara2014} preformed numerical simulations of jets breaking out from a progenitor star and showed how the Lorentz factor varies depending on the activity of the central engine. In particular, they showed that even though the central engine is modeled to have a constant luminosity, the Lorentz factor initially varies with a pulse-like structure changing with a factor of $\lesssim 2$. Similarly, \citet{Harrison2018} performed 3D simulations and find that the $\eta$-value typically is largest at the head of the outflow and then decreases due to increased mixing. These simulations thus indicate that the value of $\eta$ at the photosphere evolves and pulse-like variations of it is a natural outcome of the interaction between the jet and the progenitor star.

A majority of the bursts are consistent with this simple scenario. From Table 1 it can be seen that in $\sim 70\%$ of the pulses are consistent with $k=3$, since it is included in the $1\sigma$ error bars (68\% HPDI). The fraction increases to $\sim 90\%$ using the $3 \sigma$ error bars, instead. Most of the other pulses have smaller values of $k$, which corresponds to a smaller change in the flux compared to what is expected from the adiabatic cooling. In only one case (GRB160530) the $k$-values is significantly larger than 3. These deviations from $k=3$ indicate that the presented scenario in its simplest form is insufficient for $\sim 30\%$ (for $1\sigma$) or  $\sim 10\%$ (for $3\sigma$) of the pulses and that evolution in other flow properties need to be taken into account. For instance, it is not expected that the magnetisation, dissipation rate, nor luminosity are necessarily constant, as is assumed in the simple scenario. Other possibilities that would cause deviations include that the selected pulses actually might consist of many overlapping, unresolved pulses, each of which have different properties, or that additional spectral components are significant (see \S \ref{sec:pulsewise}). These possibilities would lead to that the observed correlations do not reflect the emission process directly.

We note that, within the scenario presented in \S \ref{sec:scenario}, the time bins in Fig.~\ref{fig:figure1} which have $\alpha < -1$, would have to be explained by a higher magnetisation of the jet, since this would cause an excess of soft photons, which would decrease the values of $\alpha$ found from fitting such a spectrum with, e.g., Eq.~(\ref{eq:cpl}). 

\subsection{Effects of the limited energy range}
\label{sec:BBwindow}

\begin{figure}
 \includegraphics[width=\columnwidth]{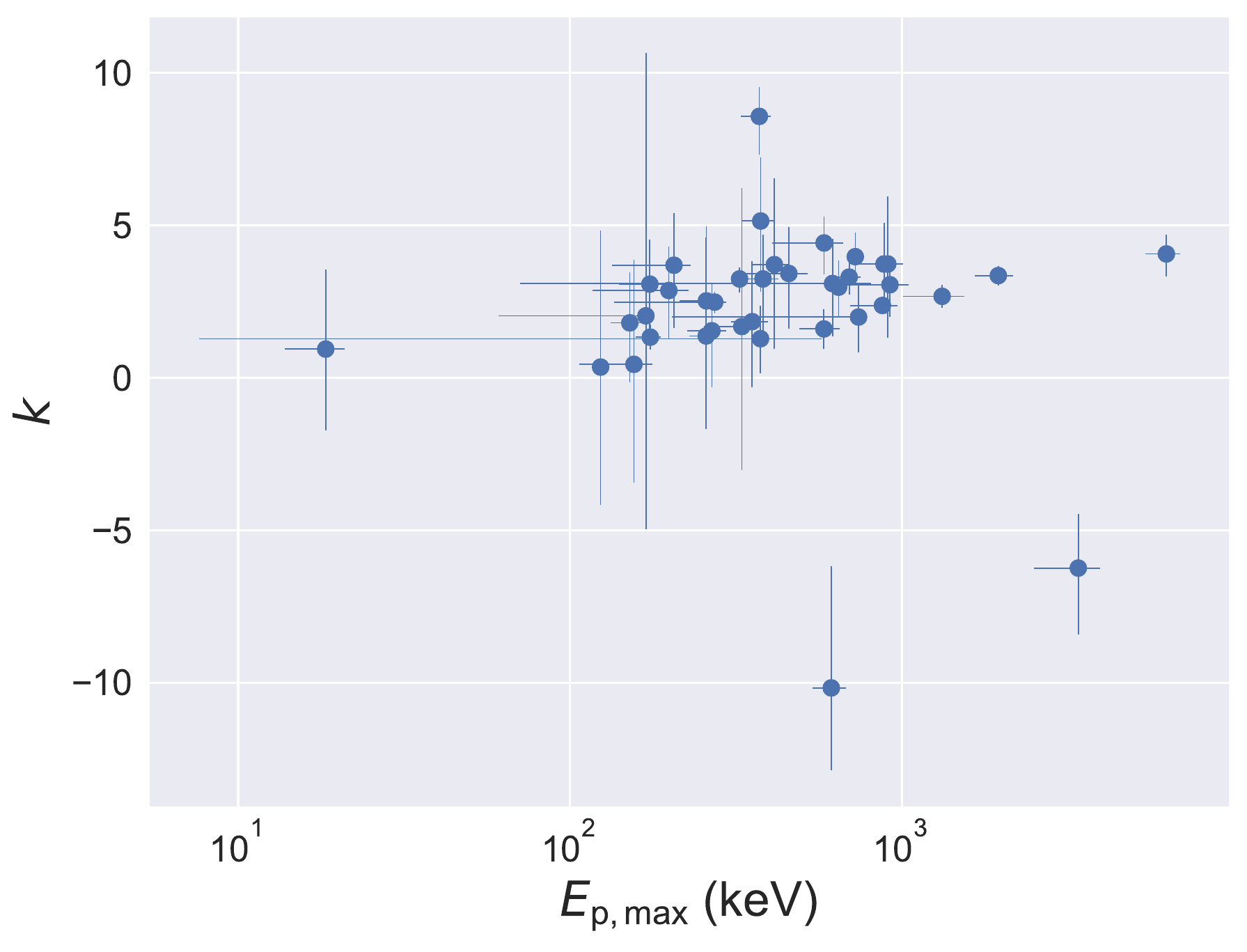}
 \caption{The dependency of the value of $k$ with maximal value of $E_{\rm pk}$. A very slight trend of smaller $k$-values at lower $E_{\rm pk}$ (Spearman's rank correlation coefficient $\rho=0.39$ with a $p$-value of $p \approx 0.01$, i.e., how likely the observed trend is produced by pure chance).}
 \label{fig:Fk}
\end{figure}

In this paper, we have argued for a photospheric emission model in order to explain the spectral evolutions. As noted in \S \ref{sec:instrumental} the window effect in this case cannot a priori be compensated for since the effect depends on the true spectral shape and to what extent Eq.~(\ref{eq:cpl}) can account for it. In the photospheric emission model the spectra can have a variety of shapes and cannot be  known before fitting the data \citep{Peer2006,ahlgren2015confronting}. However, in order to assess if there is an effect on the sample, we investigate if there is a correlation between the determined $k$-value and the maximal value of $E_{\rm pk}$. The reason is that  the window effect will have largest impact on bursts that have low  $E_{\rm pk}$-values throughout their evolution. Furthermore, assuming that the dispersion in $k$ is mainly due to the window effect, the $k$-values should mainly differ for bursts with low $E_{\rm pk, max}$. As shown in Fig.~\ref{fig:Fk} there is a weak trend of smaller $k$-values at lower at $E_{\rm pk}$. {The Spearman's rank correlation coefficient is $\rho =0.39$ for the full sample and $\rho =0.56$ when the two negative cases are removed (motivated in \S \ref{sec:aI}).} Even though the correlation is weak, this effect might lead to an increased dispersion towards lower $k$-values, causing the median $k$-value we find to be slightly underestimated.

{ In order to minimize any effects of the limited energy range, we therefore study a subsample of the pulses by ignoring all bursts that have $E_{\rm pk}< 100$ keV during a significant fraction of their evolution. These are the bursts that would be impacted the most of the  window effect \citep[][Acuner et al. 2019, in prep.]{Lloyd2002,Burgess2015_alpha}. For instance, requiring that more than half of the data points have $E_{\rm pk}< 100$ keV, will remove 7 pulses\footnote{Pulses removed are GRBs 081009140, 081009140 (episode 2), 090530760, 090804940, 100122616, 141205763, 150213001.} from our sample. After removing these pulses the median values determined in \S \ref{sec:analysis} will become slightly larger; 
{$k^{\rm Cpl}_{\rm median} = 3.09$ from previously determined $2.80$ and 
$k^{\rm Band}_{\rm median} = 3.77$ from previously determined  $3.67$.} These changes do, however, not change the overall conclusion that $k\sim 3$ for the pulses investigated.
}



\subsection{Correlations including $E_{\rm pk}$}

The correlations including $E_{\rm pk}$ (the $F$-$E_{\rm pk}$ and the $E_{\rm pk}$-$\alpha$ correlations) are mainly valid during the decay phase of a pulse and show a great diversity \citep[see ][]{Yu2018}. However, the $F$-$\alpha$ correlation is (largely) independent of rise and decay phases in the light curve. In the scenario presented above, the intensity and $\alpha$ are mainly related through the same quantity $\eta$. Therefore, any temporal variation will cause the $F$-$\alpha$ relation to move along the correlation.
On the other hand, the position of the peak of the spectrum, $E_{\rm pk}$,
also depends on properties of the flow below the Wien radius, $r_{\rm w}$, where the optical depth, $\tau_{\rm w} \gtrsim 100$ \citep{Beloborodov2013, Lopez-Camara2014}.  Depending on the photon production efficiency (due to dissipation rates and the magnetisation of the flow) the peak energy can vary between 40 keV and 15 MeV \citep{Vurm2016}. 
This additional dependency could explain why the correlations involving $E_{\rm pk}$ are not valid over the whole pulse {  and show a diversity}. It also suggests that the $\alpha$-intensity correlation is the most fundamental correlation to be studied in GRB pulses.

\section{Conclusions}

We have studied the correlation between the instantaneous values of the energy flux, $F$, and the sub-peak power-law index, $\alpha$, in a sample of pulses observed by {\it Fermi}/GBM.  We find  significant correlations in most pulses, which  are even valid through out the pulse, both during the rise and the decay phases. We explain the correlation within a qualitative photospheric emission scenario, in a flow where the dimensionless entropy $\eta$ varies.  Around the peak of the light curve a large entropy causes the photosphere to approach to the saturation radius. This leads to an intense emission with a narrow spectrum.  When the entropy decreases the photosphere secedes from the saturation radius, and weaker emission is expected. At the same time a spectrum becomes broader, since then heating can easily affect the radiation spectrum. This simple scenario gives a qualitative physical description that describes the observed correlated variation of the intensity and spectral shape and their observed value ranges. This motivates further development of a full physical model exploring the functional relationship between the intensity and spectral shape in GRBs.

\section*{Acknowledgements}

We thank Drs. B\'egu\'e and Burgess for enlightening discussions. This research made use of High Energy Astrophysics Science Archive Research Center Online Service HEASARC at NASA/Goddard Space Flight Center. We acknowledge support from the Swedish National Space Agency and the Swedish Research Council (Vetenskapsr{\aa}det). FR is supported by the  G\"oran Gustafsson Foundation for Research in Natural Sciences and Medicine. 




\bibliographystyle{mnras}
\bibliography{ref2017} 




\appendix

\section{The full sample of $\alpha$-intensity correlations}
\label{sec:fullsmaple}

Below we present the full sample of the $\alpha$-intensity correlations for the 38 pulses. The Bayesian posteriors of fits to Eq.~(\ref{eq:2}) are overlaid the data points, with the best fit indicated by the green line.

\begin{figure*}
\centering

\subfigure{\includegraphics[width=0.33\linewidth]{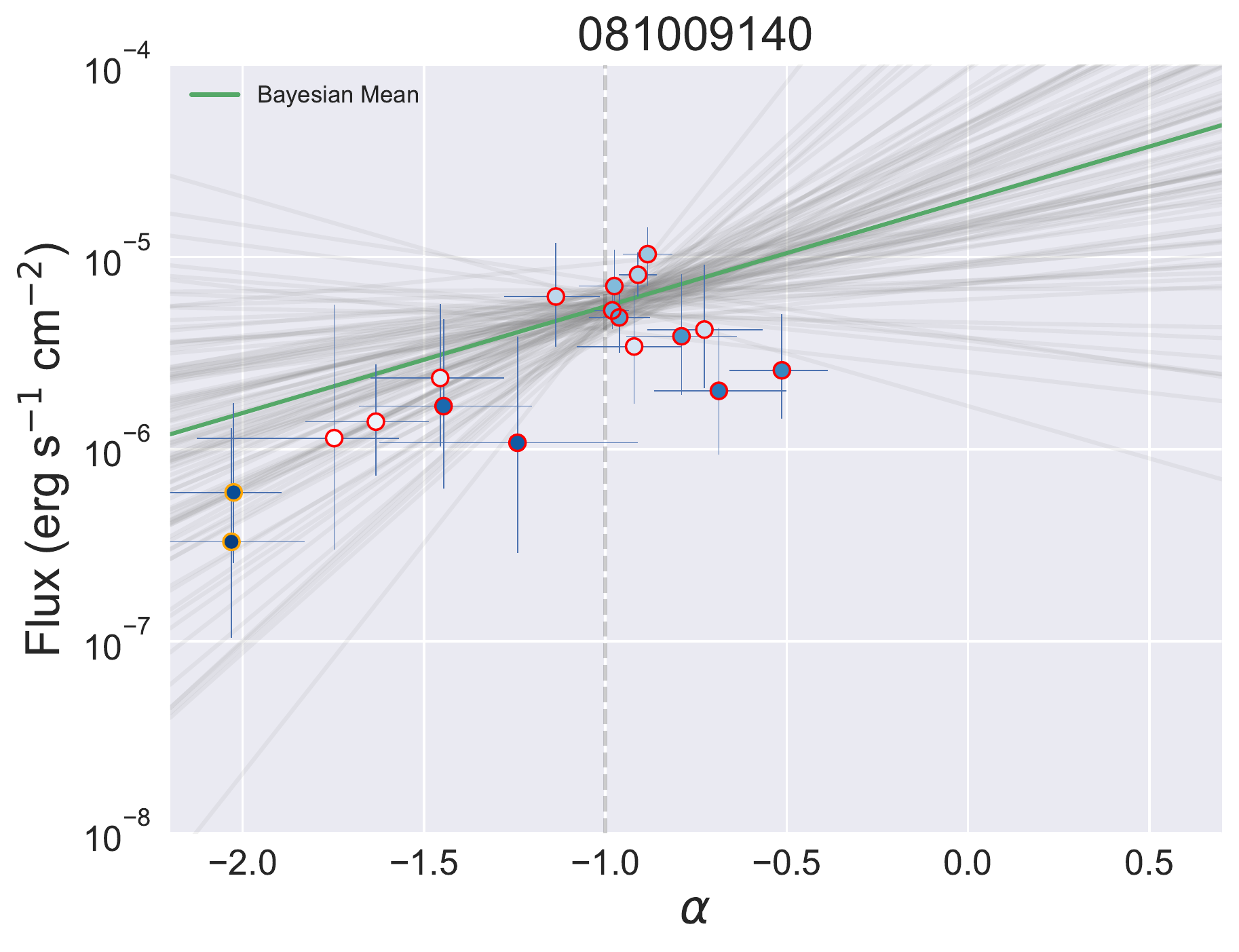}}
\subfigure{\includegraphics[width=0.33\linewidth]{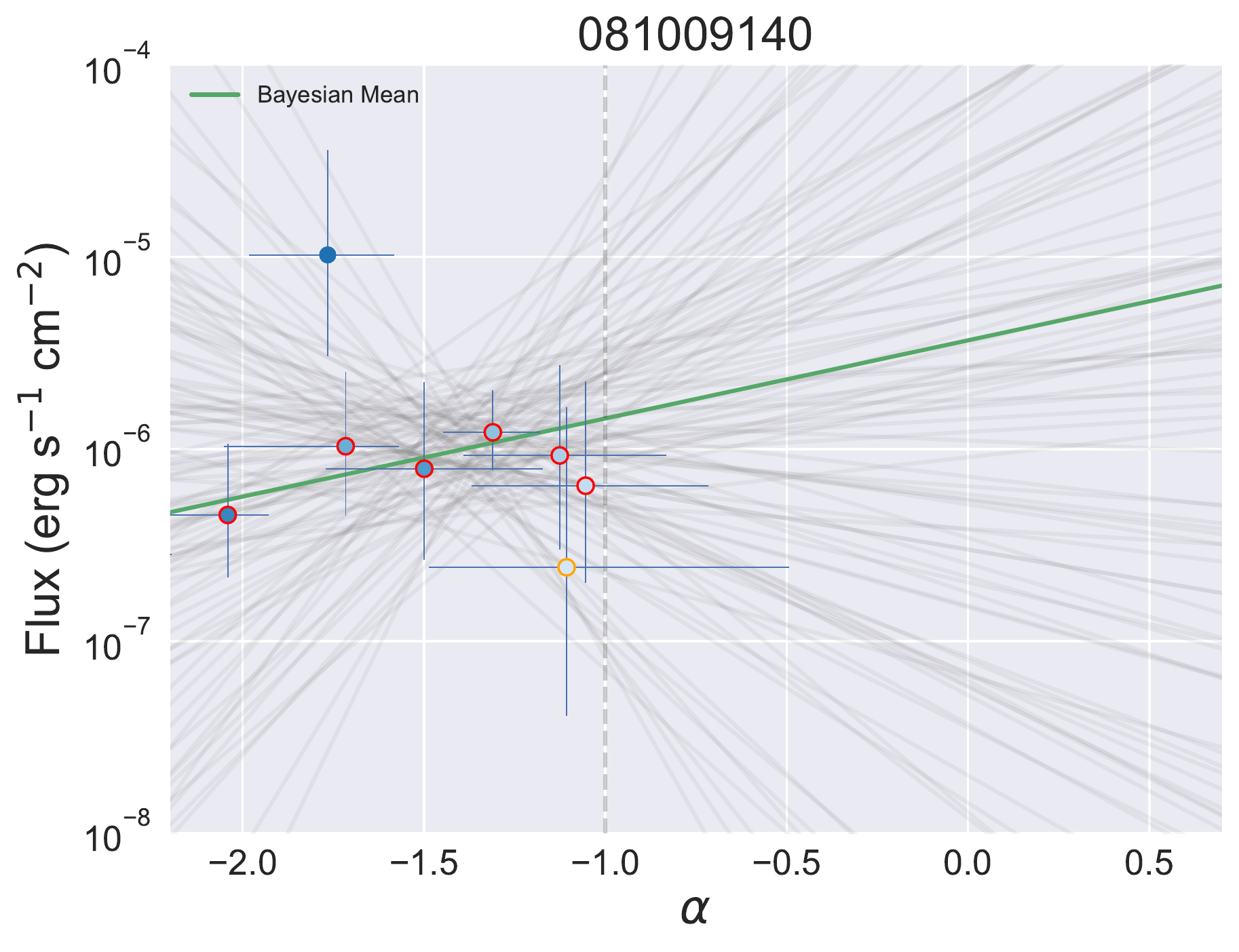}}
\subfigure{\includegraphics[width=0.33\linewidth]{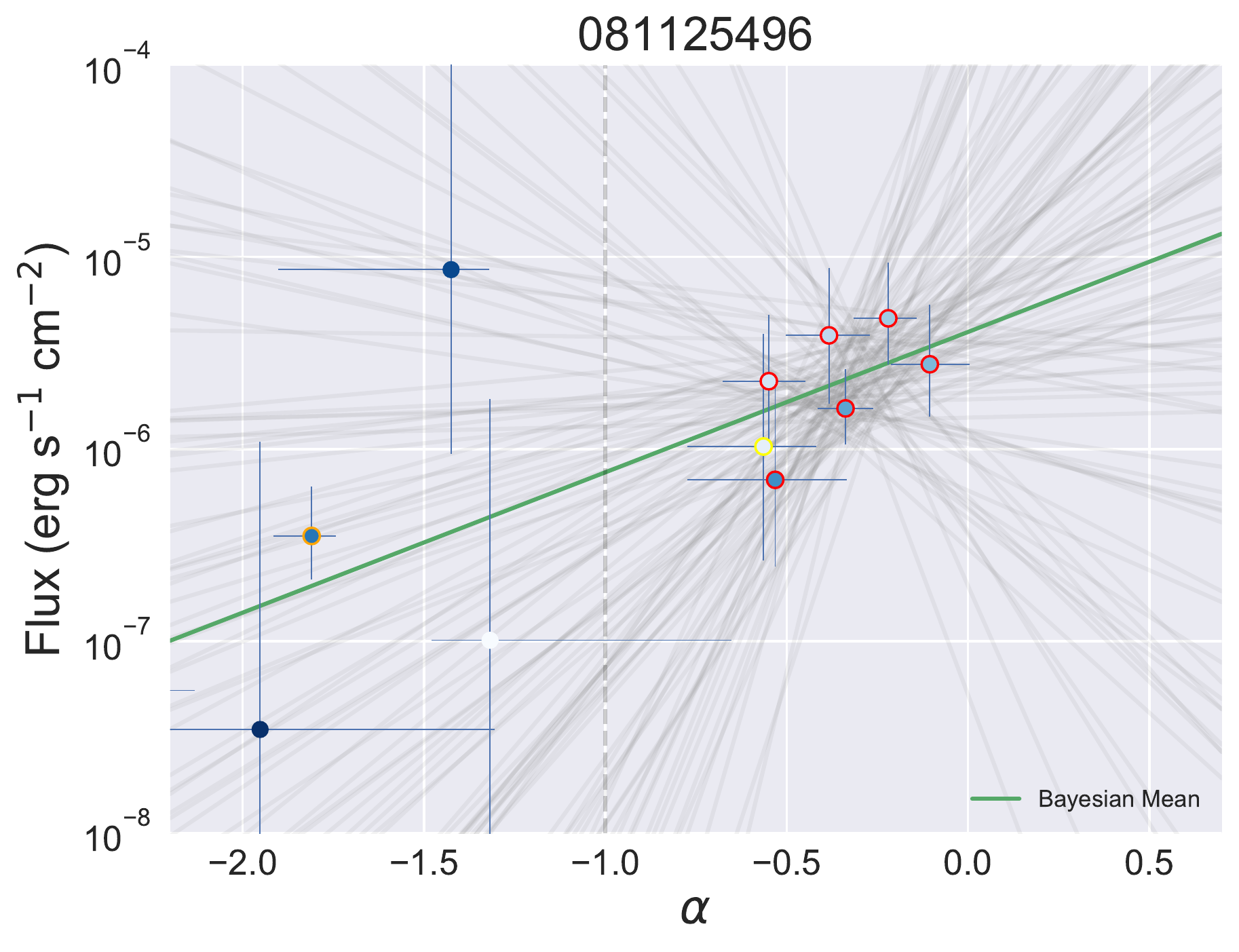}}

\subfigure{\includegraphics[width=0.33\linewidth]{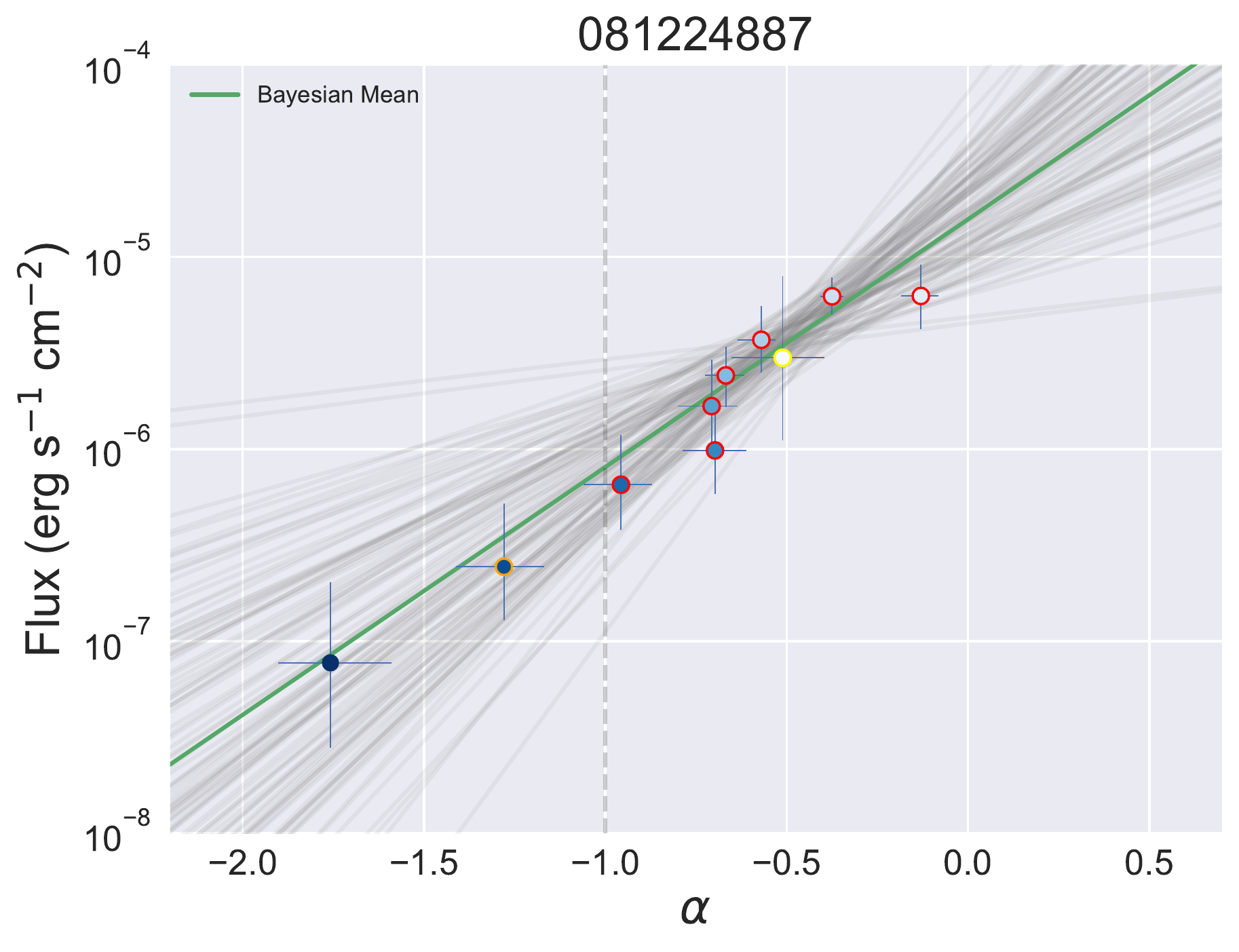}}
\subfigure{\includegraphics[width=0.33\linewidth]{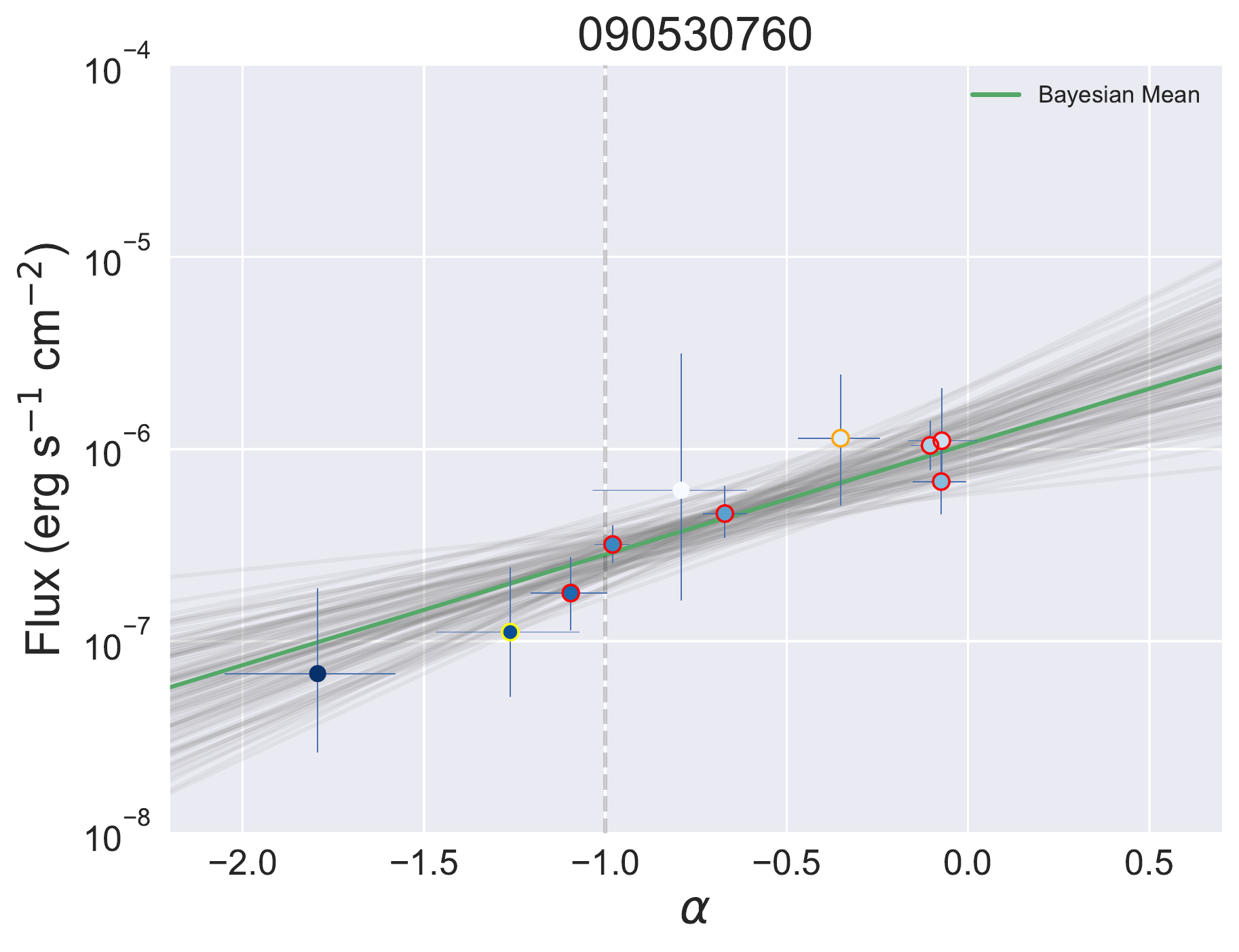}}
\subfigure{\includegraphics[width=0.33\linewidth]{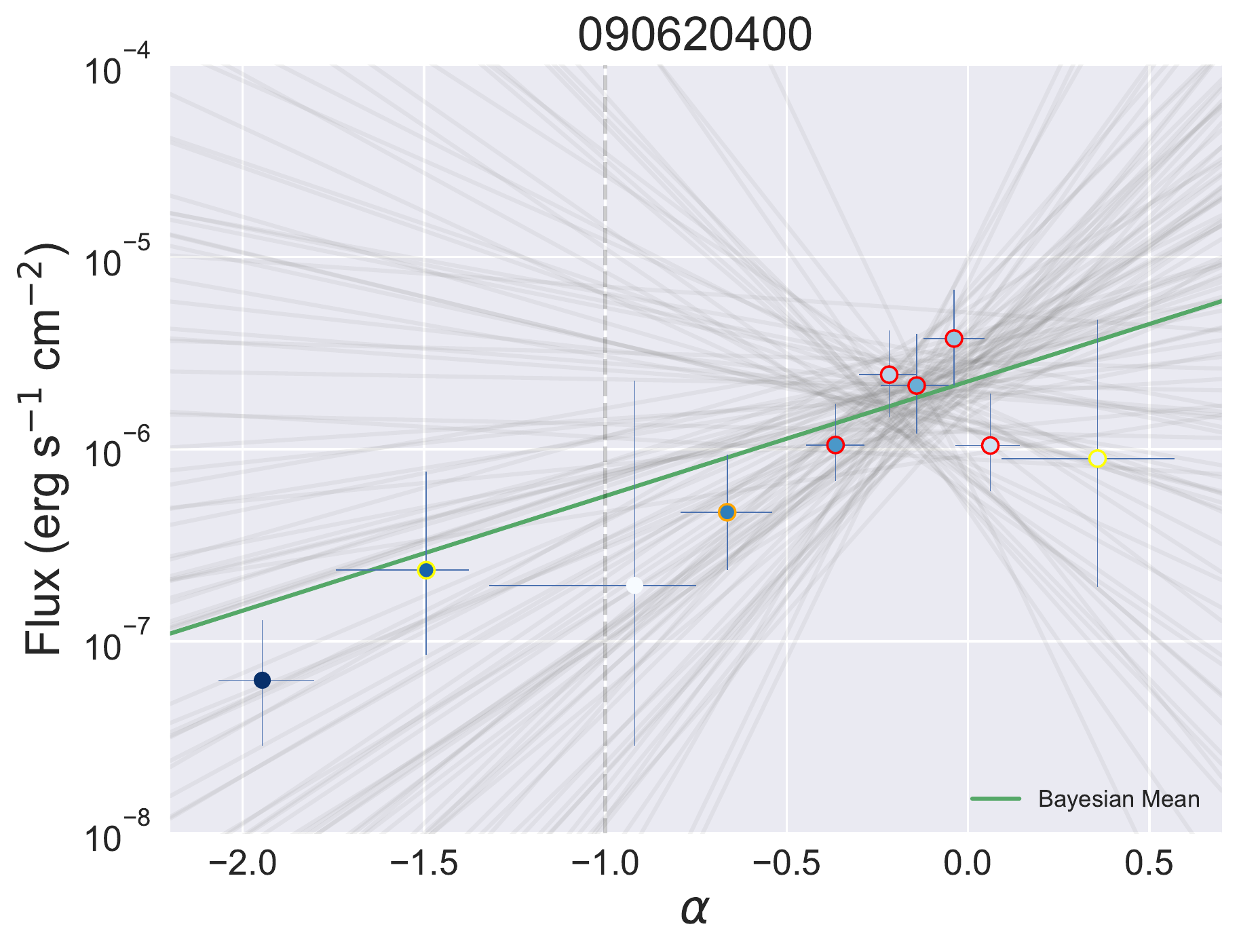}}

\subfigure{\includegraphics[width=0.33\linewidth]{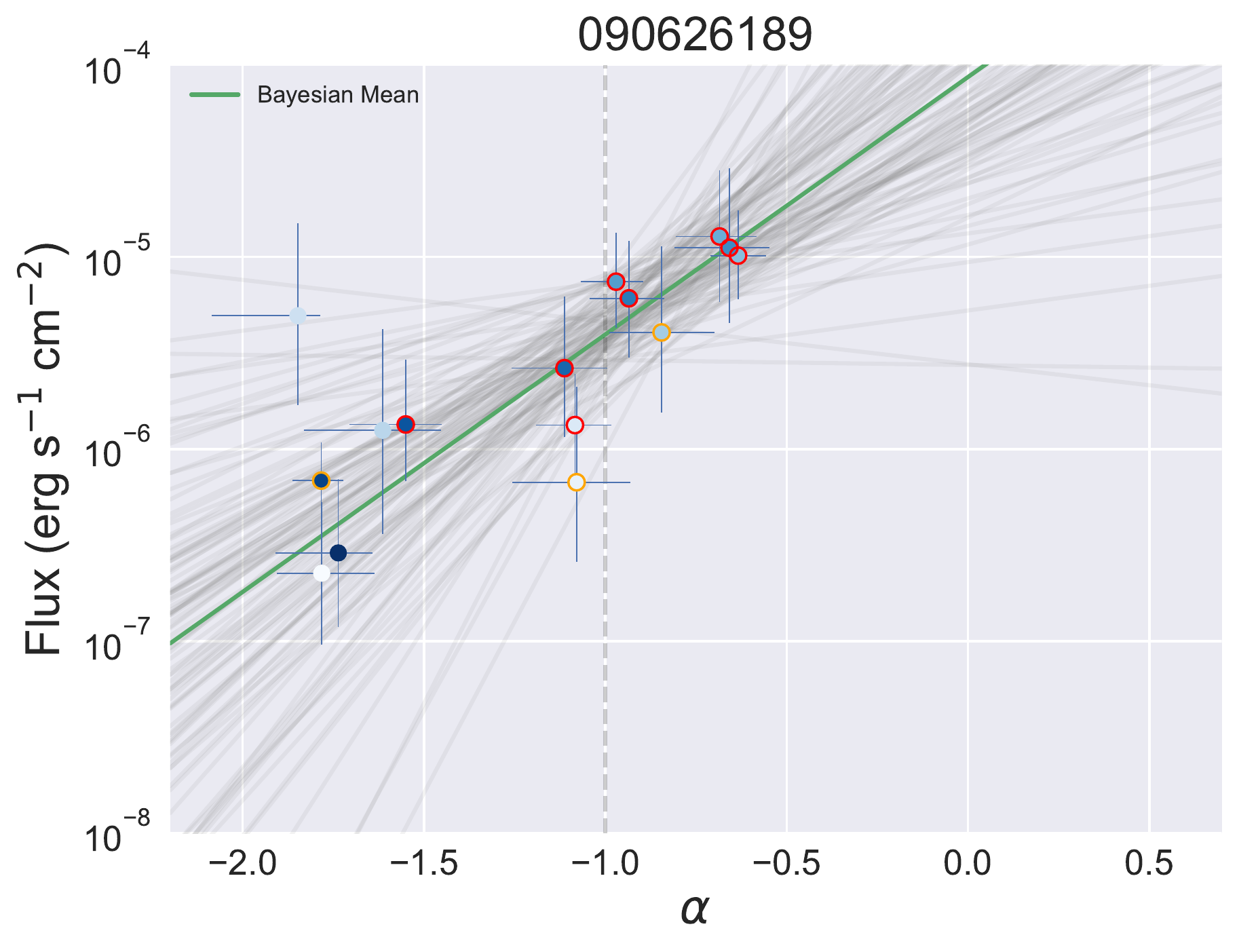}}
\subfigure{\includegraphics[width=0.33\linewidth]{BayesianFit_bn090719063.pdf}}
\subfigure{\includegraphics[width=0.33\linewidth]{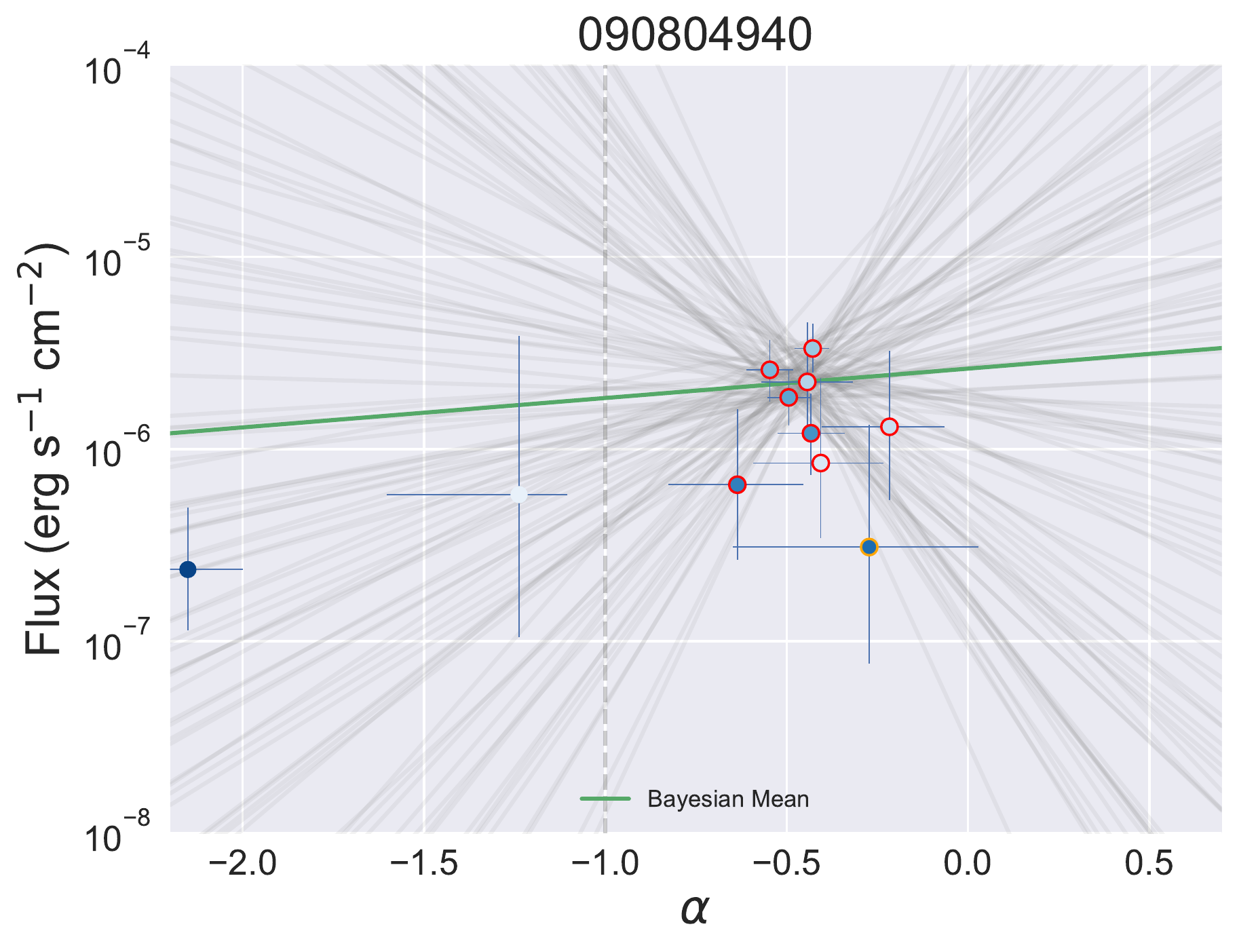}}

\subfigure{\includegraphics[width=0.33\linewidth]{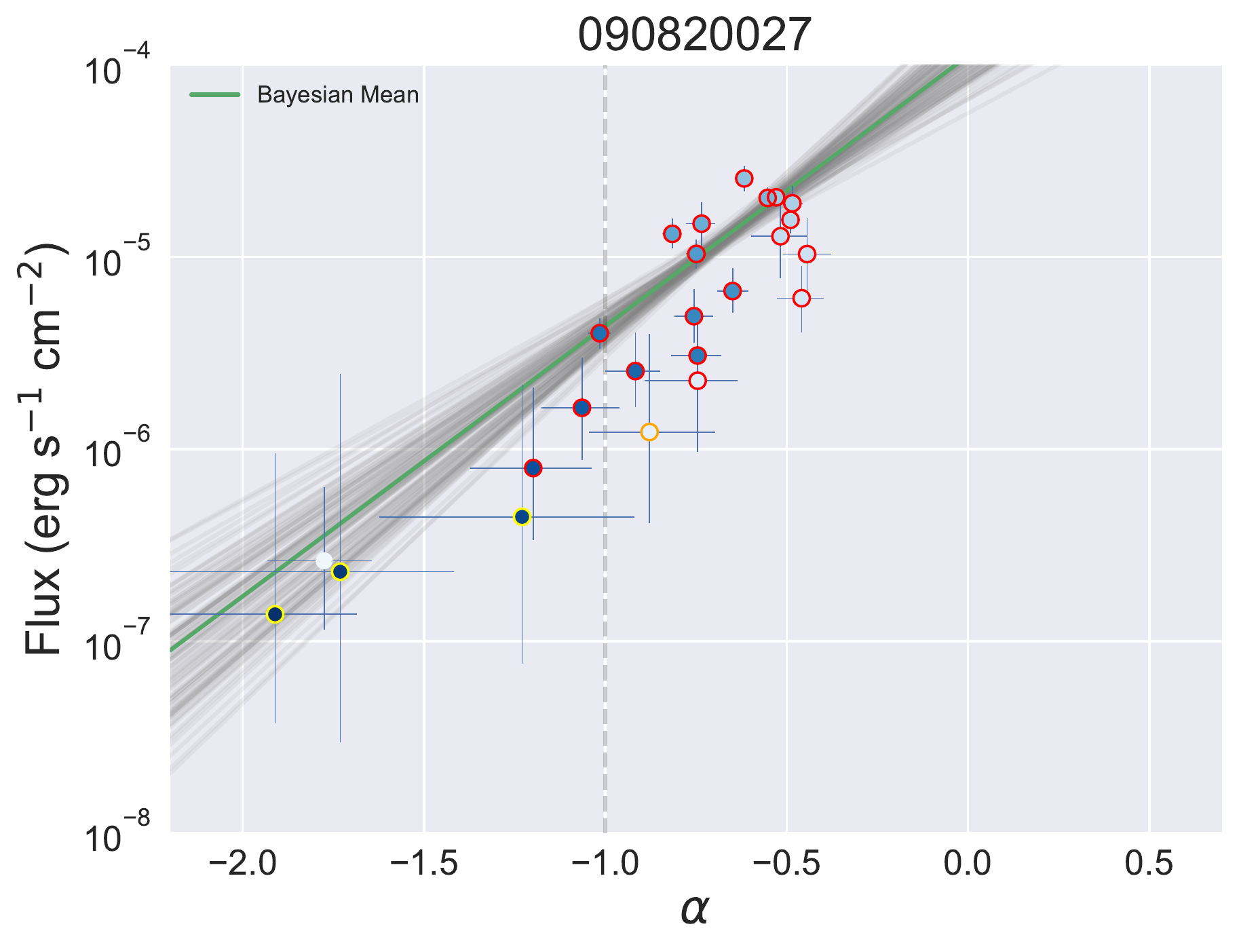}}
\subfigure{\includegraphics[width=0.33\linewidth]{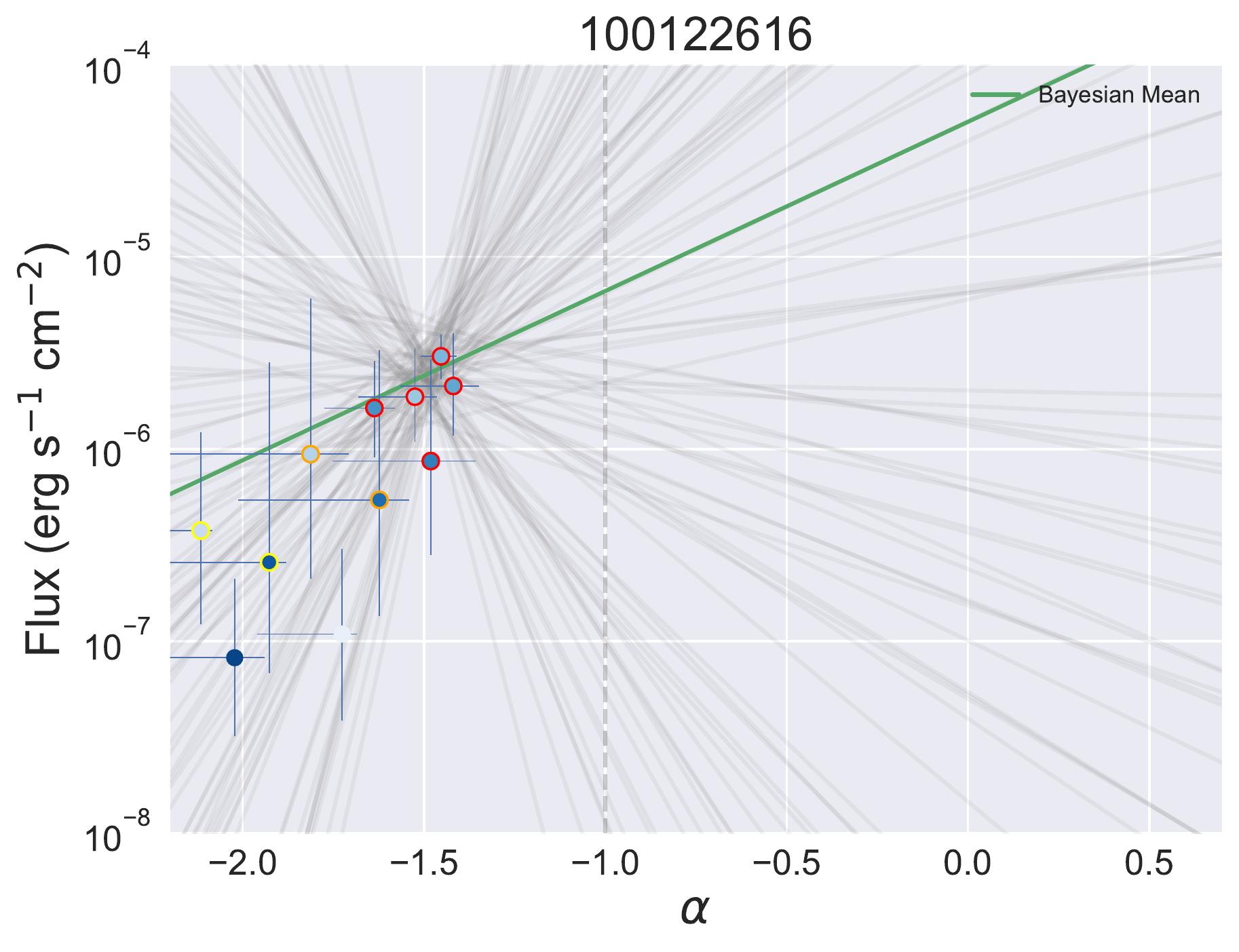}}
\subfigure{\includegraphics[width=0.33\linewidth]{BayesianFit_bn100528075.pdf}}

\subfigure{\includegraphics[width=0.33\linewidth]{BayesianFit_bn100612726.pdf}}
\subfigure{\includegraphics[width=0.33\linewidth]{BayesianFit_bn100707032.pdf}}
\subfigure{\includegraphics[width=0.33\linewidth]{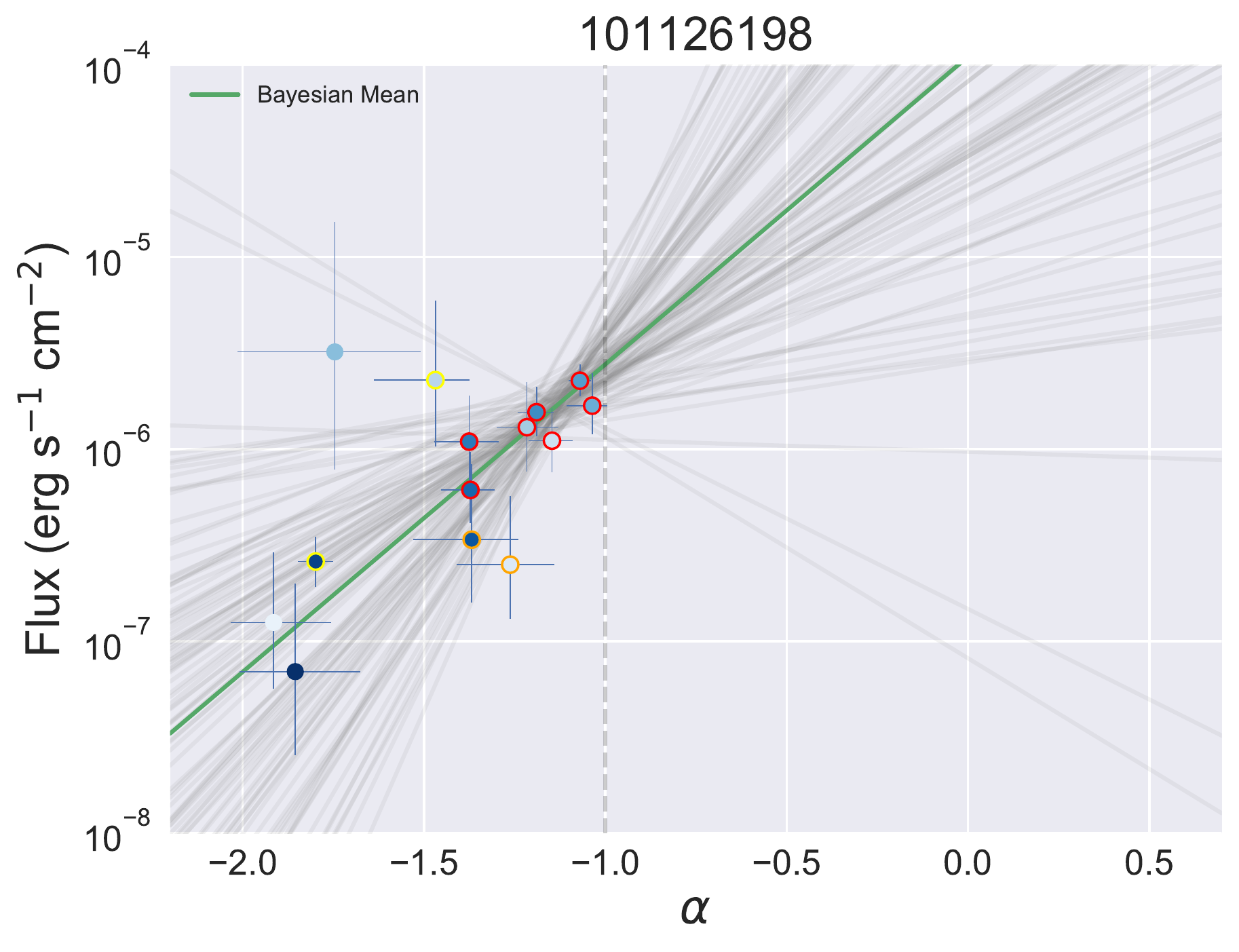}}

\caption{The $\alpha$-intensity correlation in GRB pulses. Same as Figure \ref{fig:figure1} but for the full sample of 38 pulses. The middle panel of the first row is for the second emission episode of GRB081009140.
\label{fig:figureB1}}
\end{figure*}

\begin{figure*}
\centering

\subfigure{\includegraphics[width=0.33\linewidth]{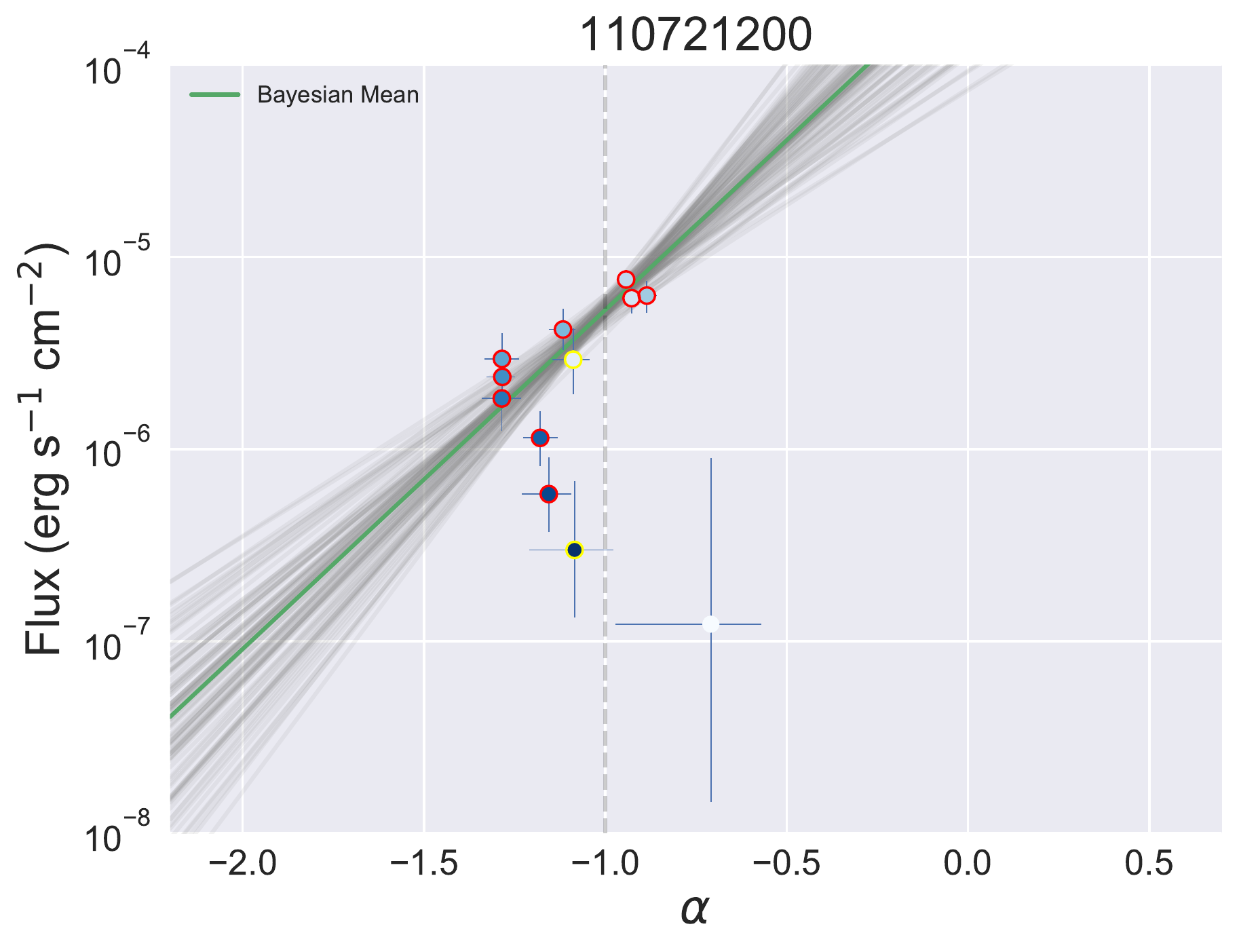}}
\subfigure{\includegraphics[width=0.33\linewidth]{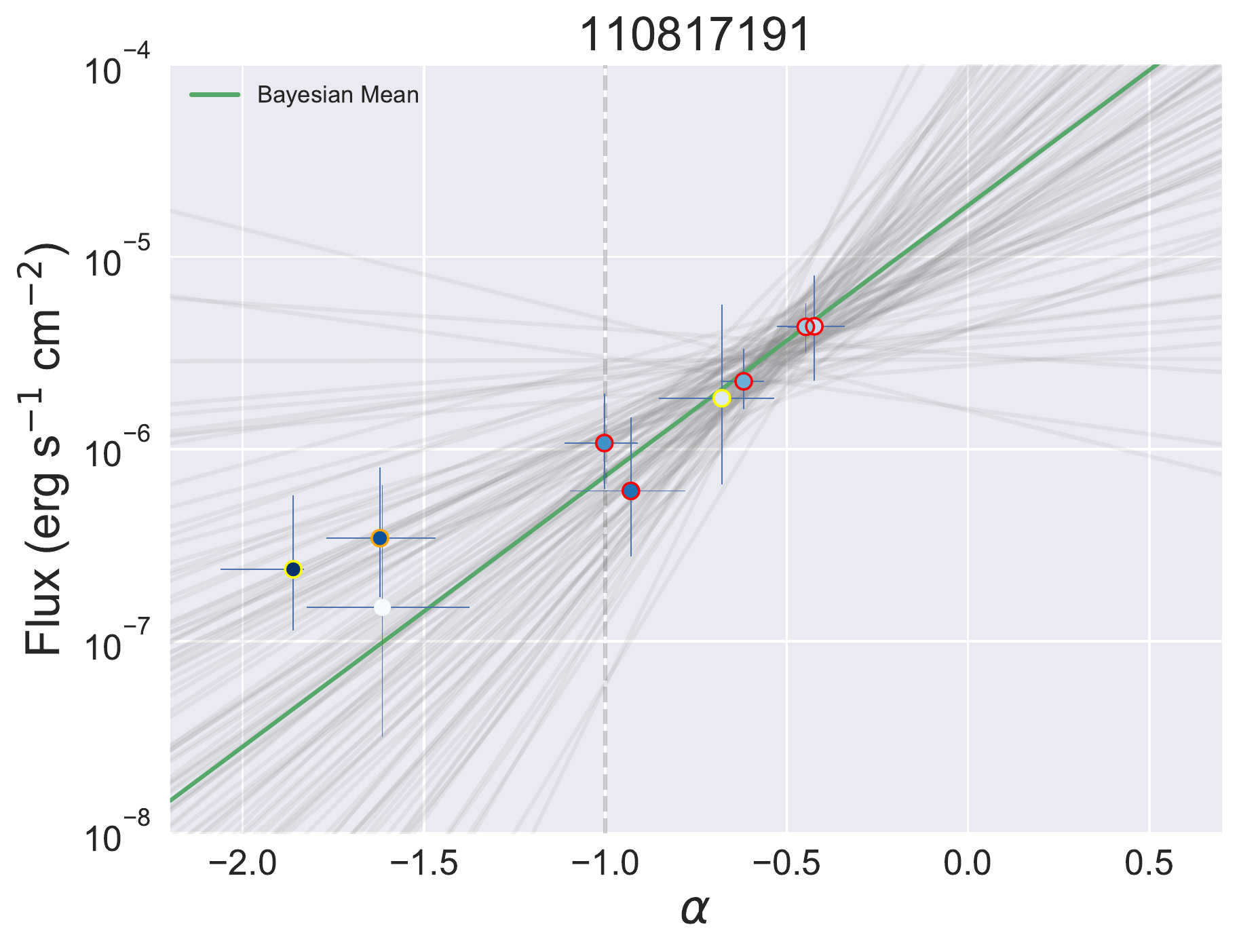}}
\subfigure{\includegraphics[width=0.33\linewidth]{BayesianFit_bn110920546.pdf}}

\subfigure{\includegraphics[width=0.33\linewidth]{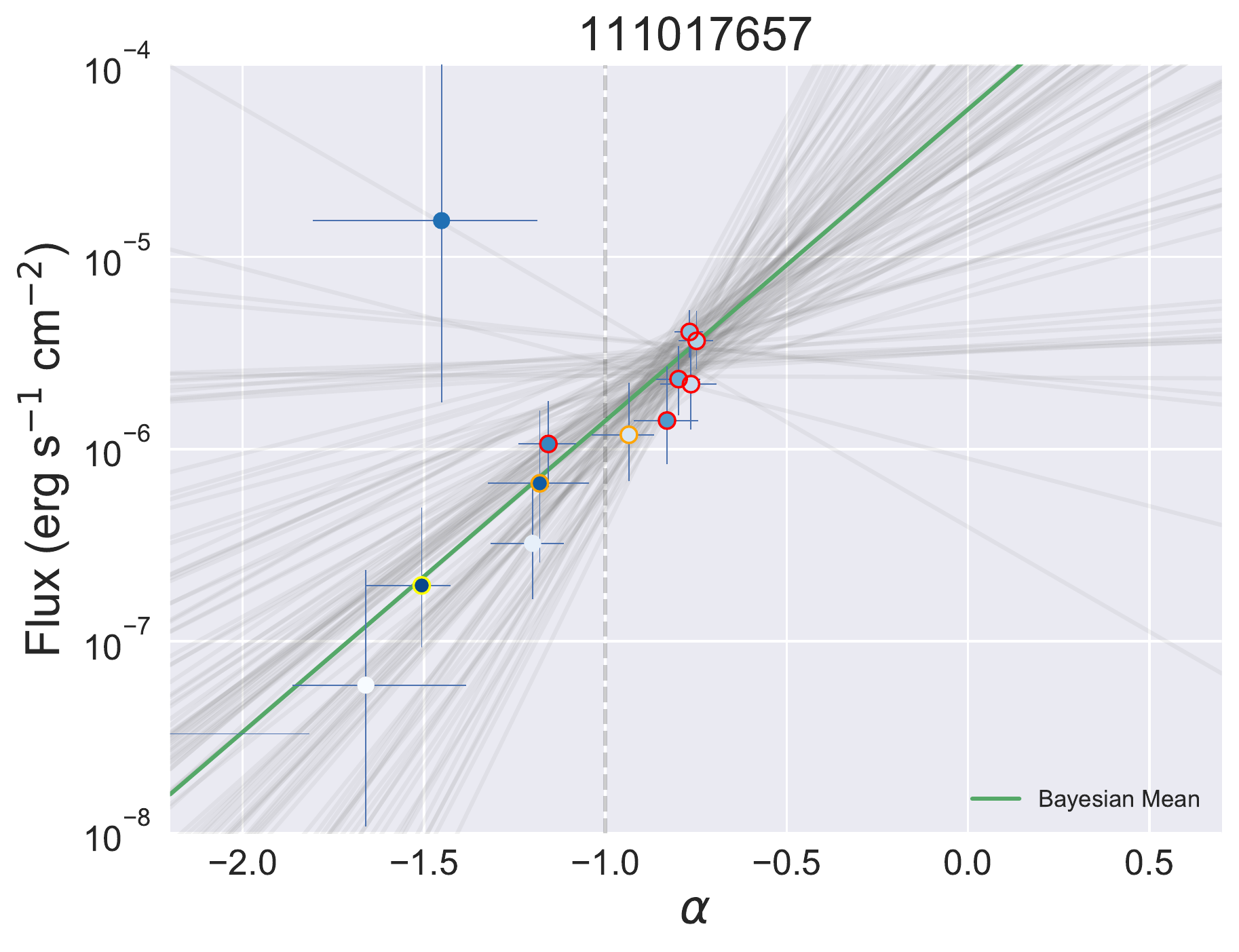}}
\subfigure{\includegraphics[width=0.33\linewidth]{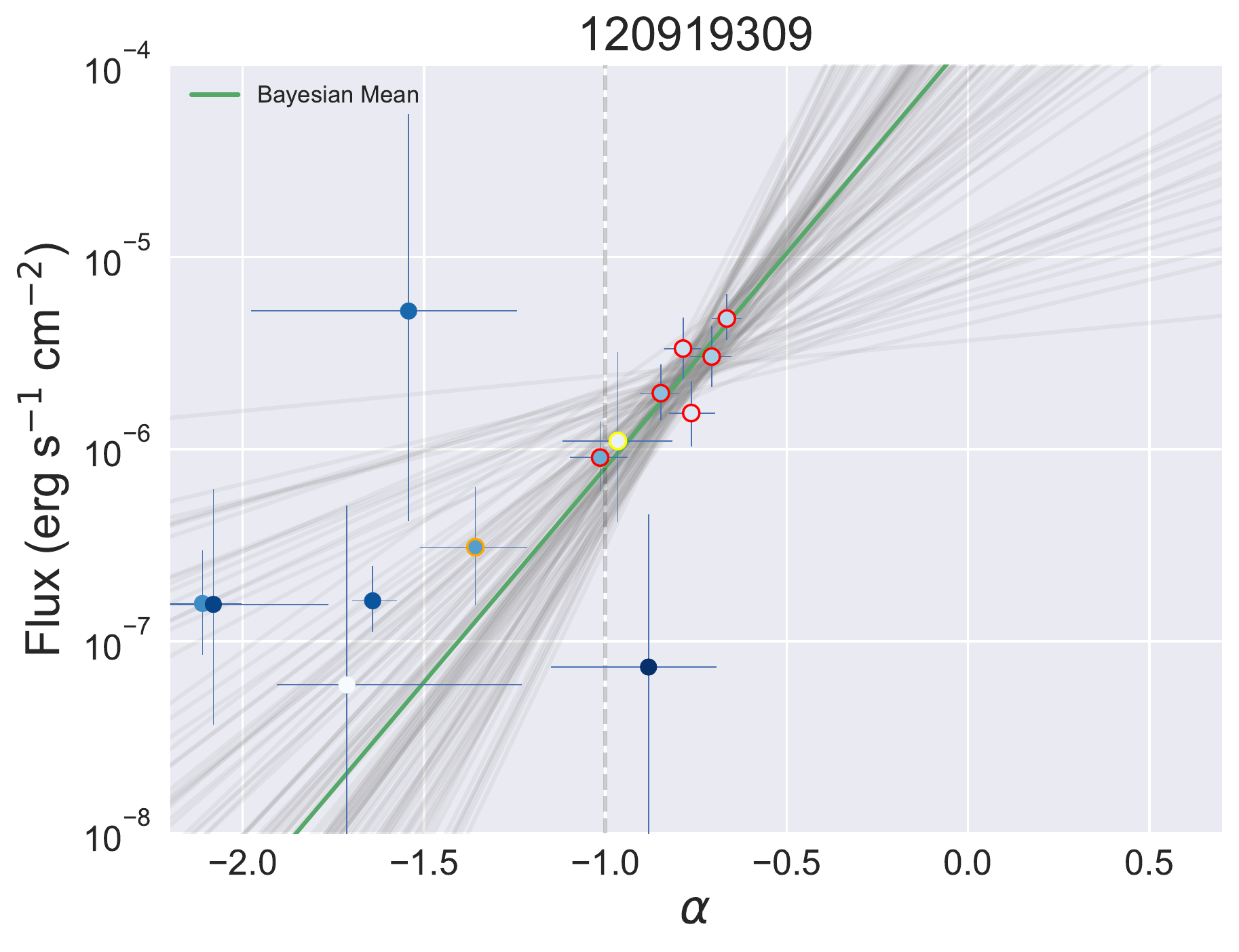}}
\subfigure{\includegraphics[width=0.33\linewidth]{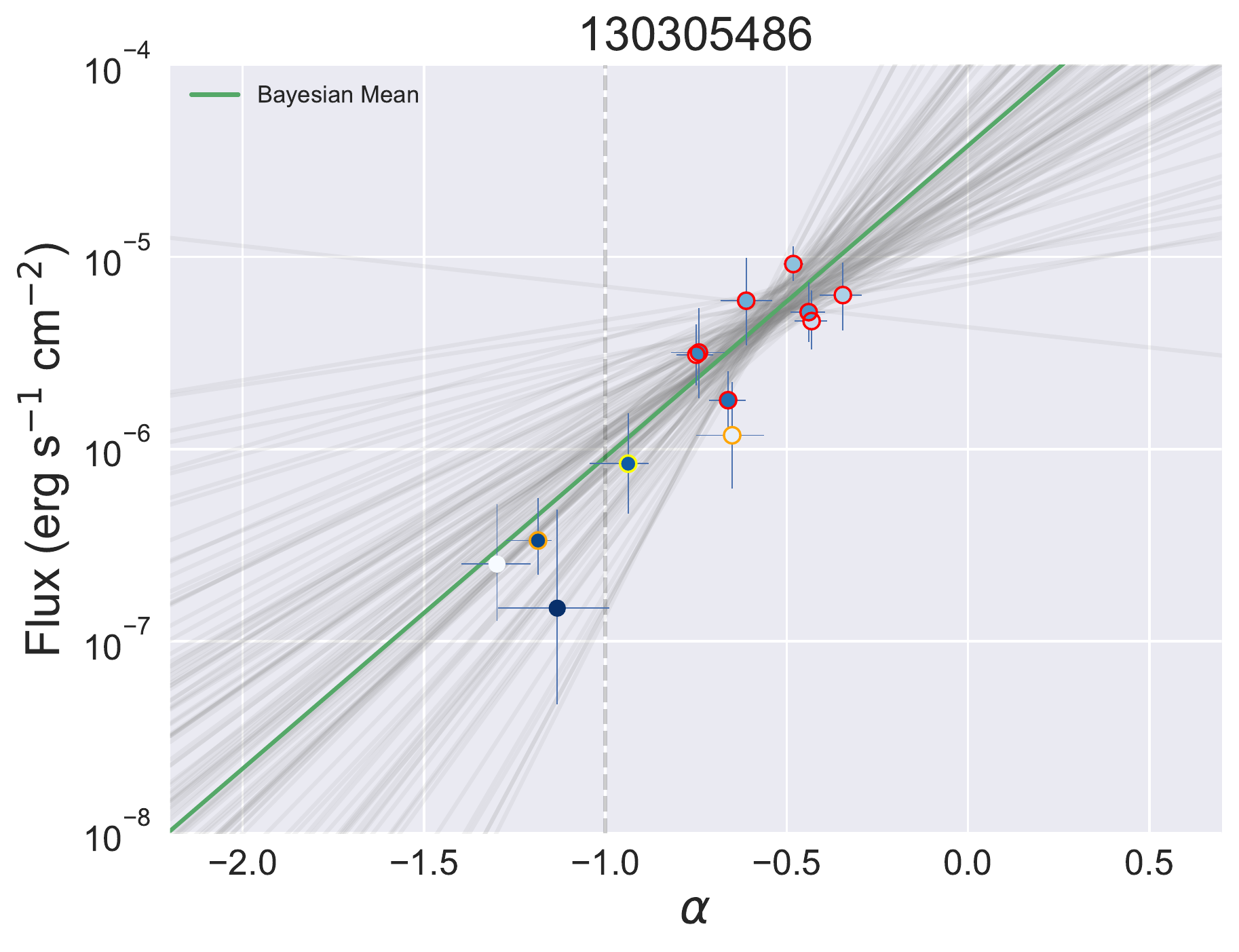}}

\subfigure{\includegraphics[width=0.33\linewidth]{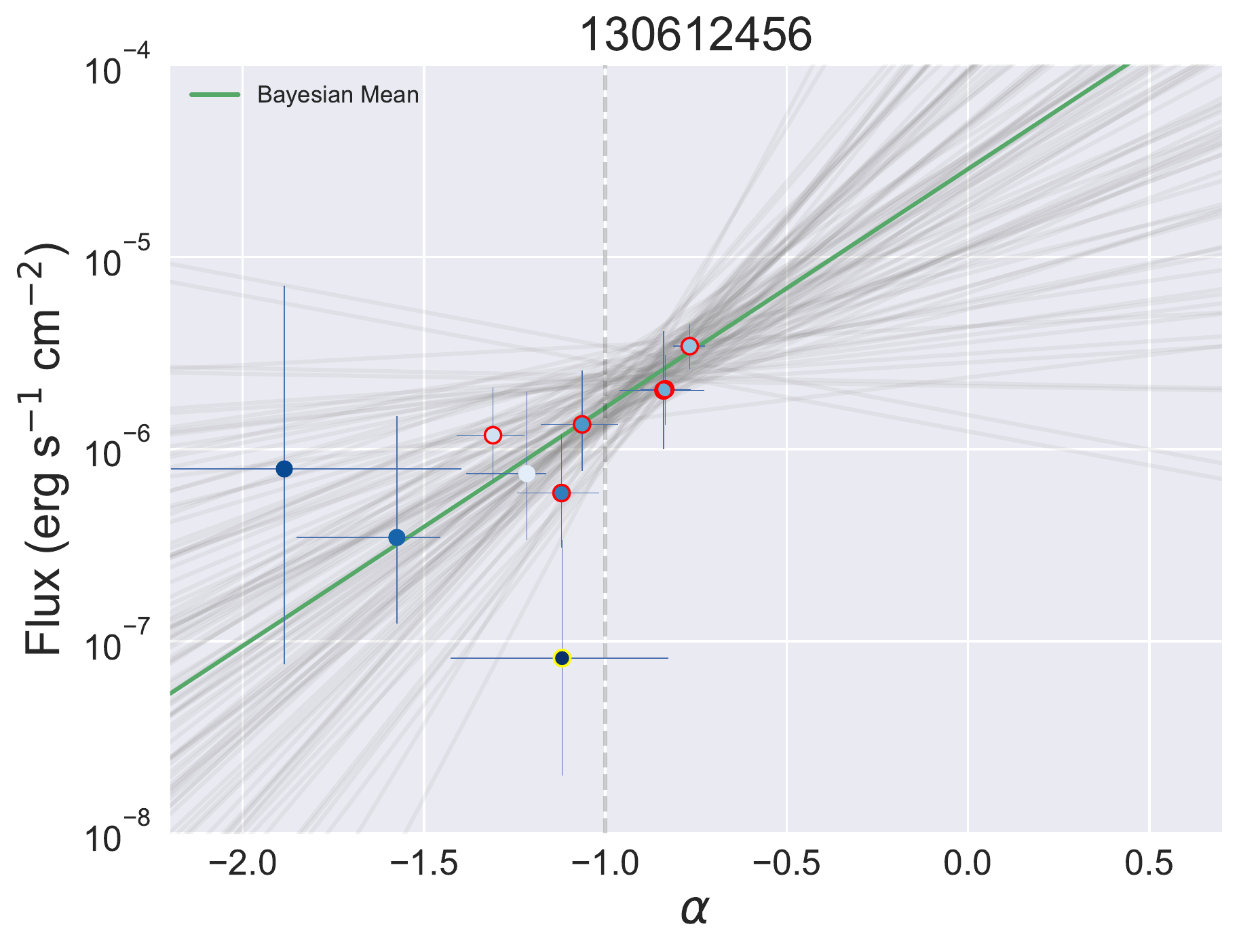}}
\subfigure{\includegraphics[width=0.33\linewidth]{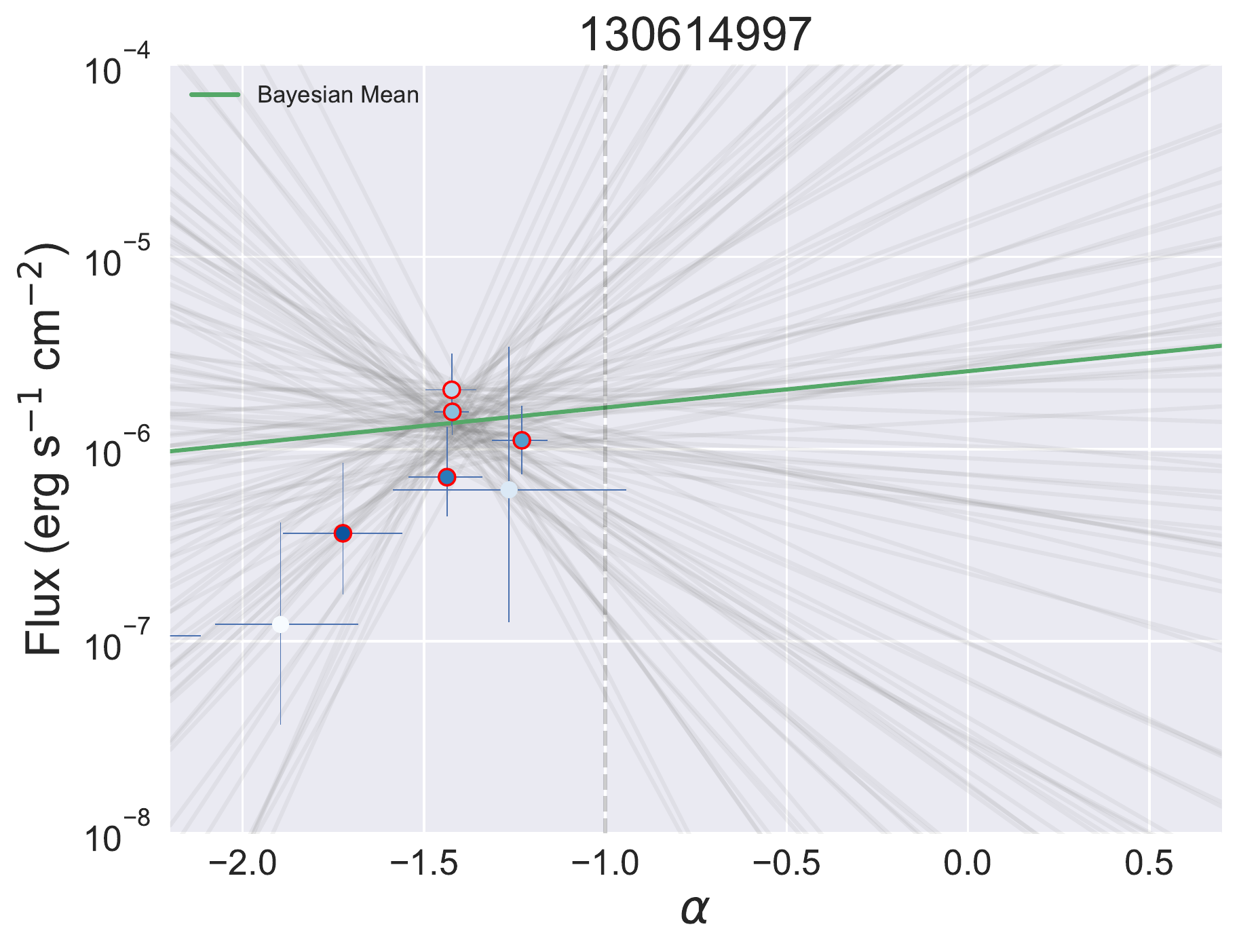}}
\subfigure{\includegraphics[width=0.33\linewidth]{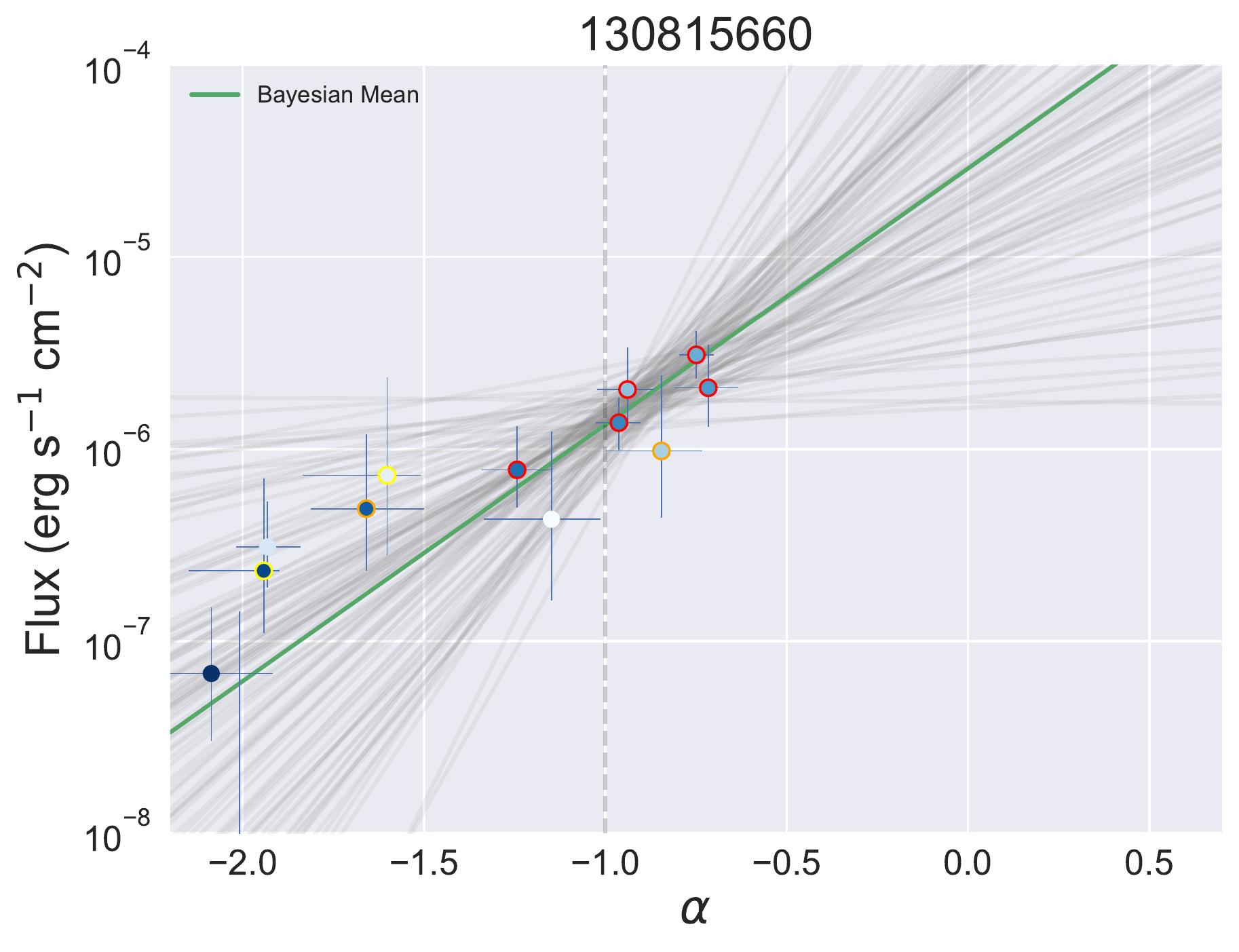}}

\subfigure{\includegraphics[width=0.33\linewidth]{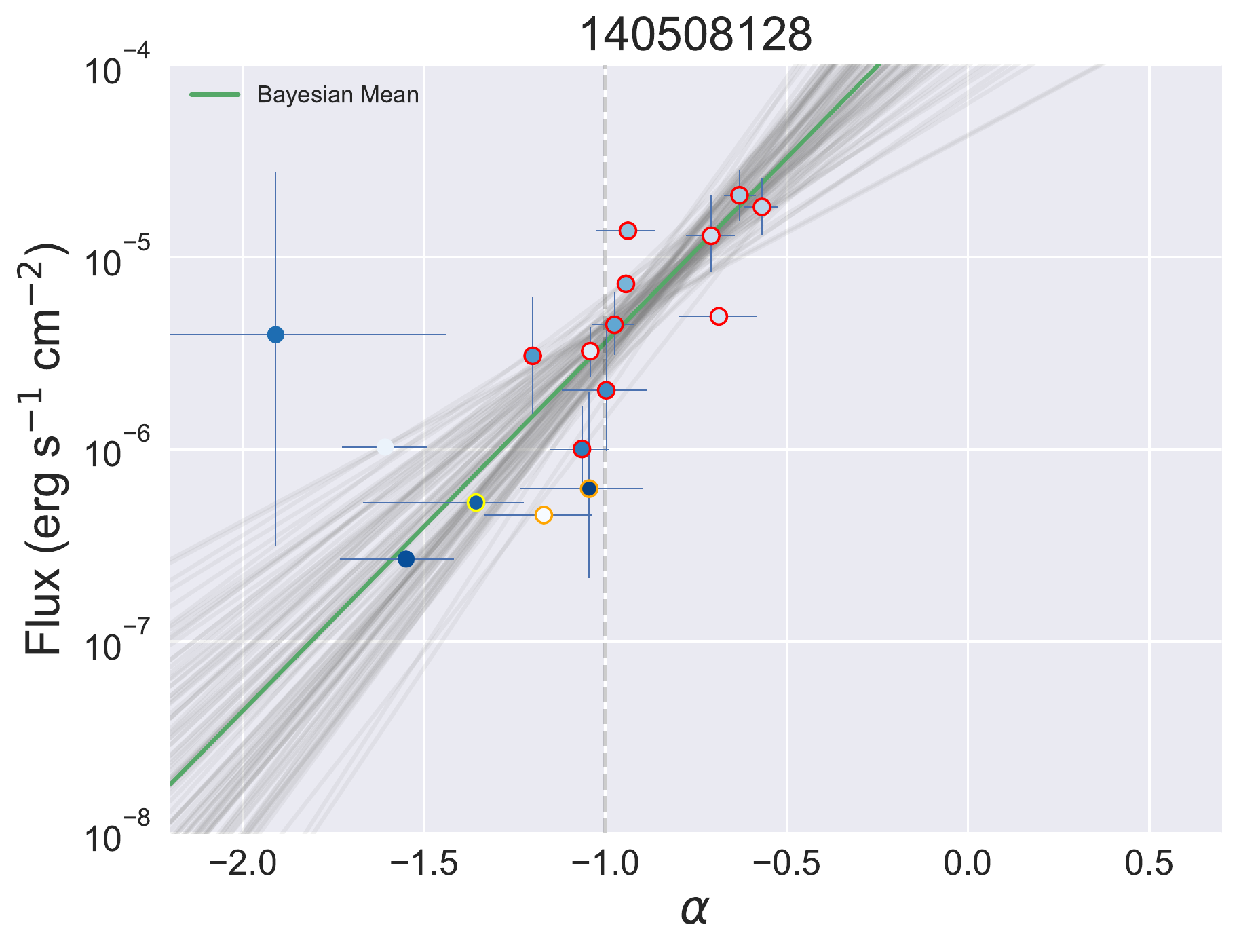}}
\subfigure{\includegraphics[width=0.33\linewidth]{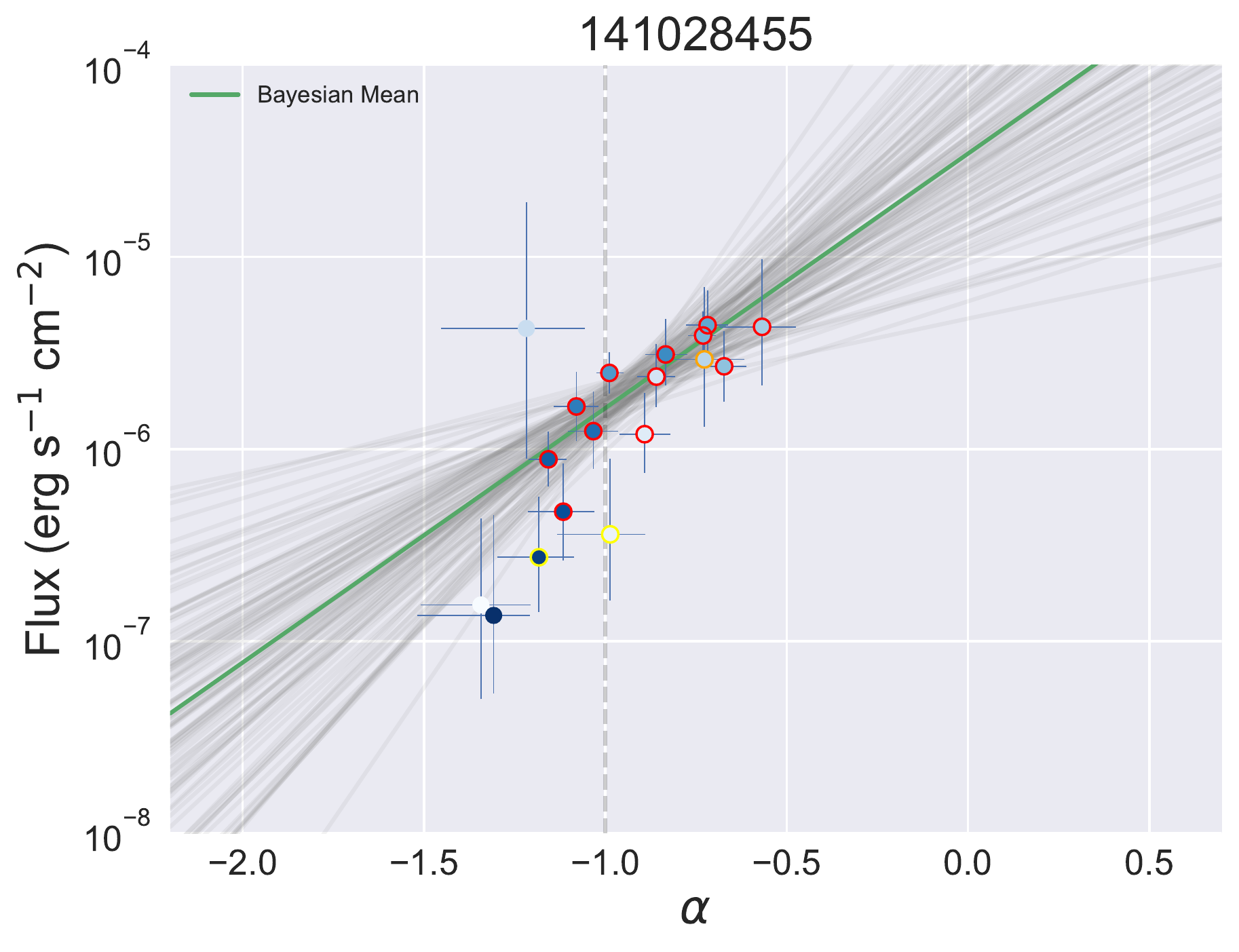}}
\subfigure{\includegraphics[width=0.33\linewidth]{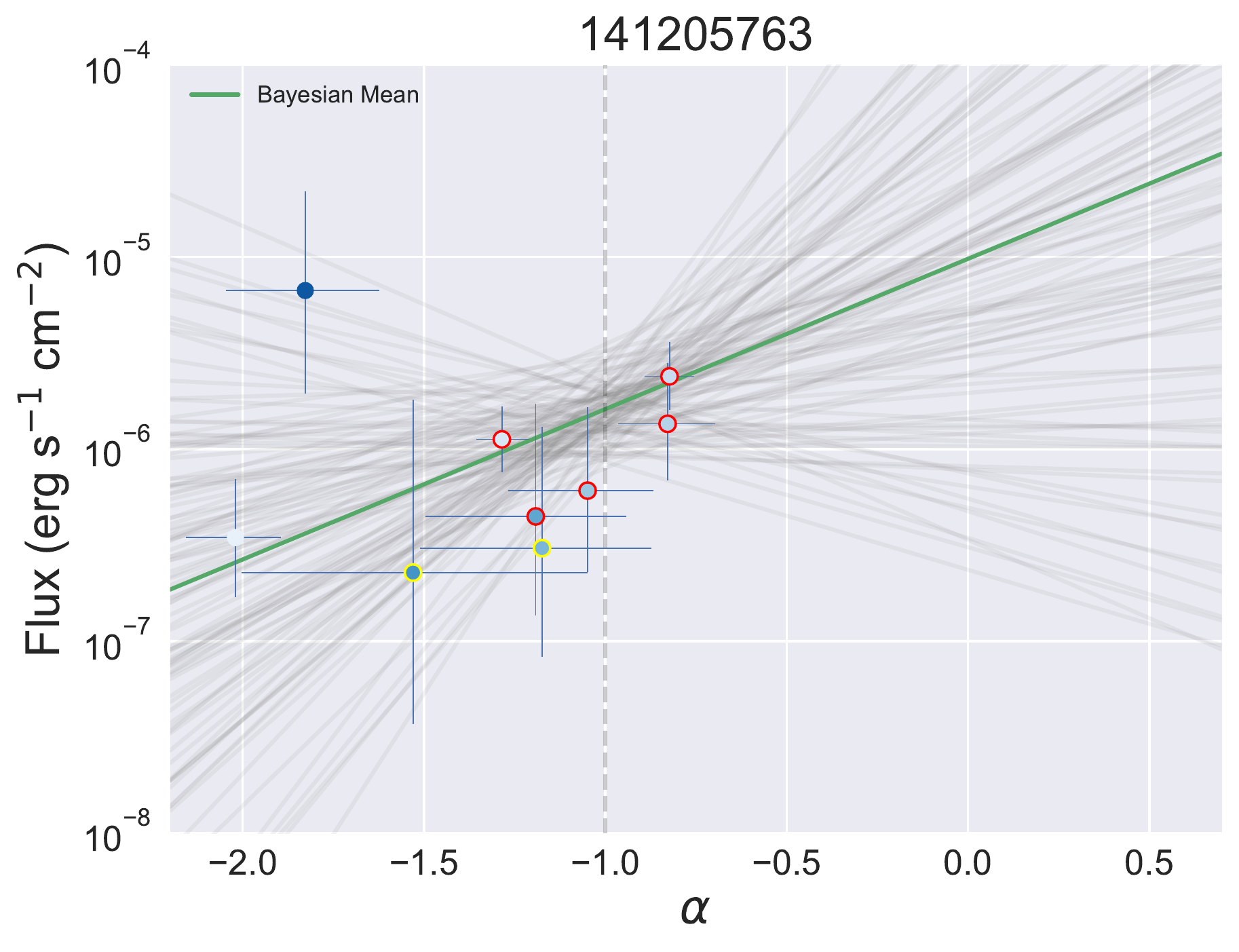}}

\subfigure{\includegraphics[width=0.33\linewidth]{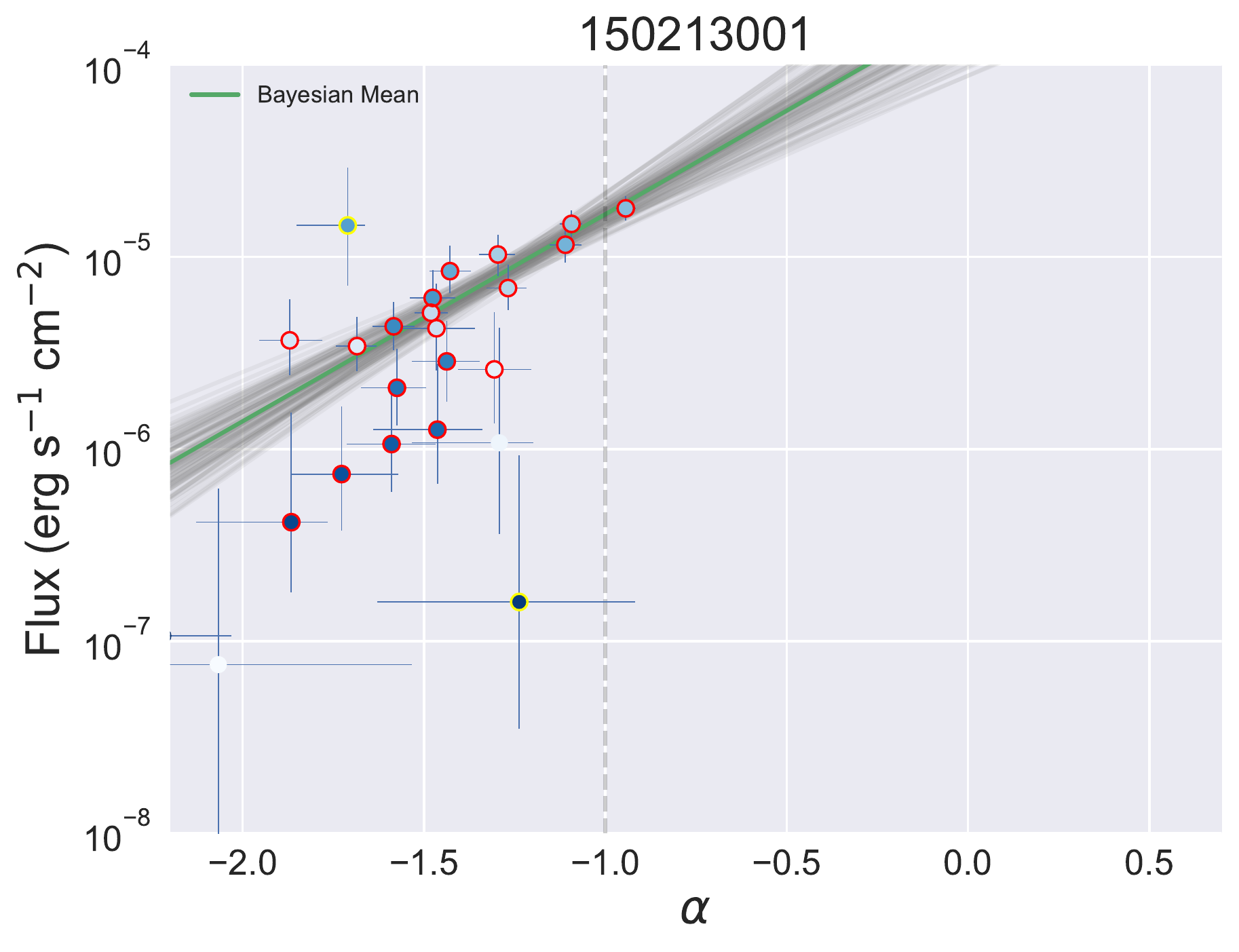}}
\subfigure{\includegraphics[width=0.33\linewidth]{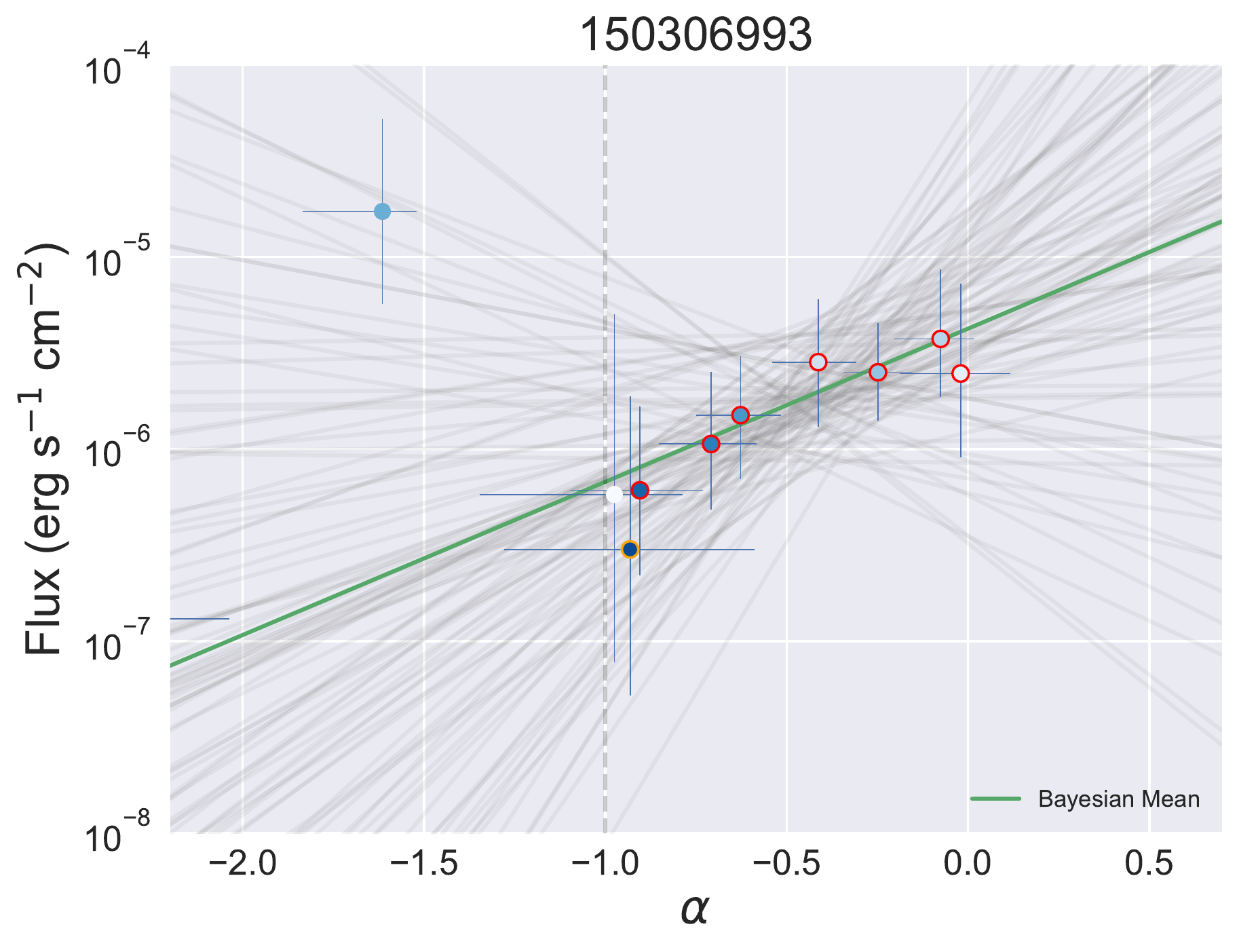}}
\subfigure{\includegraphics[width=0.33\linewidth]{BayesianFit_bn150314205.pdf}}

\caption{The $\alpha$-intensity correlation in GRB pulses. Same as Figure \ref{fig:figureB1}.
\label{fig:figureB2}}
\end{figure*}

\begin{figure*}
\centering

\subfigure{\includegraphics[width=0.33\linewidth]{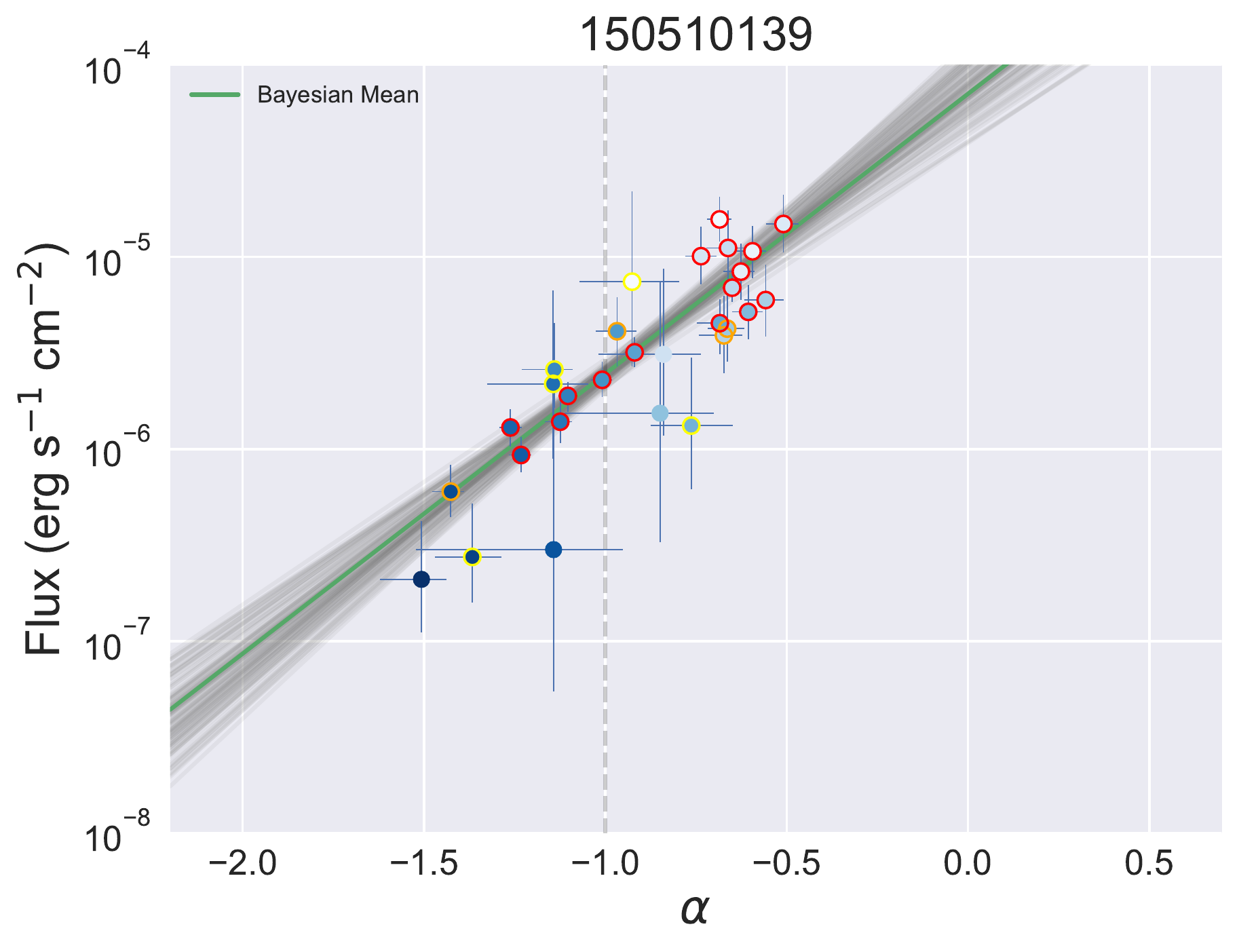}}
\subfigure{\includegraphics[width=0.33\linewidth]{BayesianFit_bn150902733.pdf}}
\subfigure{\includegraphics[width=0.33\linewidth]{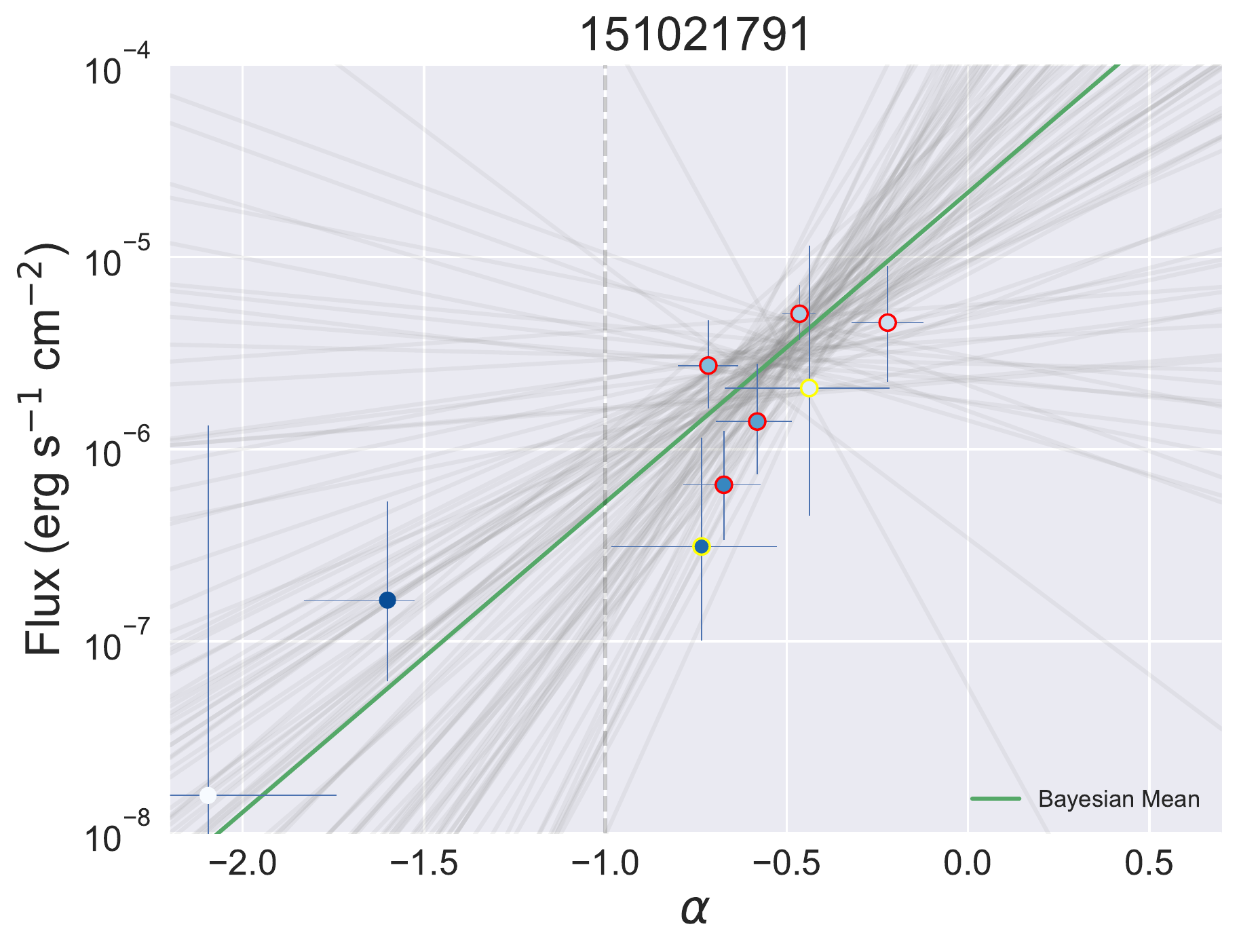}}

\subfigure{\includegraphics[width=0.33\linewidth]{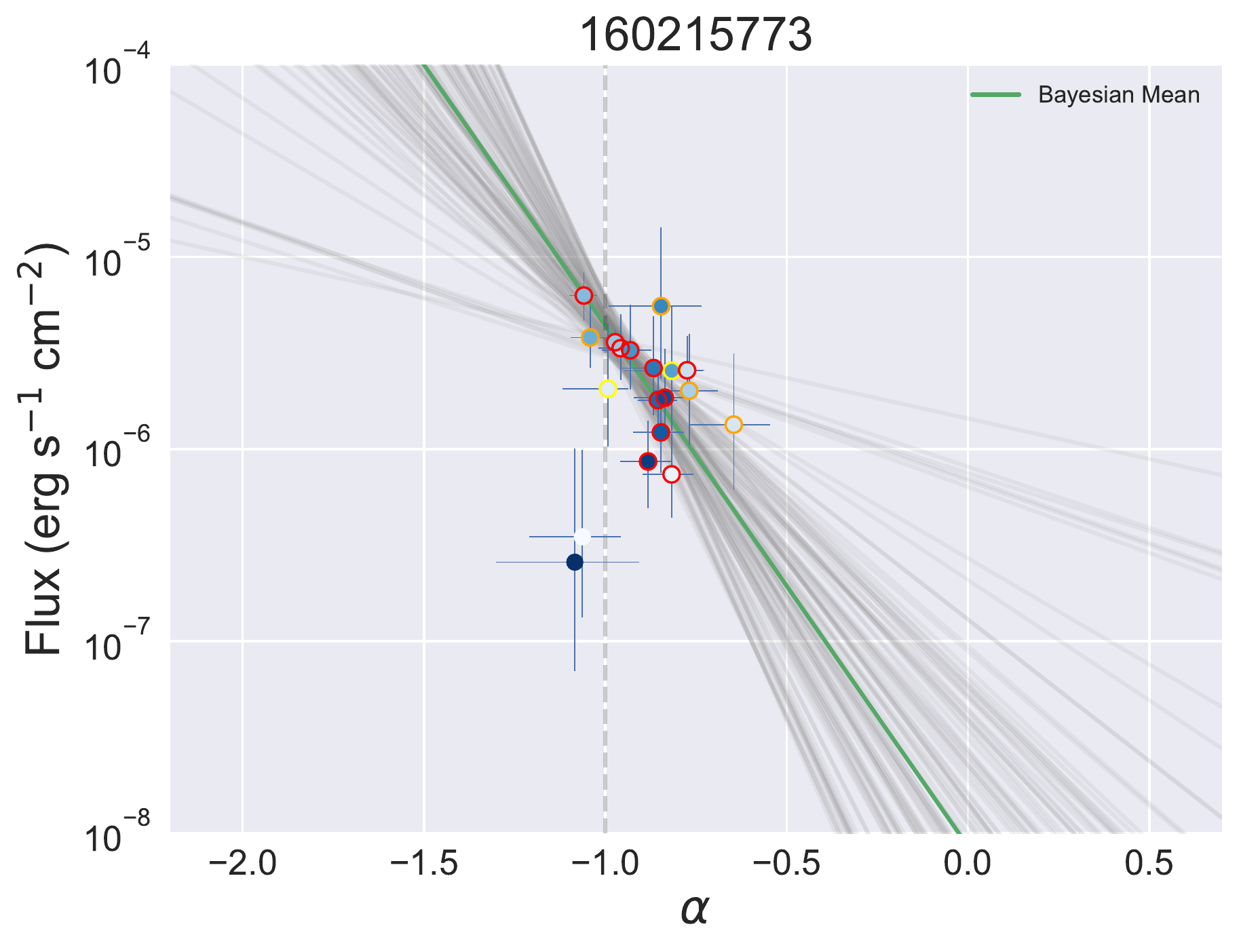}}
\subfigure{\includegraphics[width=0.33\linewidth]{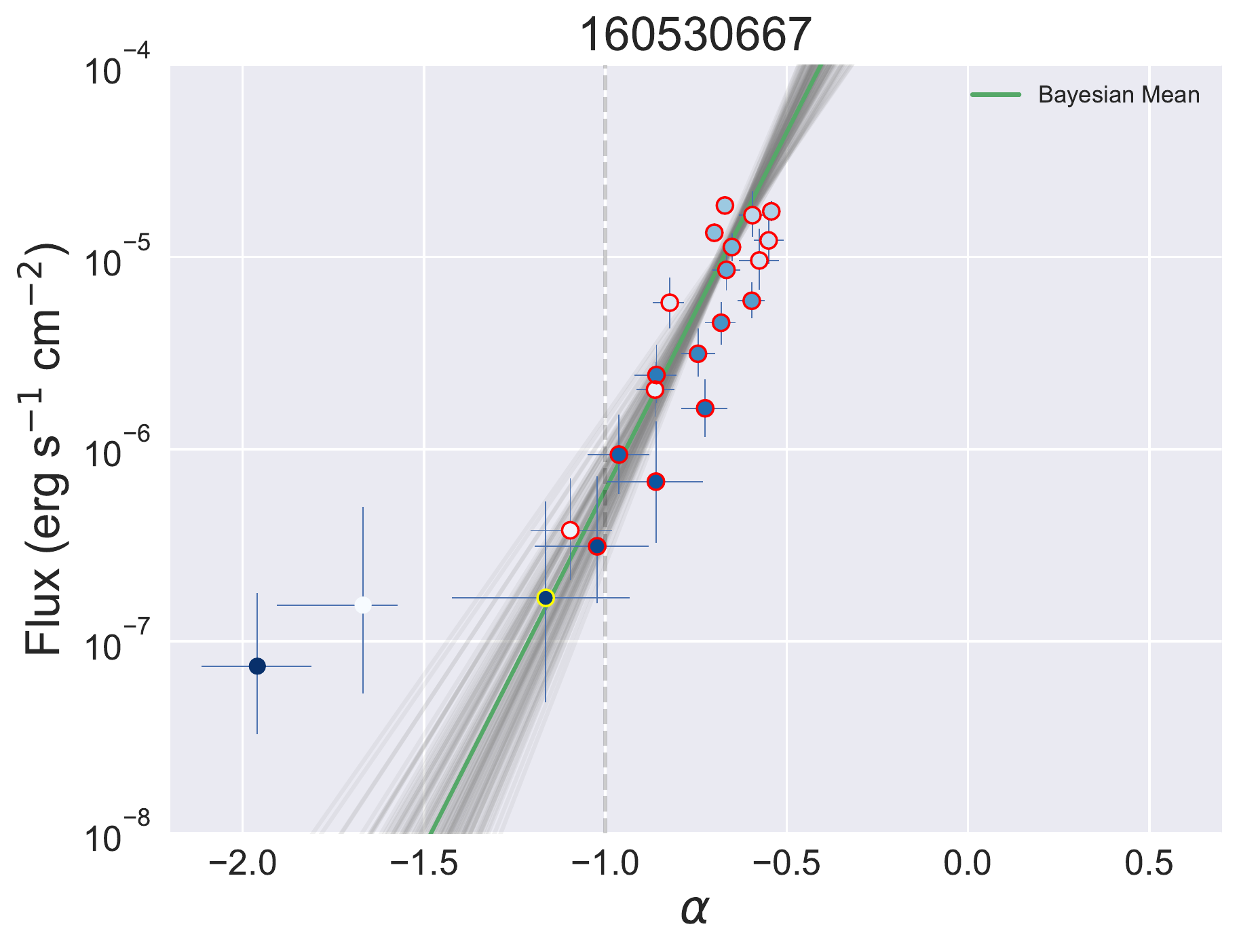}}
\subfigure{\includegraphics[width=0.33\linewidth]{BayesianFit_bn160910722.pdf}}

\subfigure{\includegraphics[width=0.33\linewidth]{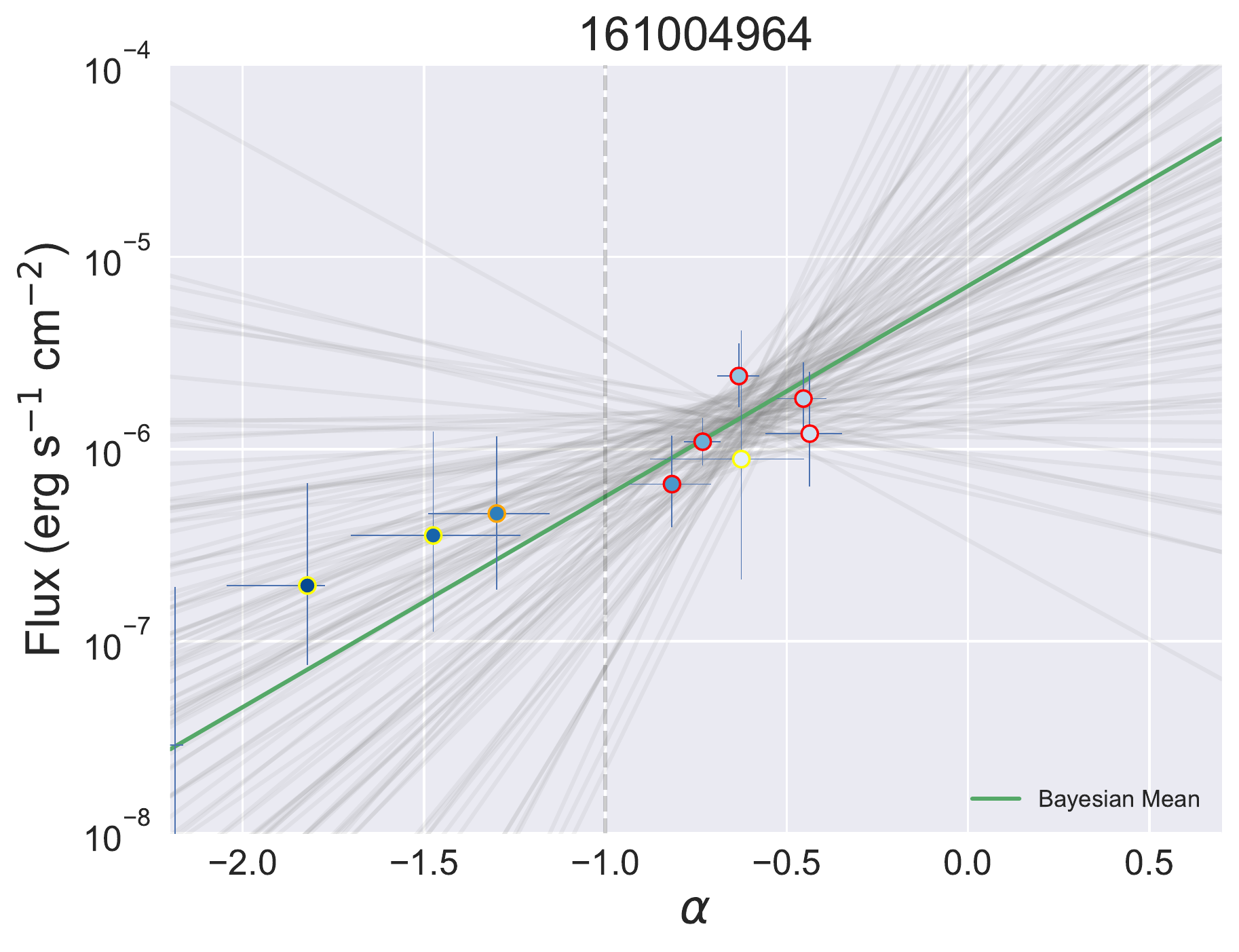}}
\subfigure{\includegraphics[width=0.33\linewidth]{BayesianFit_bn170114917.pdf}}

\caption{The $\alpha$-intensity correlation in GRB pulses. Same as Figure \ref{fig:figureB1}.
\label{fig:figureB3}}
\end{figure*}



\bsp	
\label{lastpage}
\end{document}